\documentclass[apj]{emulateapj}

\def\lapp{\ifmmode\stackrel{<}{_{\sim}}\else$\stackrel{<}{_{\sim}}$\fi}
\def\gapp{\ifmmode\stackrel{>}{_{\sim}}\else$\stackrel{>}{_{\sim}}$\fi}

\usepackage{graphicx}
\usepackage{float}
\usepackage{color}
\usepackage{enumitem}
\usepackage{amsmath}
\usepackage{amsfonts}
\definecolor{darkblue}{rgb}{0,0,0.7}
\usepackage[colorlinks,citecolor=darkblue]{hyperref}
\usepackage{natbib}
\newcommand{\ditto}{$-$\mbox{\tiny $\mid\mid$}$-$}

\submitted{Submitted to ApJ March 20, 2017; Accepted June 21, 2017}
\shorttitle{Formation of Double Neutron Star Systems}
\shortauthors{Tauris~et~al.}

\begin{document}

\title{FORMATION OF DOUBLE NEUTRON STAR SYSTEMS}

\author{
T.M.~Tauris\altaffilmark{1,2}, 
M.~Kramer\altaffilmark{1},
P.C.C.~Freire\altaffilmark{1},
N.~Wex\altaffilmark{1},
H.-T.~Janka\altaffilmark{3}, 
N.~Langer\altaffilmark{2}, 
Ph.~Podsiadlowski\altaffilmark{4,2}, \\
E.~Bozzo\altaffilmark{5},
S.~Chaty\altaffilmark{6,7}, 
M.U.~Kruckow\altaffilmark{2}, 
E.P.J.~van~den~Heuvel\altaffilmark{8}, \\
J.~Antoniadis\altaffilmark{1,9},
R.P.~Breton\altaffilmark{10},
D.J.~Champion\altaffilmark{1}
}

\altaffiltext{1}{Max-Planck-Institut f\"ur Radioastronomie, Auf dem H\"ugel 69, 53121 Bonn, Germany}
\altaffiltext{2}{Argelander-Institut f\"ur Astronomie, Universit\"at Bonn, Auf dem H\"ugel 71, 53121 Bonn, Germany}
\altaffiltext{3}{Max~Planck~Institut f\"ur Astrophysik, Karl-Schwarzschild-Str. 1, 85748~Garching, Germany}
\altaffiltext{4}{Department of Astronomy, Oxford University, Oxford OX1~3RH, UK}
\altaffiltext{5}{ISDC Data Centre for Astrophysics, Chemin dEcogia 16, CH-1290 Versoix, Switzerland}
\altaffiltext{6}{Laboratoire AIM (UMR\,7158 CEA/DSM-CNRS-Universit\'e Paris Diderot), Irfu/Service d'Astrophysique, Centre de Saclay, FR-91191 Gif-sur-Yvette Cedex, France}
\altaffiltext{7}{Institut Universitaire de France, 103 boulevard Saint-Michel, FR-75005 Paris, France}
\altaffiltext{8}{Astronomical Institute Anton Pannekoek, University of Amsterdam, P.O.~Box~94249, 1090~GE Amsterdam, The Netherlands}
\altaffiltext{9}{Dunlap Institute for Astronomy and Astrophysics, University of Toronto, 50~St George~Street, Toronto, ON~M5S~3H4, Canada}
\altaffiltext{10}{Jodrell Bank Centre for Astrophysics, School of Physics and Astronomy, The University of Manchester, Manchester M13 9PL, UK}

\begin{abstract}
Double neutron star (DNS) systems represent extreme physical objects and the endpoint of an exotic journey of stellar evolution and binary interactions.
Large numbers of DNS systems and their mergers are anticipated to be discovered using the Square-Kilometre-Array searching for radio pulsars and high-frequency gravitational wave detectors (LIGO/VIRGO), respectively.
Here we discuss all key properties of DNS systems, as well as selection effects, and combine the latest observational data with new theoretical progress on various physical processes with the aim 
of advancing our knowledge on their formation. We examine key interactions of their progenitor systems and evaluate their accretion history during the high-mass X-ray binary stage, 
the common envelope phase and the subsequent Case BB mass transfer, and argue that the first-formed NSs have accreted at most $\sim 0.02\;M_{\odot}$.
We investigate DNS masses, spins and velocities, and in particular correlations between spin period, orbital period and eccentricity.  
Numerous Monte~Carlo simulations of the second supernova (SN) events are performed to extrapolate pre-SN stellar properties and probe the explosions. 
All known close-orbit DNS systems are consistent with ultra-stripped exploding stars.
Although their resulting NS kicks are often small, we demonstrate a large spread in kick magnitudes which may, in general, 
depend on the past interaction history of the exploding star and thus correlate with the NS mass.
We analyze and discuss NS kick directions based on our SN simulations.
Finally, we discuss the terminal evolution of close-orbit DNS systems until they merge and possibly produce a short $\gamma$-ray burst.
\end{abstract}

\email{email: tauris@astro.uni-bonn.de}

%%%%%%%%%%%%%%%%%%%%%%%%%%%%%%%%%%%%%%%%%%%%%%%%%%%%%%%%%%%%%%%%%%%%%%%%%%%
\section{Introduction}\label{sec:intro}
\setcounter{footnote}{0}
The evolution of massive binary stars and the subsequent production of pairs of compact objects in tight orbits
play a central role in many areas of modern astrophysics, including the origin of different types of supernova (SN) explosions \citep{wi73,it84,ywl10}, 
possibly including long~gamma-ray bursts \citep[GRBs,][]{woo93,cyll07}, the modelling of accretion processes in X-ray binaries \citep{lv06} 
and the formation of millisecond pulsars \citep[MSPs,][]{bv91,tv06,pw12}. In some cases, we even expect massive binary systems to terminate as 
spectacular collisions between neutron stars (NSs) and/or black holes (BHs). These events may lead to 
short~GRBs \citep{elps89} and thereby chemical enrichment of the interstellar medium by heavy $r$-process elements \citep[e.g.][]{ros15,jbp+15},
aside from powerful emission of gravitational waves (GWs) as recently detected by LIGO \citep{aaa+16,aaa+16b}.
For a general review of pre-SN massive star evolution in binaries, see \citet{lan12}. 

Double neutron star (DNS) systems are of special interest since some of them are observable as radio pulsars \citep{lk04} for several millions, if not billions,  
of years before the NSs eventually collide as a result of orbital GW damping, in case their orbital periods are small enough (i.e. less than about 1~day, depending on the orbital eccentricity).
Moreover, given that the detected DNS systems represent the endpoint of exotic binary stellar evolution in which progenitor systems have survived multiple stages of mass transfer,
one or more common-envelope (CE) episodes, as well as two SN explosions \citep[e.g.][]{fv75,bkb02,vt03,dp03,plp+04,pdl+05,dpp05,tv06}, 
their observed properties are fossil records of their past evolutionary history. Therefore, DNS systems can be used as key probes of binary stellar astrophysics. 

Furthermore, the combination of compact objects in relativistic orbits and radio pulsars being ultra stable clocks  
allows for unprecedented tests of gravitational theories in the strong-field regime \citep{dt92,sta03,ksm+06,kw09,dam09,fwe+12,wex14,kk16}.
Finally, DNS systems help to constrain the long-sought-after equation-of-state (EoS) 
of nuclear matter at high densities \citep{lp07,kwkl16,of16}, although tighter constraints have so far been
obtained from massive NSs with an orbiting white dwarf (WD) companion \citep{dpr+10,afw+13}. 

In recent years, the discovery rate of DNS systems has increased and there are now more than 15 DNS systems known. 
The coming decade is expected to reveal a large number of discoveries of new DNS systems, as well as their progenitors and merger remnants. 
The Square-Kilometre Array (SKA) is predicted to increase the number of known radio pulsars by a factor of 5 to 10, thus resulting in a total of more than 100 known DNS systems \citep{kbk+15}. 
The Five-hundred-meter Aperture Spherical Telescope (FAST) is also expected to contribute with a significant number of new radio pulsars \citep{slk+09}, including new discoveries of DNS binaries. 
Indirect evidence of DNS mergers is thought to come from detections of short~GRBs and their optical afterglows \citep{ber14,met16}. 
Moreover, following the first detections of high-frequency GWs from merging double BH binaries \citep[e.g.][]{aaa+16,aaa+16b},
LIGO/VIRGO is also expected to deliver direct evidence of merging DNS systems in a near future \citep{aaa+10}. 
The progenitors of DNS systems, the high-mass X-ray binaries (HMXBs), are continuously being discovered with ongoing X-ray missions \citep[INTEGRAL, Swift, XMM and Chandra, see e.g.][]{cha13}
and will be supported by upcoming X-ray telescopes like eROSITA, eXTP and Athena.
Hence, we are currently in an epoch where a large wealth of information on NS binaries is available. In light of this, it is important
to explore and understand the formation of DNS systems in more detail. 

Before presenting the scope of this paper in more detail (Section~\ref{subsec:scope}), in the following subsections we summarize the
background, the current state-of-the-art and the main motivations for this investigation.

\subsection{Galactic and Magellanic Cloud populations of NSs}\label{subsec:local_NSs}
In our galaxy, the Milky Way, it is estimated that some $10^8-10^9$ NSs are present. As a rough estimate, one can simply assume a constant NS 
formation rate of 2 per century for 10~Gyr. However, we can only detect a tiny fraction of these --- 
namely those which are rapidly spinning and possess strong magnetic fields (thereby producing radio pulsars which may beam in our direction), 
those which are accreting material from a companion star in a close binary (giving rise to emission of X-rays), or
those which are nearby, young and hot (giving rise to thermal emission). 
According to the ATNF Pulsar Catalogue\footnote{http://www.atnf.csiro.au/research/pulsar/psrcat/} \citep{mhth05}, there are $\sim\!2600$ Galactic radio pulsars detected at present (June~2017), 
out of a total known population of roughly 3000 NSs in the Milky Way.
The two Magellanic Clouds host some $\sim\!30$ known radio pulsars \citep{ckm+01,rcl+13}, besides from an impressive number of $\sim\!140$ HMXBs \citep{ck15,az16},
some of which are potential progenitors of DNS systems. 

\subsection{Recent DNS discoveries}\label{subsec:recentDNS}
Among the new DNS systems recently discovered, the most notably are PSRs~J1930$-$1852 \citep{srm+15}, J0453+1559 \citep{msf+15} and J1913+1102 \citep{lfa+16}. 
PSR~J1930$-$1852 has an orbital period of 45~days and is by far the widest DNS system known.
At the same time this radio pulsar has the slowest spin period (185~ms) of any recycled pulsar in a DNS system. This system thus provides
important data for understanding the formation of wide-orbit DNS systems.
PSR~J0453+1559 is a DNS system with an asymmetric pair of NS masses (1.56 and $1.17\;M_{\odot}$), 
deviating significantly from the ``canonical'' value of $\sim\!1.35\;M_{\odot}$ \citep{tc99}. This system is the first to finally break out of the 
narrowly limited observed NS mass interval, which was previously found to be $1.23-1.44\;M_{\odot}$ for all NSs in DNS systems,    
thereby alleviating a theoretical conundrum on NS birth masses \citep[e.g.][and references therein]{spr10,tlk11}.
A third new DNS binary is PSR~J1913+1102 \citep{lfa+16} which has the largest total mass of any known DNS system, thus potentially yielding another asymmetric-mass system.
Its GW merger timescale ($\sim\!480\;{\rm Myr}$) is on the short side. 

\subsection{Main characteristics of DNS systems}\label{subsec:characteristics}
To understand the formation of DNS systems from a binary stellar evolution point of view, it is important to consider all data provided by observations
and their derived parameters, e.g. 
NS masses ($M_{\rm NS}$), spin periods ($P$), their time derivatives ($\dot{P}$) and hence surface magnetic fields ($B$), 
orbital periods ($P_{\rm orb}$), eccentricities ($e$), as well as kinematic properties such as systemic velocities ($v_{\rm sys}$) and their Galactic locations. 

Geometric constraints (e.g. misalignment angles between the spin axis of the recycled radio pulsar and the orbital angular momentum vector) are also important for
probing the SN explosion. Direct relations between DNS systems and potential SN remnants (none detected so far) would be helpful to constrain their ages.
In addition, any correlations between, for example, $P_{\rm orb}$, $P$ and eccentricity are crucial for understanding the
recycling process and needs to be investigated throughly --- one of the major aims of this paper. 

In the vast majority of DNS systems, only the (mildly) recycled radio pulsar is seen. The reason
being that the mass accretion process decreases the B-field strength of the NS significantly \citep{bha02}, 
whereby the loss rate of rotational energy becomes much lower due to weaker braking torques from pulsar winds, magnetic dipole radiation 
and plasma currents in the magnetosphere, thus making the NS observable as a radio pulsar on a much longer timescale \citep{sv82}.
This is particularly so because the B-fields of recycled pulsars apparently do not decay further \citep{kul86,vvt86}. 
For these recycled DNS pulsars, the values of $P$ ($23-185\;{\rm ms}$) and their 
derivatives ($\dot{P}=2.2\times 10^{-20}-1.8\times 10^{-17}\;{\rm s\,s}^{-1}$) are direct measures of the efficiency of the 
recycling process. The corresponding derived surface dipole B-fields are approximately in the range $0.3-18\times 10^{9}\;{\rm G}$.
Table~\ref{table:obs_char} provides the observed ranges of key parameters of DNS systems. 

The observed orbital periods now span between $P_{\rm orb}=0.10-45\;{\rm days}$. As we shall discuss in this paper, 
selection effects make it somewhat difficult to detect radio pulsars in DNS systems at the lower end of this range, as well as second-born pulsars with slow spin periods of a few seconds. 
The measured eccentricities ($e=0.08-0.83$) and estimated systemic velocities, in the local standard of rest reference frame ($v_{\rm sys}=v^{\rm LSR}\approx 25-240\;{\rm km\,s}^{-1}$), 
are important for constraining the SN physics of the second-born NS 
and reflect the combined effects of the amount of mass ejected and the kick velocity imparted onto the newborn NS.

\begin{table}[b]
  \caption[]{Observed ranges of key properties of DNS systems.}
  \begin{center}
  \begin{tabular}{ll}
        \hline
        \hline
	\noalign{\smallskip}
	\noalign{\smallskip}
	   Properties of recycled (old) NSs: & \\
           $\qquad$Spin period, $P$	        	  	 & $23-185\;{\rm ms}$ \\
           $\qquad$Period derivative, $\dot{P}$ 		 & $(0.027-18)\times10^{-18}\;{\rm s\,s}^{-1}$ \\
           $\qquad$Surface dipole B-field, $B$                   & $(0.29-18)\times10^{9}\;{\rm G}$ \\
           $\qquad$Mass, $M_{\rm NS,1}$     			 & $1.32^{\ast}-1.56\;M_{\odot}$ \\ 
	   Properties of young NSs: & \\
           $\qquad$Spin period, $P$  		      		 & $144-2773\;{\rm ms}$ \\
           $\qquad$Period derivative, $\dot{P}$ 		 & $(0.89-20)\times10^{-15}\;{\rm s\,s}^{-1}$ \\
           $\qquad$Surface dipole B-field, $B$                   & $(2.7-5.3)\times10^{11}\;{\rm G}$ \\
           $\qquad$Mass, $M_{\rm NS,2}$              		 & $1.17-1.39\;M_{\odot}$ \\ 
	   Orbital properties: & \\
           $\qquad$Orbital period, $P_{\rm orb}$                 & $0.10-45\;{\rm days}$ \\
           $\qquad$Eccentricity, $e$                             & $0.085-0.83$\\
           $\qquad$Merger time, $\tau_{\rm gwr}$                 & $86\;{\rm Myr} \rightarrow \infty$\\
           $\qquad$Systemic velocity, $v_{\rm sys}$              & $25-240\;{\rm km\,s}^{-1}$\\
	\noalign{\smallskip}
        \hline
        \hline
  \end{tabular}
\end{center}
%\begin{flushleft}
       {\bf Notes} --- Data taken from the ATNF Pulsar Catalogue \citep{mhth05} --- see Table~\ref{table:DNS} for further details.
       Only DNS systems in the Galactic disk are listed.
       The systemic recoil velocity, $v_{\rm sys}=v^{\rm LSR}$ is quoted with respect to the local standard of rest (Section~\ref{subsec:vsys}). 
       $^{\ast}$ marks an upper limit to the lowest mass of the first-born NS.
%\end{flushleft}
\label{table:obs_char}
\end{table}

\begin{figure}[h]
  \begin{center}
    \includegraphics[width=1.04\columnwidth, angle=0]{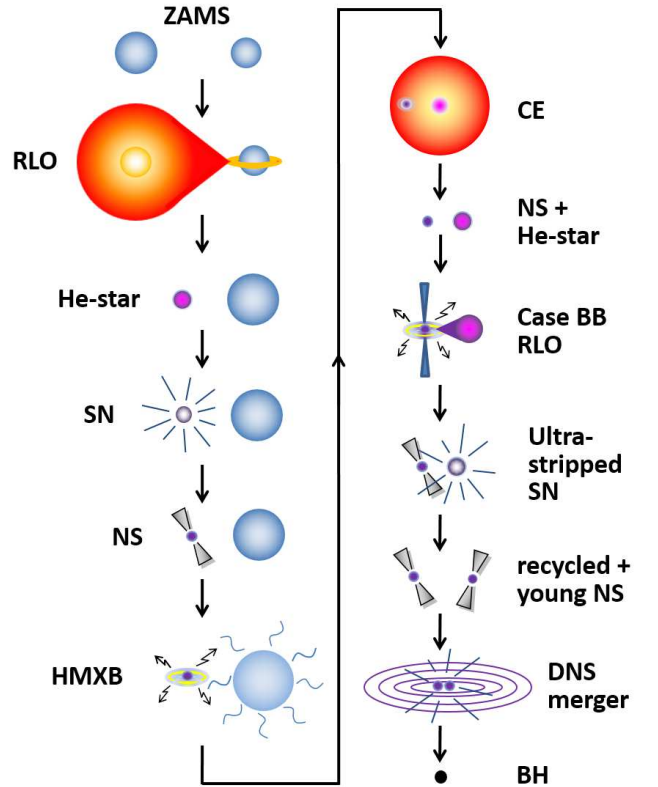}
    \vspace{0.2cm}
    \caption{Illustration of the formation of a DNS system which merges
             within a Hubble time and produces a single BH, following a powerful burst of GWs and a short~GRB.
             Acronyms used in this figure: ZAMS: zero-age main sequence; RLO: Roche-lobe overflow (mass transfer); He-star: helium star;
             SN: supernova; NS: neutron star; HMXB: high-mass X-ray binary; CE: common envelope; BH: black hole.}
  \label{fig:vdHcartoon}
  \end{center}
\end{figure}

\subsection{R\'{e}sum\'{e} of DNS formation}\label{subsec:std-scenario}
Previous theoretical work on the physics of DNS formation includes (here disregarding general population synthesis studies): 
\cite{bk74,wml74,fv75,sv82,vdh94a,ibk+03,dp03,plp+04,vdh04,dpp05}. 
From these papers, a {\it standard scenario}\footnote{See brief discussion given in Section~\ref{subsec:CE} 
for an alternative ``double core scenario'' \citep{bro95,dps06} in which CE evolution with a NS is avoided.}
has emerged \citep[see, for example,][]{bv91,tv06} which we now summarize in more detail.

In Fig.~\ref{fig:vdHcartoon}, we show an illustration of the formation of a DNS system. 
The initial system contains a pair of OB-stars which are massive enough\footnote{The secondary (initially least massive) star may be a $5-7\;M_{\odot}$ star which 
accretes mass from the primary (initially most massive) 
star to reach the threshold limit for core collapse at $\sim\!8-12\;M_{\odot}$ \citep[][see also Section~\ref{subsec:preHMXB}]{jhn+13,wh15}.} 
to terminate their lives in a core-collapse SN (CCSN). To enable formation of a tight DNS system in the end,
the two stars must initially be in a binary system close enough to ensure interactions via either stable or unstable mass transfer. 
If the binary system remains bound after the first SN explosion \citep[which is of type~Ib/c,][]{ywl10}, the system eventually becomes
observable as a HMXB. Before this stage, the system may also be detectable as a radio pulsar
orbiting an OB-star, e.g. as in PSRs~B1259$-$63 \citep{jml+92} and J0045$-$7319 \citep{kjb+94}.
When the secondary star expands and initiates full-blown RLO during the HMXB stage, the system eventually becomes dynamical unstable. For wide systems, where the donor star has a deep 
convective envelope at the onset of the mass transfer (i.e. during so-called Case~B RLO, following termination of core hydrogen burning), the timescale on which the system becomes
dynamically unstable might be as short as a few 100~yr \citep{sav78}. 
This leads to the formation of a CE \citep{pac76} where dynamical friction
of the motion of the NS inside the giant star's envelope often causes extreme loss of orbital angular
momentum and (in some cases) ejection of the hydrogen-rich envelope. 
If the binary system survives the CE phase, it consists of a NS orbiting a helium star (the
naked core of the former giant star). 
Depending on the orbital separation and the mass of the helium star, an additional phase
of mass transfer \citep[Case~BB RLO,][]{hab86a,tlp15} may be initiated. This stage of
mass transfer is important since it enables a relatively long phase of accretion onto
the NS, whereby the NS is recycled, and it allows for extreme stripping of the
helium star prior to its explosion \citep[as a so-called {\it ultra-stripped} SN,][]{tlm+13,tlp15,sys+15,mmt+17}.
Whether or not the system survives the second SN depends on the orbital separation and
the kick imparted onto the newborn NS \citep{fv75,hil83,tt98}. As we shall argue in this paper,
we expect most systems to survive the second SN explosion.  
If the post-SN orbital period is short enough (and especially if the eccentricity is large)
the DNS system will eventually merge due to GW radiation and produce a strong high-frequency GW signal and possibly a short~GRB \citep[e.g.][]{elps89}.
The final remnant is most likely a BH although, depending on the EoS, a NS (or, at least, a meta-stable NS) may be left behind instead \citep{vs98}.

\subsection{Major uncertainties of DNS formation}\label{subsec:uncertainties}
Besides from the pre-HMXB evolution, which is discussed in Section~\ref{subsec:preHMXB}, the most important and uncertain aspects of our current understanding of DNS formation are related to:
\begin{itemize}
  \item[i)] CE evolution and spiral-in of the NS 
  \item[ii)] momentum kicks imparted onto newborn NSs 
  \item[iii)] the mass distribution of NSs 
\end{itemize} 
In the following subsections we provide more details on each of these three aspects.

\subsubsection{CE evolution}\label{subsubsec:CE}
A CE phase \citep[see][for reviews]{tas00,ijc+13} develops when the donor star in a HMXB fills its Roche lobe. The large mass ratio between the
OB-star donor and the accreting NS causes the orbit to shrink significantly upon mass transfer, whereby the NS is captured inside the envelope of the donor star.  
As a consequence of the resulting drag force acting on the orbiting NS, efficient loss of orbital angular momentum often leads to a huge reduction in 
orbital separation prior to the second SN explosion, thus explaining the tight orbits of the observed DNS systems.

In a recent study on CE evolution with massive stars, it was demonstrated by \citet{ktl+16} that an in-spiralling NS may indeed be able to eject the envelope of its 
massive star companion, at least from an energy budget point of view. However, they also showed that it is difficult to predict the final post-CE separation for several reasons.
First of all, it is difficult to locate the bifurcation point \citep{td01} within the massive star, separating the remaining core from the ejected envelope. Secondly,  
while additional energy sources, like accretion energy, may play a significant role in helping facilitate the CE-ejection process,
models of time-dependent energy transport in the convective envelope are needed to quantify this.
Finally, the effect of inflated envelopes of the remaining helium core --- as a consequence of the stellar luminosity reaching 
the Eddington limit in their interiors \citep[e.g.][]{sglb15} --- makes it non-trivial to map the core size of a massive donor star model
to the size of the naked helium core (Wolf-Rayet star) left behind. 
Until future 3D hydrodynamical simulations might succeed in ejecting the CE of massive stars, the estimated post-CE separation (and thus the LIGO/VIRGO detection
rates from population synthesis of merging BH/NS binaries) will remain highly uncertain.

\subsubsection{NS kicks}
The second major remaining issue that needs to be solved is the magnitude and direction of
the momentum kick added to a NS during the SN explosion \citep{jan12}. 
In particular for the application to population synthesis modelling of DNS formation, it is important to identify any differences between the first and the 
second SN explosion \citep[e.g.][and references therein]{be16}. 
Whereas there is ample observational evidence for large NS kicks (typically $400-500\;{\rm km\,s}^{-1}$) in observations of young radio pulsars \citep{ggo70,crl93,ll94,kbm+96,hllk05,cvb+05}
it also seems evident that the second SN explosion forming DNS systems involves, on average, significantly smaller kicks \citep{plp+04,vdh04,spr10,bp16}.
Furthermore, there is evidence from observations of pulsars in globular clusters (which have small escape velocities of less than 
$40\;{\rm km\,s}^{-1}$) that some of them are born with very small kicks. This could suggest a bimodal kick velocity distribution, 
which might be connected to a bimodality in the formation mechanism of NSs. 
This picture is supported by observations of HMXBs \citep{prps02}. 

Due to the continuous growth of supercomputing power and the development of efficient and highly parallelized numerical simulation tools, 
2D and 3D simulations of stellar core collapse and explosions with sophisticated treatment of the complex microphysics have become 
feasible and demonstrate the viability of the neutrino-driven and MHD explosion mechanisms in principle \citep{jan12}. Initiating the 
SN blast wave with an approximate neutrino transport description in 2D and 3D simulations, it could be demonstrated \citep{wjm13} that mass-ejection 
asymmetries imprinted by hydrodynamic instabilities in the SN core during the initiation of the explosion can lead to NS (and BH) kicks 
compatible with the measured velocities of young pulsars. Moreover, a first systematic attempt to establish the progenitor--explosion--remnant 
connection based on the neutrino-driven mechanism has revealed the interesting possibility that the NS--BH transition might 
occur in stars below $20\;M_{\odot}$ \citep{ujma12}, which seems to be in line with conclusions drawn from observational SN-progenitor 
associations \citep{sma09}.
This result should have consequences for the predicted ratio of the number of compact objects (NSs vs BHs) of the GW mergers that LIGO/VIRGO will detect. 

It has recently been demonstrated \citep{tlm+13,tlp15} that close-orbit DNS systems (i.e. those which are tight enough to merge within a Hubble time due to GWs) must form via 
ultra-stripped SNe when the last star explodes. The reason being that Case~BB RLO, 
i.e. mass transfer via Roche-lobe overflow from a naked helium star in the post-HMXB/post-CE system, 
causes the NS to significantly strip its evolved helium star companion, almost to a naked metal core prior to its explosion.
This has an important effect on the number and properties of surviving DNS systems --- in particular, in terms of their kinematic properties ---
as we shall discuss rigorously in more detail.

To learn about the progenitor binaries that formed DNS systems and the conditions of the SN explosions which 
produced the second-born NSs, many studies in the literature have analysed or simulated a large number of SNe and compared the outcome 
with observations of DNS systems \citep{fv75,hil83,kl84,rs85,dc87,bp95,kal96,tt98,wkk00,afkw15}.
We therefore conclude our investigation presented in this paper with an extended analysis of the dynamical impact of SNe, taking into account the latest observational constraints
on many DNS systems. Our analysis leads to firm results regarding the kick magnitude and direction,
as well as the mass of the exploding star in the second SN events. 

\subsubsection{Mass distribution of NSs}
A third main aspect of DNS systems which is far from being understood --- and possibly related to the above discussion
on SNe and binary stellar evolution in general --- is the mass distribution of NSs, 
and why it differs from the NSs measured in systems with WD companions. \citet{ato+17} recently
analysed the mass distribution of 32 MSPs in orbit with (mostly) WD companions.
They found evidence for a bimodal mass distribution with a low-mass component centred
at $1.39\pm0.03\;M_{\odot}$ and dispersed by $0.06\;M_{\odot}$, and a high-mass component with a mean of about
$1.81\pm0.10\;M_{\odot}$ and dispersed by $0.18\;M_{\odot}$.
The diversity in spin and orbital properties of high-mass NSs suggests that this mass difference is most
likely not a result of the recycling process, but rather reflects differences in the NS birth masses.
For the case of DNS systems, we investigate in this work the total amount of mass accreted by the first-born NS at various phases of evolution.

\subsection{The structure and aims of this paper}\label{subsec:scope}
Following this introduction, we present the latest observational data on DNS systems in Section~\ref{sec:data}, and discuss their observational
selection effects and biases, with a special focus on DNS masses and the DNS systemic velocities. 
In Section~\ref{sec:evol1}, we discuss the binary evolution from the ZAMS stage until the HMXB stage.  
We derive in detail in Section~\ref{sec:Mrecycled} the amount of mass accreted by the first-born NS during its evolution in a binary system,
with the aim of inferring the birth masses of these first-born NSs in DNS systems. This allows us to directly compare our results with the
masses of the second-born NSs and also NS masses in other systems such as HMXBs.
In Section~\ref{sec:PorbPspinEcc}, we analyse DNS correlations between $(P_{\rm orb},\,P)$, and $(P_{\rm orb},\,e)$,
and derive theoretical correlations, based on stellar evolution models and accretion physics, which we compare with observational data. We argue that the suggested $(P,\,e)$ correlation
is a natural consequence of the two aforementioned correlations, and simulate the spread in all three correlations as a result of
a kick velocity added to the second-born NS.
We analyse and discuss the dynamical consequences of asymmetric SNe (kicks) during SN explosions in Section~\ref{sec:kicks}, and also investigate a possible connection
between final stellar structure, kick magnitude and NS mass. 
In Section~\ref{sec:mapping}, we discuss the mapping of observed DNS systems to simulated DNS systems (for which we compute properties right after the second SN), 
in terms of parameter evolution over time and observational selection effects.
Monte Carlo simulations for all known DNS systems are presented in Section~\ref{sec:sim} in order to derive their
pre-SN properties and place interesting constraints for some of these systems, which are useful to investigate the nature of the second SN explosion
--- one of the prime aims of this paper.
The details from the simulations of each individual DNS system are given in the Appendix. The ramifications of these SN simulations are discussed further in Section~\ref{sec:ramifications}. 
Finally, in Section~\ref{sec:merger}, we discuss DNS mergers (with applications to LIGO/VIRGO and observations of short~GRBs) in the context of this work. 
Our conclusions are summarized in Section~\ref{sec:summary}.

%%%%%%%%%%%%%%%%%%%%%%%%%%%%%%%%%%%%%%%%%%%%%%%%%%%%%%%%%%%%%%%%%%%%%%%%%%%
\section{Observational data of DNS systems}\label{sec:data}
In Table~\ref{table:DNS}, we list 15 DNS systems, including a couple of sources for which the DNS nature is not confirmed. 
The vast majority of these DNS systems are found in the Galactic disk; only two sources are found in a globular cluster. These two latter sources
will be disregarded for further discussions in this paper related to the recycling process and impact of the second SN explosion. 
The reason being that these two systems almost certainly formed by secondary exchange encounters, i.e. where already recycled pulsars change stellar companions or are being exchanged into new binaries 
because of close encounters. In the cases of PSRs~B2127+11C and B1807$-$2500B, such arguments were given by \citet{pakw91,lfrj12,vf14}. 
In these dynamical processes, information on all traces of the past evolutionary links to their former companions is lost and cannot be recovered from observations of the present binary systems. 

In Table~\ref{table:DNS}, we indicate whether an observed NS is a recycled pulsar or a young 
(second-born) NS component. Only in the PSR~J0737$-$3039 system are both NS components detected as radio pulsars.
This system is therefore known as the {\it double pulsar}.

\begin{table*}[b]
     \caption[]{Properties of 15 DNS systems with published data (including a few unconfirmed candidates).}
        \begin{center}
        \begin{tabular}{lcrccrcrrlrrrrr}
           \hline
           \noalign{\smallskip}
                        &      & $P\;\;$  & $\dot{P}$     & $B$               & $P_{\rm orb}$ &  $e$ & $M_{\rm psr}$ & $M_{\rm comp}$ & $M_{\rm total}$ & $\delta\quad$ &  Dist. & $v^{\rm LSR\,**}$    &  $\tau_{\rm gwr}$
 \\
           Radio Pulsar & Type & (ms)     & $(10^{-18})$  & $(10^9\;{\rm G})$ & (days)        &      & $(M_{\odot})$ & $(M_{\odot})$  & $(M_{\odot})$   & (deg)         & (kpc)  & $({\rm km\,s}^{-1})$ & (Myr) \\
           \noalign{\smallskip}
           \hline
           \noalign{\smallskip}
           J0453+1559$^a$    & recycled &   45.8 &     0.186  &  0.92 & 4.072 & 0.113 &     1.559 &     1.174 & 2.734& --         &1.07 & 52  & $\infty$ \\
           J0737$-$3039A$^b$ & recycled &   22.7 &     1.76   &  2.0  & 0.102 & 0.088 &     1.338 &     1.249 & 2.587& $<3.2$     &1.15 & 33  &   86     \\ 
           J0737$-$3039B$^b$ & young    & 2773.5 &   892      &  490  &\ditto &\ditto &     1.249 &     1.338 &\ditto& $130\pm1$  &\ditto&\ditto&\ditto \\ 
           J1518+4904$^c$    & recycled &   40.9 &     0.0272 &  0.29 & 8.634 & 0.249 & $^{***}$ & $^{***}$ & 2.718& --           &0.63 & 28  & $\infty$ \\ 
           B1534+12$^d$      & recycled &   37.9 &     2.42   &  3.0  & 0.421 & 0.274 &     1.333 &     1.346 & 2.678&$27\pm3$    &1.05 & 136 &   2730   \\ 
           J1753$-$2240$^e$  & recycled &   95.1 &     0.970  &  2.7  & 13.638 & 0.304 &     --   &      --   &  --  & --         &3.46 & --     & $\infty$ \\ 
           J1755$-$2550$^{f*}$ & young  &  315.2 &  2430      &  270  &  9.696 & 0.089 &     --    & $>\!0.40$&  --  & --         &10.3 & --     & $\infty$ \\ 
           J1756$-$2251$^g$  & recycled &   28.5 &     1.02   &  1.7  & 0.320 & 0.181 &     1.341 &     1.230 & 2.570& $<34$      &0.73 & 41$^{****}$ &   1660 \\ 
           J1811$-$1736$^h$  & recycled &  104.2 &     0.901  &  3.0  & 18.779 & 0.828 & $<\!1.64$ & $>\!0.93$& 2.57 & --         &5.93 & --     & $\infty$ \\ 
           J1829+2456$^i$    & recycled &   41.0 &     0.0525 &  0.46 & 1.176 & 0.139 & $<\!1.38$ & $>\!1.22$ & 2.59 & --         &0.74 & --     & $\infty$ \\ 
           J1906+0746$^{j*}$ & young    &  144.1 & 20300      &  530  & 0.166 & 0.085 &     1.291 &     1.322 & 2.613& --         &7.40 & --     & 309      \\ 
           J1913+1102$^k$    & recycled &   27.3 &     0.161  &  0.63 & 0.206 & 0.090 & $<\!1.84$ & $>\!1.04$ & 2.88 & --         & --  & --     & $\sim\!480$   \\ 
           B1913+16$^l$      & recycled &   59.0 &     8.63   &  7.0  & 0.323 & 0.617 &     1.440 &     1.389 & 2.828&$18\pm6$    &9.80 & 240 & 301      \\ 
           J1930$-$1852$^m$  & recycled &  185.5 &    18.0    &  18   & 45.060 & 0.399& $<\!1.32$ & $>\!1.30$ & 2.59 & --         &1.5  & --     & $\infty$ \\ 
           \noalign{\smallskip}
           \hline
           \noalign{\smallskip}
           J1807$-$2500B$^{n*}$& GC     &    4.2 &     0.0823 &  0.18 & 9.957 & 0.747 &     1.366 &     1.206 & 2.572& --         &3.0  & --  & $\infty$ \\ 
           B2127+11C$^p$     & GC       &   30.5 &     4.99   &  3.8  & 0.335 & 0.681 &     1.358 &     1.354 & 2.713& --         &12.9 & --  & 217      \\ 
           \noalign{\smallskip}
           \hline
        \end{tabular}
        \end{center}
%\begin{flushleft}
  $\dot{P}$ values are not corrected for the (in most cases negligible) Shklovskii effect \citep{shk70}. 
  B-field values are estimated from Eq.~(\ref{eq:Bmodified}). All the NS masses quoted with 4 significant figures have uncertainties of $\la 0.010\;M_{\odot}$.
  All values of $\tau _{\rm gwr}>50\;{\rm Gyr}$ are shown with $\infty$. \\Notes: 
  $^*$ Not a confirmed DNS system --- could also be a WD+NS binary. 
  $^{**}$ Based on the median value of $v^{\rm LSR}$, cf. Section~\ref{subsec:vsys} and Table~\ref{table:DNS_v_LSR}. 
  $^{***}$ See Section~\ref{subsec:1518} for updated NS masses of PSR~J1518+4904. 
  $^{****}$ Based on the 2D proper motions estimated in \citet{fsk+14}.\\
  References: 
  $^a$~\citet{msf+15}. 
  $^b$~\citet{ksm+06,bkk+08,fsk+13}. 
  $^c$~\citet{jsk+08}.
  $^d$~\citet{fst14}. 
  $^e$~\citet{kkl+09}.
  $^f$~\citet{ncb+15,nkt+17}. 
  $^g$~\citet{fkl+05,fsk+14}. 
  $^h$~\citet{cks+07}. 
  $^i$~\citet{clm+04}. 
  $^j$~\citet{lsf+06,vks+15}.
  $^k$~\citet{lfa+16}. 
  $^l$~\citet{ht75,kra98,wnt10}.
  $^m$~\citet{srm+15}.
  $^n$~\citet{lfrj12}.
  $^p$~\citet{agk+90,jcj+06}.
%\end{flushleft}
\label{table:DNS}
\end{table*}

\begin{table*}[b]
\caption[]{Estimated 3D systemic LSR velocities for Galactic disk DNS systems with proper motion measurements.}
\begin{center}
\begin{tabular}{lrrlrrcrccr}
\hline
\noalign{\smallskip}
             & $l\quad$ & $b\quad$ & Distance & $\mu _\alpha\qquad$          & $\mu _\delta\qquad$          & $v_T^{\rm SSB}$      &  $v^{\rm LSR}_{\rm median}$ & $v^{\rm LSR}_{1\sigma}$ & $v^{\rm LSR}_{2\sigma}$ & $v^{\rm LSR, max}_{90\%~{\rm CL}}$ \\
Radio Pulsar &  (deg) & (deg) & (kpc)    & (${\rm mas\,yr}^{-1}$) & (${\rm mas\,yr}^{-1}$) & (${\rm km\,s}^{-1}$) & (${\rm km\,s}^{-1}$)        & (${\rm km\,s}^{-1}$)    & (${\rm km\,s}^{-1}$)    & (${\rm km\,s}^{-1}$) \\ 
\noalign{\smallskip}
\hline
\noalign{\smallskip}
J0453+1559    & 184.12 & -17.14 & 1.07     & -5.5$\qquad$ & -6.0$\qquad$ & 41$\pm$ 8 &  52$\quad$ & 36$-$85   & 26$-$197  & 102$\quad$\\
J0737$-$3039A & 245.24 &  -4.50 & 1.15$^a$ & -3.8$\qquad$ &  2.1$\qquad$ & 25$\pm$ 4 &  33$\quad$ & 18$-$57   & 10$-$124  &  68$\quad$\\
J1518+4904    &  80.81 &  54.28 & 0.63     & -0.7$\qquad$ & -8.5$\qquad$ & 26$\pm$ 5 &  28$\quad$ & 20$-$49   & 15$-$123  &  61$\quad$\\
B1534+12      &  19.85 &  48.34 & 1.05$^b$ &  1.5$\qquad$ & -25.3$\qquad$&127$\pm$ 1 & 136$\quad$ & 118$-$224 & 115$-$512 & 276$\quad$\\
J1756$-$2251  &   6.50 &   0.95 & 0.73$^c$ & -2.4$\qquad$ &  5.5$\qquad$ & 21$\pm$ 7 &  41$\quad$ & 21$-$71   & 11$-$123  &  82$\quad$\\
B1913+16      &  49.97 &   2.12 & 9.80$^d$ & -1.2$\qquad$ & -0.8$\qquad$ & 69$\pm$22 & 240$\quad$ & 174$-$273 &  83$-$401 & 290$\quad$\\
\noalign{\smallskip}
\hline
\noalign{\smallskip}
J0453+1559    & 184.12 & -17.14 & 0.52 & -5.5$\qquad$ & -6.0$\qquad$ & 20$\pm$ 4 &  27$\quad$ & 20$-$43  & 15$-$99  &  52$\quad$\\
J1518+4904    &  80.81 &  54.28 & 0.96 & -0.7$\qquad$ & -8.5$\qquad$ & 39$\pm$ 8 &  39$\quad$ & 28$-$70  & 21$-$184 &  89$\quad$\\
B1913+16      &  49.97 &   2.12 & 5.25 & -1.2$\qquad$ & -0.8$\qquad$ & 37$\pm$ 7 & 132$\quad$ & 98$-$168 & 67$-$229 & 179$\quad$\\
\noalign{\smallskip}
\hline
\end{tabular}
\end{center}
%\begin{flushleft}
The upper part of the table uses NE2001 for the DM distances to PSRs~J0453+1559 and J1518+4904; otherwise distances are taken from: $^a$~\citet{dbt09}, $^b$~\citet{fst14}, $^c$~\citet{fsk+14}, and $^d$~\citet{wnt10}.
The lower part of the table uses \citet{ymw16} for DM distance determinations. See Section~\ref{subsec:vsys} for further details. 
%\end{flushleft}
\label{table:DNS_v_LSR}
\end{table*}

\subsection{Selection effects}\label{subsec:selection}
Before engaging in an analysis of DNS systems, we must first assess any selection effects associated with discovering DNS systems. 
Most of the recycled pulsars in DNS systems have spin periods of a few tens of ms. This implies that they are near the optimal sensitivity of current surveys for dispersion measure values
(DMs) of the order of $325\;{\rm cm}^{-3}\,{\rm pc}$ and lower \citep[see e.g.][]{lbh+15}. This is not the case for slow, non-recycled radio pulsars in DNS systems like PSR~J0737$-$3039B,
for which \citet{lbh+15} determined a significant DM-dependent loss of sensitivity caused by gain and system temperature fluctuations in radio receivers, due mostly to
radio frequency interference. This is particularly dramatic at low DMs where the loss of sensitivity can be one order of magnitude.

The main selection effect against the detection of a recycled pulsar in a tight DNS system is caused by the orbital acceleration of the pulsar (this does not affect
the detection of slow pulsars). The fast-changing Doppler shift then means that the spin period is perceived (on Earth)
to change significantly within the time of a single observation. Unless corrected, this effect smears the pulsar signal in the Fourier domain, where the searches are performed; particularly for
the higher harmonics, it becomes increasingly severe for cases where the observation length ($T_{\rm obs}$) is more than a few percent of the orbital period ($P_{\rm orb}$).
Several strategies have been adopted to cope with this. These include mostly ``acceleration searches'' \citep{agk+90,clf+00,rem02},
which assume a constant acceleration during the observation. However, these surveys are computationally expensive. While they increase the number of trials
necessary for detecting a binary pulsar, they necessarily cause a reduction in sensitivity even if the acceleration is nearly constant.
The reason for this is that the much larger number of trials increases the number of candidates that originate from the Gaussian noise in the data. 
As a result, the signal-to-noise (S/N) threshold for a detection in an acceleration search is generally larger than in a survey with no acceleration. 
Furthermore, these surveys start losing  sensitivity when $T_{\rm obs}/ P_{\rm orb} \, \geq \, 0.1$ (this number becomes smaller for larger accelerations), i.e. 
when the assumption of a constant acceleration no longer holds \citep{ncb+15}. Thus also systems with large eccentricities can be detrimental. Acceleration searches are now
widely used in blind surveys \citep[e.g.][]{lbh+15}, because computing improvements make them feasible as demonstrated by their successes in discovering new DNS systems. 

Several strategies have been pursued to search for pulsars in even tighter systems, these are too numerous to describe here \citep[see e.g.][and references therein]{eklk13}; 
they are not yet widely used in large-scale blind surveys because of their computational cost.

To summarize, there is generally a bias against the detection of pulsars in DNS systems, particularly in very tight systems, but this depends crucially on the length of the
observations in each survey (the shorter the integration the smaller is the bias) and the particular search algorithms being used in each survey.

\subsection{Estimating the systemic velocities of DNS systems relative to their local standard of rest}
\label{subsec:vsys}

When comparing the outcome of our SN simulations (cf.\ Section~\ref{sec:sim}) with observational data 
it is necessary to calculate the systemic velocities of the DNS systems relative to their local standard 
of rest (LSR) in the Galaxy. 

From pulsar timing one often gets an accurate value for the proper motion of the DNS system at the sky. 
Combined with an estimation of the pulsar distance, a transverse velocity, 
$v_T^{\rm SSB}$ with respect to the Solar system barycenter (SSB), can be found. Using a Monte~Carlo (MC) simulation, 
we account for the uncertainties in distance and proper motion. Concerning the pulsar distance, in some cases 
the only available estimation comes from the DM~distance, which is based 
on a model for the free electron density in our Galaxy \citep[e.g.][]{cl02,ymw16}. In these cases, we assume in our 
MC simulation a 20\% uncertainty, and use this uncertainty to be $1\,\sigma$ in a Gaussian distribution for the 
DNS distance. 

For DNS systems one does not have a measurement of the radial velocity, unlike the case for some nearby 
pulsar+WD systems \citep[see e.g.][]{cgk98,avk+12} for which phase-resolved spectroscopy of the optically 
bright WD is possible. Therefore, one does not have a 3D velocity vector of the DNS system. At this stage we 
make the assumption that the spatial orientation of the velocity vector has no preferred direction with 
respect to the SSB. Adding a random orientation to our MC~simulation, one then obtains a probability 
distribution for the radial component, $v_R^{\rm SSB}$ (based on the calculated $v_T^{\rm SSB}$): 
$v_R^{\rm SSB} = v_T^{\rm SSB} \cot\chi$, where $\chi$ denotes the angle between the direction to the 
pulsar and the orientation of the 3D velocity vector. 

Finally, to convert the SSB velocity vector to the LSR we have used the Galactic model of \citet{mcm17}. 
This conversion is done for every single MC realization in our simulation. Consequently, we get a probability 
distribution for the magnitude of the systemic velocity with respect to the LSR, $v^{\rm LSR}$, which we then 
compare with our SN simulations later in this paper. 

The results for the systemic velocities, for those DNS systems near the Galactic plane with well-measured 
proper motion, are given in Table~\ref{table:DNS_v_LSR}. In the last four columns we state the median value, 
the $1\sigma$ and $2\sigma$ ranges, as well as the 90\% confidence limit for the upper limit of the 3D systemic 
velocity, $v^{\rm LSR, max}$, based on our MC simulations (Strictly speaking, since the $v^{\rm LSR}$ 
distributions are non-Gaussian, these $1\sigma$ and $2\sigma$ values refer to 68\% and 95\% confidence intervals, 
respectively). We notice that we make the assumption that the LSR is a reasonable representation of the birth 
reference frame of each DNS system. That might not necessarily be the case, as for instance discussed for the 
Hulse-Taylor pulsar in \citet{wkk00}. In cases where no precise pulsar distance 
is known from parallax measurements or orbital decay, we have applied the DM distances based on the NE2001 
model \citep{cl02}. We adopt this model as the default\footnote{In the case of PSR~B1913+16, we adopted a default distance based on H{\rm I} absorption measurements \citep{wsx+08}.} 
in our analysis of DNS kinematics discussed further in this work. 

The lower part of Table~\ref{table:DNS_v_LSR} shows our results using the DM distances based on 
\citet{ymw16}, who suggest that their model is better than NE2001. However, more thorough studies of this new model are needed 
before any firm conclusions can be drawn. 
For some DNS systems, like PSR~J0737$-$3039, the results obtained from the different DM distance models are quite 
similar, whereas in other cases (PSRs~J0453+1559 and B1913+16) the value of $v^{\rm LSR}$ can differ by a factor of two.
One particular example of a DNS pulsar with a discrepancy is PSR~J1756$-$2251 where \citet{ymw16} find a DM distance of 2.81~kpc, although an
upper value of 1.2~kpc for its distance can be derived \citep{fsk+14} using orbital period decay ($\dot{P}_{\rm orb}$). 

To further check our results against the sensitivity of the DM distance, we recalculated the values of $v^{\rm LSR}$ 
for the \citet{ymw16} DM distances, now assuming in our MC simulation a 50\% uncertainty in distance using a flat probability range centered on the DM distance.
The resulting ranges of values for $v^{\rm LSR}_{1\sigma}$ and $v^{\rm LSR}_{2\sigma}$ only changed with a few ${\rm km\,s}^{-1}$ for PSRs~J0453+1559 and J1518+4904,
whereas for PSR~B1913+16 these interval ranges were widened on each end by $\sim\!18\;{\rm km\,s}^{-1}$ and $\sim\!12\;{\rm km\,s}^{-1}$, respectively.  
Thus, these are relatively small changes compared to the values stated in Table~\ref{table:DNS_v_LSR} and therefore not significant for our NS kick discussions in Section~\ref{sec:kicks}.

\begin{figure}[t]
  \begin{center}
     \includegraphics[width=0.72\columnwidth, angle=-90]{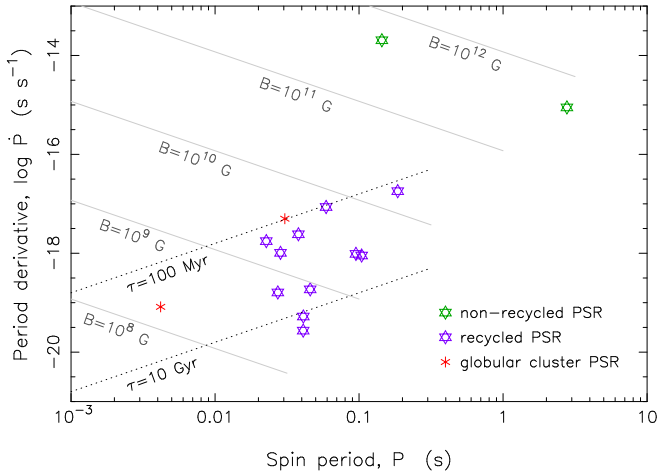}
    \vspace{0.2cm}
    \caption{($P,\dot{P}$)--diagram of all radio pulsars detected in DNS systems. 
       Data are given in Table~\ref{table:DNS}. 
       The solid grey lines represent constant surface dipole B-fields and the dotted black lines represent constant characteristic ages. 
       The $\dot{P}$ values are not corrected for the (mostly negligible) Shklovskii effect \citep{shk70}.
       For the globular cluster pulsars, $\dot{P}$ must also be corrected for the (potentially significant) NS acceleration in the cluster potential.}
  \label{fig:PPdot}
  \end{center}
\end{figure}

\subsection{Derived surface B-fields}\label{subsec:Bsurf}
To estimate the (equatorial) surface B-fields of pulsars, the standard method is to use the classic vacuum magnetic dipole expression \citep[e.g.][]{lk04}: 
$B\simeq 3.2\times 10^{19}\;{\rm G}\;\sqrt{P\dot{P}}$. 
The incompleteness of the vacuum magnetic dipole model, however, is particularly evident after the discovery
of intermittent pulsars by \citet{klo+06} and demonstrates the need for including the plasma term in the spin-down torque.
The classical expression does not include the rotational energy loss contribution from the pulsar wind nor the spin-down torque caused by the $\vec{j}\times\vec{B}$ 
force exerted by the plasma current in the magnetosphere \citep[e.g.][]{gj69,klo+06,spi08}.  
Hence, the classic expression does not predict any spin-down torque for an aligned rotator ($\alpha = 0^{\circ}$),
which is not observed. 
A combined model was derived by \citet{spi06} \citep[see also][]{lst12} and applying his relation between $B$ and $\alpha$, \citet{tlk12} obtained\footnote{Using slightly larger values for
the NS radius and moment of inertia than the standard values of $R=10\;{\rm km}$ and $I=10^{45}\;{\rm g\,cm}^2$.}: 
\begin{eqnarray}\label{eq:Bmodified} 
     B   & = & \sqrt{\frac{c^3IP\dot{P}}{4\pi ^2 R^6}\;\frac{1}{1+\sin^2\alpha}}\\ \nonumber 
         & \simeq & 1.3\times 10^{19}\,G\quad\sqrt{P\dot{P}}\;\left( \frac{M}{1.4\,M_{\odot}}\right) ^{3/2} \sqrt{\frac{1}{1+\sin^2\alpha}}\,.
\end{eqnarray}
To estimate the surface B-fields of our DNS pulsars we use the above expression for an assumed magnetic inclination angle, $\alpha =60^{\circ}$.
This yields roughly the following relation: $B\simeq 1.0\times 10^{19}\;{\rm G}\;\sqrt{P\dot{P}}$. 
Fig.~\ref{fig:PPdot} shows the distribution of DNS radio pulsars in the ($P,\dot{P}$)--diagram and their estimated B-fields. 

\subsection{Measuring NS masses}\label{subsec:NSmeasure}
Precise measurements of NS masses serve as an important diagnostic tool for understanding NS 
formation. In the following subsection, we therefore outline in detail how to determine such NS 
masses from observations.

There are a wide variety of methods available for measuring NS masses \citep{of16}. For DNS systems 
specifically, all mass measurements to date result from the measurement of relativistic 
post-Keplerian (PK) parameters via high-precision radio pulsar timing. In DNS binaries, where 
classical tidal or magnetic effects are absent, PK parameters describe relativistic corrections to 
the timing and pulse-structure data. Most notable are effects related to the orbital motion and time 
dilation of the pulsar and the propagation of the pulsar signals in the curved spacetime of the 
binary system \citep{dt92}. Generally, for a given theory of gravity, the PK parameters are (to leading order) 
functions only of the (well known) Keplerian orbital parameters and the masses of the two components of 
the system. Hence, a measurement of two PK parameters determines the masses of the pulsar and 
companion uniquely. In several DNS systems, more than two PK parameters have been measured, which 
then also allows for a consistency test of the applied gravity theory. So far, general relativity 
(GR) has passed all these tests with flying colors \citep{wex14}, and has therefore been confirmed 
to be the correct theory to determine the masses from PK parameters in DNS systems. 

The validity of this approach has been particularly well demonstrated for the double pulsar system, 
PSR~J0737$-$3039A/B, where six PK parameters have been measured so far, plus (uniquely in a DNS) the 
component mass ratio provided by the detection of the second NS as a radio pulsar. Crucially, all 21
pairwise combinations of mass ratio and PK parameters yield consistent (and in most cases very 
precise) mass values for the two pulsars in this system. These agreements can be reframed as a 
total of five high-precision tests of GR, with one PK parameter and the mass ratio providing the 
pulsar masses, and the remaining PK parameters representing one test of GR each 
\citep{ksm+06,bkk+08}. Apart from confirming the validity of GR, these measurements also confirm 
that there are no unknown additional classical effects perturbing the timing of the pulsar in a 
measurable way. 

Although the vast majority of MSP+WD systems generally have higher timing precision than DNS 
systems, they often have poorly measured PK parameters. One of the reasons for this is their orbital 
eccentricity, which is many orders of magnitude smaller than for most DNS systems.
A large eccentricity implies that the system has a well-defined periastron, allowing for the 
measurement of its rate of advance, $\dot{\omega}$, which immediately yields the sum of the masses 
of the two components of the system in GR. The orbital eccentricity also allows the measurement of the 
``Einstein delay'', $\gamma$ (being a combination of a variation in the special relativistic time 
dilation, due to the changing orbital velocity of the pulsar, and a variation in the gravitational 
redshift for the pulsar due to a changing distance to the companion). However, the latter effect 
can be re-absorbed in the definition of the size of the Keplerian orbit. For it to be separately 
measurable, the orbit has to precess by at least a few degrees --- something that would take many 
decades to millennia for the wider DNS systems to be measured with present radio telescopes.

The above two PK parameters ($\dot{\omega},\gamma$) provide most of the precise NS mass measurements in tight DNS 
systems. The remaining PK parameters can be observed even for circular orbits. The emission of GWs 
gives rise to an orbital decay, $\dot{P}_{\rm orb}$. Like $\gamma$, this can only be measured for 
the more compact systems. A large orbital eccentricity magnifies the effect significantly \citep{pet64}.

Finally, the Shapiro delay is a relativistic effect that yields two PK parameters: $r$ and $s$ in 
the \citet{dd86} parameterization, or $h_3$ and $\varsigma$ in the \citet{fw10} parameterization. 
These allow for a direct determination of the mass of the pulsar's companion ($r=GM_{\rm comp}/c^2$) 
and the orbital inclination $i$, more precisely $\sin i$. The precision of this method is 
independent of the orbital period, but it requires a fortunate combination of high orbital 
inclination and good timing precision.

For the wide-orbit DNS systems, $\gamma$ and $\dot{P}_{\rm orb}$ are generally not measurable and a 
non-detection of the Shapiro delay implies that we cannot separate the two NS masses. In such cases 
only the total mass of the system is known from $\dot{\omega}$.

\subsection{Observed distribution of masses in DNS systems}\label{subsec:masses}
\begin{figure}
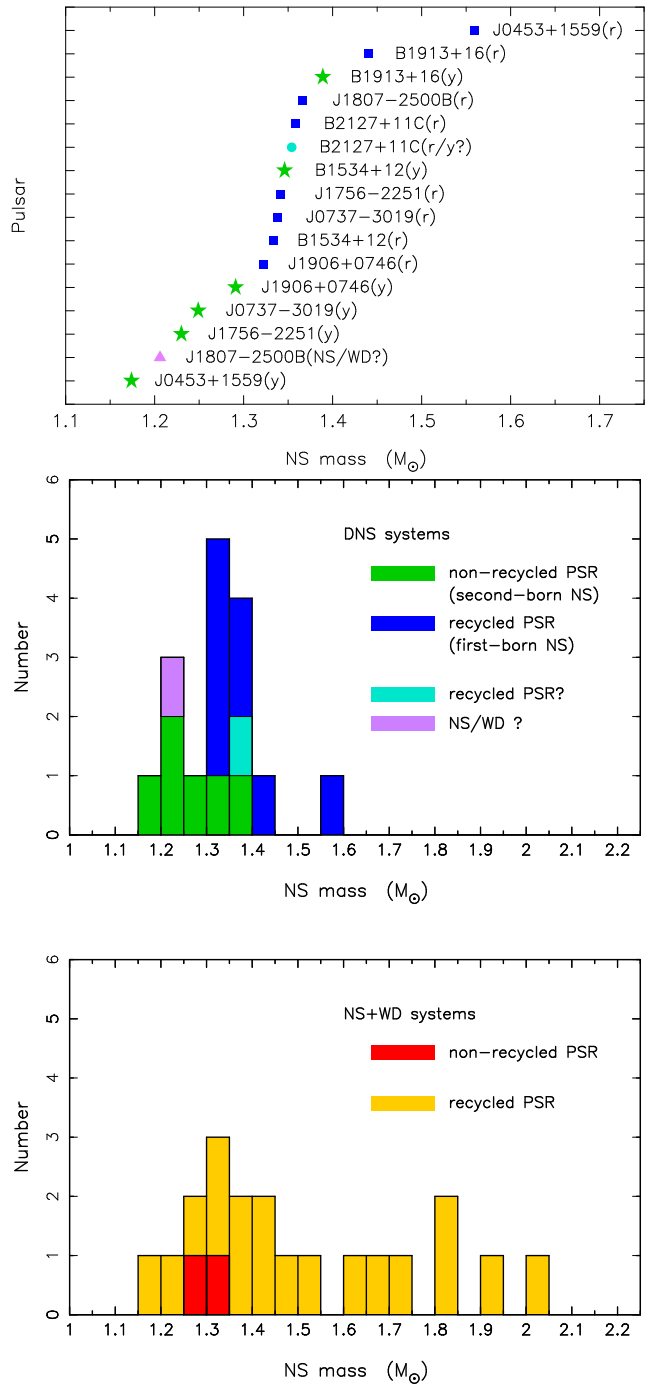

 \centering
 \includegraphics[width=0.72\columnwidth,angle=-90]{Figures/DNS_masses_plot3.ps}
 \includegraphics[width=0.97\columnwidth,angle=0]{Figures/DNS_histo.ps}
 \caption{Distribution of NS star masses in DNS systems (top, central) and NS+WD systems (bottom panel).
          Recycled pulsars are indicated by (r) and blue or yellow color; 
          young pulsars are indicated by (y) and green or red color.
          It is evident that NSs formed second in these binaries (i.e. the non-recycled, young pulsars)
          in general have a smaller mass than the recycled NSs. See Section~\ref{subsec:masses}.  
          Note, none of the colored bars overlap -- all NS data are visible.}
 \label{fig:NSmass}
\end{figure}

In Fig.~\ref{fig:NSmass}, we have plotted the distribution of NS masses in both DNS and NS+WD systems. 
The data are taken from \citet{of16} and table~1 in \citet{ato+17}.  
Whereas DNS systems descend from HMXBs, NS+WD systems are usually produced in LMXBs.
The error bars on NS masses from the precisely measured DNS systems are $\la 0.01\;M_{\odot}$, whereas the error bars for
the pulsar masses in the NS+WD systems plotted here range between $0.1-10$\%.
Two NS+WD systems (PSRs~B2303+46 and J1141$-6545$) are unique in the sense that here the NS is believed to have formed {\it after} the WD \citep{ts00}.
These NSs are observed with the typical characteristics of young and non-recycled radio pulsars (i.e. large values of $P$ and $\dot{P}$),
and besides, their orbits are eccentric and not circular as in the vast majority of MSP binaries.
For these two systems we take the NS mass estimates of \citet{kk99} and \citet{bbv08}. 
It is expected that recycled pulsars will have larger masses than non-recycled pulsars
since a certain amount of accreted mass is needed to recycle a pulsar. 
Indeed, for the DNS systems (Fig.~\ref{fig:NSmass}, top and central panels) it seems that the recycled NSs might have larger masses
by an amount of the order $\sim\!0.1\;M_{\odot}$. 
However, a crucial question is how much mass these NSs actually accreted, and whether
the mass difference between second-born (non-recycled) NSs and recycled NSs is instead 
mainly caused by differences in the birth masses of NSs resulting from different conditions in the first and the second SN explosion, respectively.
We investigate this question in more detail later in this paper.

%%%%%%%%%%%%%%%%%%%%%%%%%%%%%%%%%%%%%%%%%%%%%%%%%%%%%%%%%%%%%%%%%%%%%%%%%%%
\section{Evolution of OB-star binaries from the ZAMS to the HMXB stage}\label{sec:evol1}
In the following section, we discuss the early stages of massive binary star evolution and, in particular, the observed Galactic population of HMXBs.
We examine their destiny with the aim of understanding the evolutionary links between HMXBs and DNS systems. 

\subsection{Pre-HMXB evolution}\label{subsec:preHMXB}
It has been shown recently that interacting with a binary companion during its lifetime is the rule rather than the exception 
for a massive star. The majority of them are found in close binaries that will start mass transfer before the first SN occurs \citep{sdd+12,kk12,smm+14}.

If the mass transfer (i.e. RLO) remains stable, it results in the almost complete loss of the hydrogen-rich envelope of the initially more massive star, 
a significant fraction of which will be accreted by its companion \citep{pjh92}, cf. Fig.~\ref{fig:vdHcartoon}. The accretion process leads to a spin-up of the mass gainer, 
which is expected to end up spinning close to critical rotation \citep{pac81,hab86a,dli+13}, and to a net widening of the orbit.
Most models predict that the mass donor will reach the SN stage first. If the system does not break up due to the SN (Section~\ref{sec:kicks}), its observational counterparts
include the so-called Be-star X-ray binaries (Be-HMXBs, Section~\ref{subsec:HMXBpop}), where the Be-nature of the core-hydrogen burning component 
signals its state of near-critical rotation \citep{toh04}. The major uncertainty in our understanding of this evolutionary 
path concerns the mass and angular momentum loss from the system during the mass-transfer phases, i.e. the amounts, but also the temporal behavior 
of mass and angular momentum loss \citep{lan12}.
Some Be-HMXBs may later evolve into supergiant HMXBs (sgHMXBs) similar to those observed in wide orbits (Section~\ref{subsec:HMXBpop}). 

In case the binary mass transfer becomes unstable, the system will undergo a CE phase where both components share the hydrogen-rich envelope of the primary.
There are two possible outcomes from this \citep{pjh92}. One is that the two stars merge into a single star, likely accompanied with some mass loss \citep{ggpp13}. 
Since the system is no longer a binary, this does not lead to an X-ray binary phase nor a DNS system. Alternatively, the hydrogen-rich envelope is ejected before the two 
stars coalesce, leaving the stripped primary orbiting its companion (a main sequence star in most cases).
The result is thus similar to the case of stable mass transfer, except that orbital widening is avoided, and the orbit may have decayed significantly 
during the CE evolution \citep{ijc+13}. This is possibly the formation channel to the short-orbit ($\la 10\;{\rm days}$) HMXBs, including the RLO-HMXBs (Section~\ref{subsec:HMXBpop}). 
Also, as a result of the short duration of the CE spiral-in, there is no expected spin-up of the main sequence component. For further discussions of
the CE phase, see Sections~\ref{subsubsec:CE} and \ref{subsec:CE}.

In the vast majority of cases, X-ray binary progenitors will consist of a star which is stripped of nearly all of its
hydrogen-rich envelope and a main-sequence star companion.
Binaries which avoid any mass transfer until the first SN are generally so wide that they would easily break up at this stage (Section~\ref{sec:kicks}) 
--- unless the stars evolve chemically homogeneously \citep[e.g.][]{dcl+09,mlp+16}, in which case the orbit remains tight throughout the entire evolution. 

Although depending on the initial parameters and the adopted physics of internal mixing, some models predict the non-stripped component (i.e. the secondary star, the accretor from
the first mass-transfer phase) to evolve to the SN stage first \citep{pol94,wlb01}. In such binaries --- if subsequent reverse mass transfer could be avoided --- 
the orbit would be more likely to be disrupted, since now the secondary star is more massive than its companion. In the more common situation of the stripped star
evolving to the SN stage first (Fig.~\ref{fig:vdHcartoon}), it is the less massive star exploding, and the binary would therefore not break up unless the NS receives a large kick at birth (Section~\ref{sec:kicks}). 

We note that this picture, which is expected to apply for primary masses below about $40\;M_{\odot}$, may differ at higher masses. 
Above this threshold, the star radiates close to the Eddington limit, 
which may lead to loosely bound extended envelopes even in main-sequence stars \citep{sglb15}, or envelope ejection without the assistance of a companion 
as a luminous blue variable \citep{van91} or yellow hypergiant star \citep{kby+17}. The consequences of these effects have not yet been well explored, but will also affect only a small fraction
of the massive systems evolving to HMXBs. 

For single stars at solar metallicity, the initial critical ZAMS mass to undergo a SN and produce a NS is $8-12\;M_{\odot}$ \citep[][and references therein]{lan12}.
This critical mass depends on the amount of convective core overshooting and the metallicity content of the star. For larger amounts of convective overshooting, and for lower metallicities, 
the threshold mass decreases significantly. 
In close binaries, this limit depends on the initial system parameters. Especially for mass donors in the initially closest systems, the
mass limit at solar metallicity can be as high as $15\;M_{\odot}$ \citep{wlb01}. On the other hand, as discussed by \citet{plp+04}, electron capture SNe may occur in close
binaries from stars with an initial mass below $8\;M_{\odot}$, because the second dredge-up episode, which reduces the helium core, can be avoided once
the hydrogen envelope is lost.

\subsection{Observations of HMXB systems}\label{subsec:HMXBpop}
HMXBs are composed of a compact object (NS or BH) orbiting a luminous and massive ($\ga 10\;M_{\odot}$), early spectral type companion star. 
We can distinguish between three types of HMXBs according to their accretion process:  
(i) Be-star X-ray binaries (BeHMXBs), (ii) supergiant X-ray binaries (sgHMXBs), and (iii) Roche-lobe overflow systems (RLO-HMXBs).
For a review on HMXBs, see e.g. \citet{cha13,wlbt15}.
We now discuss each of these three types in more detail. 

\subsubsection{BeHMXBs}
\begin{figure}
  \begin{center}
     \includegraphics[width=0.72\columnwidth,angle=-90]{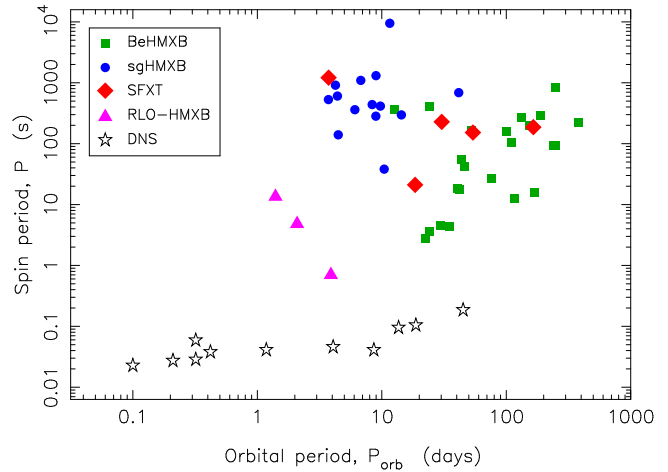}
    \vspace{0.2cm}
    \caption{Extended Corbet diagram showing the different populations of HMXBs (X-ray pulsars) and DNS systems (recycled radio pulsars) with measured values of both $P_{\rm orb}$ and $P$. 
             A linear correlation among the DNS systems is given by Eq.~(\ref{eq:PspinPorb0}).
             Interesting questions are which, and how, HMXBs evolve to become DNS systems -- see text for discussions, and 
             Section~\ref{subsec:HMXBstat} for information on data.}
  \label{fig:corbet}
  \end{center}
\end{figure}

Be-star X-ray binaries (BeHMXBs) 
host a main-sequence donor star of spectral type B0--B2e~III/IV/V, a rapid rotator surrounded by a circumstellar (so-called ``decretion'') disk of gas, as seen by the presence of a 
prominent H$\alpha$ emission line. This disk is created by a low-velocity/high-density stellar wind of $\sim 10^{-7}\;M_{\odot}\,{\rm yr}^{-1}$ \citep{lc99}. Transient and bright (Type~I) 
X-ray outbursts periodically occur each time the compact object (usually a NS in a wide and eccentric orbit) approaches periastron and accretes matter from 
the decretion disk \citep[see][and references therein]{cc06}. These systems exhibit a correlation between NS spin and orbital period, as shown by their location from the lower center  
to the upper right of the Corbet diagram (see Fig.~\ref{fig:corbet}), due to efficient transfer of angular momentum at each periastron passage: rapidly spinning NSs correspond to 
short orbital period systems, and slowly spinning NSs to long orbit systems. The BeHMXBs are generally believed to be in spin equilibrium, meaning that the outer edge of the 
NS magnetosphere (defining the inner edge of the truncated accretion disk of the NS) rotates with the Keplerian velocity \citep{do73,swr86,wvk89}.
One can notice two outliers, however, SAX~J2103.5+4545 and 1A1118-615, from this trend among BeHMXBs. \citet{rnf+04} and \citet{spd+11} explain these systems through long episodes of 
quiescence, presumably due to the lack of sufficient circumstellar matter in the decretion disk of the Be-star, whereby accretion mainly takes place from the stellar wind 
provided by the Be-star and thus causing the pulsars to spin down to the (longer) equilibrium period that is expected for wind-fed systems.
Apart from MWC\,656 possibly hosting a BH \citep{cnr+14}, the vast majority of (if not all other) known BeHMXBs seem to host a NS.

\subsubsection{sgHMXBs}
Supergiant X-ray binaries (sgHMXBs) 
host a supergiant star of spectral type O8--B1~I/II, characterized by an intense, dense and highly supersonic radiatively-driven stellar wind \citep[e.g.][]{mkb+17}. 
B-type supergiants typically possess slower winds ($v_{\infty}=1000\;{\rm km\,s}^{-1}$) than O-type supergiants \citep[$v_{\infty}=2000\;{\rm km\,s}^{-1}$,][]{neg10}.  
There are $\sim\!16$ so-called ``classical'' persistent sgHMXBs, most of them being close systems with a compact object in a short and circular orbit, directly accreting from the stellar wind 
through a Bondi-Hoyle-like accretion process (cf. Section~\ref{subsec:HMXBwind}). Such wind-fed systems exhibit a luminous and persistent X-ray emission ($L_X = 10^{36-38}\;{\rm erg \,s}^{-1}$), 
with superimposed large variations on short timescales, and a cut-off ($10-30\;{\rm keV}$) power-law X-ray spectrum. Located in the upper left part of the Corbet diagram 
(small orbital periods of $P_{\rm orb} \simeq 3-10\;{\rm days}$ and long spin periods of $P\simeq 100-10\,000\;{\rm s}$, Fig.~\ref{fig:corbet}), sgHMXBs do not show any correlation
between $P_{\rm orb}$ and $P$, 
since there is no net transfer of angular momentum via wind accretion \citep{wvk89}. 
The detection of regular pulsations implies that they host young NSs with surface magnetic fields of $B \sim\! 10^{11-12}\;{\rm G}$. 
Nearly half of the sgHMXBs exhibit a substantial intrinsic, local extinction $n_{H} \ge 10^{22}\;{\rm cm}^{-2}$, 
with a compact object deeply embedded in the dense stellar wind \citep[such as the highly obscured IGR~J16318-4848,][]{cr12}.
Likely in transition to RLO-HMXBs, these systems are characterized by slow winds which, depending on the degree of flow through the inner Lagrangian point (L1), 
may cause shrinking of the orbit and spiral-in of the compact object over time, leading to a CE phase (Sections~\ref{subsubsec:CE} and \ref{subsec:CE}).
An example of a sgHMXB where accretion takes place both through RLO and stellar wind is the BH system Cyg\,X-1. 

A significant subclass of sgHMXBs comprises some $\sim\!12$ (and a few candidate) supergiant fast X-ray transients \citep[SFXTs, see][]{nsr+06,wlbt15}. These systems, characterized by a 
compact object orbiting with $P_{\rm orb} \simeq 3 - 100\;{\rm days}$ in a circular or eccentric orbit, and typically with $P\simeq 100-1000\;{\rm s}$, span a large area in the 
Corbet diagram, mostly in-between BeHMXBs and sgHMXBs (Fig.~\ref{fig:corbet}). They are not persistent sources, but exhibit short and intense X-ray outbursts, an unusual characteristic among HMXBs, rising in tens of minutes 
up to a peak luminosity $L_X \simeq 10^{35-37}\;{\rm erg\,s}^{-1}$, lasting for a few hours, and alternating with long ($\sim\!70\;{\rm days}$) epochs of quiescence at $L_X \sim\! 10^{32-34}\;{\rm erg\,s}^{-1}$, 
with an impressive variability factor $L_{\rm max}/L_{\rm min}$ going up to $10^2-10^6$ \citep{rbm+15}. Various processes have been invoked to explain these flares, such as wind inhomogeneities, 
magneto/centrifugal accretion barriers, transitory accretion discs, etc. \citep[see e.g.][and references therein]{cha13,wlbt15}. 
To explain the distribution of sgHMXBs in the Corbet~diagram, \citet{lcy11} argued that the majority of the systems  
evolved directly from OB-type main-sequence star--NS systems (i.e. without a significant accretion history since the formation of the NS), whereas  
a minority of the systems evolved from BeHMXBs (i.e. SFXTs with an accretion history and thus located within the area of BeHMXBs).

\subsubsection{RLO-HMXBs}
Beginning (atmospheric) Roche-lobe overflow systems (RLO-HMXBs) 
host a massive star filling its Roche lobe, where accreted matter flows via the inner Lagrangian point (L1) to form an accretion disk (similarly to the situation in LMXBs). 
This process is often referred to as beginning or ``atmospheric RLO'' \citep{sav78,sav79,sav83}. They constitute the classical bright pulsing HMXBs (i.e. Cen\,X-3, SMC\,X-1 and LMC\,X-4) 
which have high X-ray luminosities ($L_X \sim 10^{38}\;{\rm erg\,s}^{-1}$) during outbursts \citep[possibly caused by accretion disk instabilities,][]{dhl01}. 
There are only a few such sources, located in the lower left part of the Corbet diagram (Fig.~\ref{fig:corbet}), characterized by short orbital and spin periods.
The lifetime of these systems before they become dynamically unstable is expected to relatively short \citep[at most a few $10^4\;{\rm yr}$,][]{sav78}. 

\subsection{HMXB population statistics}\label{subsec:HMXBstat}
A total of 114 HMXBs are reported in \citet{lvv06}, and 117 in \citet{bbm+16}. By cross-correlating both catalogues, we compute that they share 79 HMXBs in common,
although 6 of these common sources are actually now assigned to other types in the \citet{bbm+16} catalogue.
Adding new identifications by \citet{cc13} and \citet{fcc17}, we find that the total number of HMXBs currently known in our Galaxy amounts to 167. These sources thus represent $\sim\! 43\%$ of the 
total number of high-energy binary systems, i.e. adding all known LMXBs and HMXBs reported in \citet{lvv07}, \citet{cc13}, \citet{bbm+16} and \citet{fcc17}, respectively.
Among the 167 HMXBs, there are 70 firmly identified BeHMXBs, 35 sgHMXBs (including 12 SFXTs), 3 RLO-HMXBs, and 59 HMXBs of unidentified nature \citep{fcc17}.
Thus, in relative fractions, the subpopulations of HMXBs are: 42\% BeHMXBs, 21\% sgHMXBs (7\% SFXTs), 2\% RLO-HMXBs and 35\% unidentified HMXBs. 

Contrary to LMXBs (which have lost traces of their precise birth), HMXBs are concentrated in the Galactic plane, toward tangential directions of Galactic arms, rich in star-forming regions 
and stellar complexes \citep[][and references therein]{cc13}. The distribution of HMXBs in any galaxy is thus a good indicator of star formation activity, and their collective X-ray luminosity has been used 
to compute the star-formation rate of their host galaxy \citep{ggs03}. \citet{cc13} have correlated the position of Galactic HMXBs (including BeHMXBs and sgHMXBs) with the 
position of star-forming complexes (SFCs), and showed that HMXBs are clustered within $0.3\;{\rm kpc}$ of a SFC, with an inter-cluster distance of $1.7\;{\rm kpc}$, 
thus showing that HMXBs remain close to their birth place \citep[see also][who found a slightly larger average offset between HMXBs and OB associations, 
but still consistent within error bars of \citet{cc13}]{btrj12}. 
As we shall discuss later, in connection to the kinematics of DNS systems (Section~\ref{sec:kicks}),
this is an important finding for constraining the magnitude of the kick velocity imparted on the resulting NS formed in the first SN explosion. In addition, \citet{cc13} showed  
that the HMXB distribution is offset by an age difference of $\sim\!10^7\;{\rm yr}$ with respect to the spiral arms, corresponding to the delay between the star birth and the HMXB formation. 
By taking into account the galactic arm rotation, they could derive parameters such as the age, migration distance and systemic velocities for a number of BeHMXBs and sgHMXBs.

\subsection{Evolution and destiny of HMXB systems}\label{subsec:HMXBevol}
In Fig.~\ref{fig:corbet}, we have plotted an ``extended Corbet diagram'', which includes the locations of the known DNS systems.
An interesting question is how the different types of HMXBs and DNS systems are connected in an evolutionary context. 

As the massive star in the different types of HMXBs evolves, there will be a point in most systems (except for those with very long
orbital periods) when it starts to fill its Roche lobe and transfer mass by RLO. For NS systems, the ensuing mass
transfer is generally unstable because of the large mass ratio which causes the orbit to shrink efficiently \citep[e.g.][and references therein]{tv06}. 
An additional effect that strongly contributes to unstable mass transfer is the Darwin instability\footnote{A NS orbiting a massive
star induces a tide in the massive star which will try to spin it up and ultimately make it co-rotate with the NS's orbit. However, for a
sufficiently large mass ratio, there is not enough angular momentum in
the NS's orbit to bring the massive star into co-rotation. The transfer of angular momentum from the orbit to the
star causes continued shrinking of the orbit.} \citep{dar79}. This leads to runaway accretion onto the NS and the build-up of a CE  
engulfing the whole system (cf. Sections~\ref{subsubsec:CE} and \ref{subsec:CE}). The further evolution of the system then depends on whether the CE 
can be ejected and on the fate of the NS inside the massive envelope. The following discussion assumes that the NS does
not experience hypercritical accretion inside the CE, in which case it could potentially be converted to a BH \citep{che93},
which requires a different channel for forming DNSs (see the more detailed discussion in Section~\ref{subsec:CE}).

\subsubsection{Close-orbit HMXBs and formation of T\.ZOs}
If the orbital period of the binary at the time of RLO is relatively short \citep[$\la 1\;{\rm yr}$,][]{tbo78,tth95,taa96}, 
then the binding energy of the donor star envelope remains too large to allow for a successful ejection via
orbital energy released in the spiral-in \citep[e.g.][]{ktl+16}. In this case the NS is expected to spiral to the centre, leading to
the complete coalescence of the system. The resulting objects are referred to as Thorne-\.Zytkow objects \citep[T\.ZOs,][]{tz75,tz77}.
These objects will appear as very cool red supergiants. Since these are difficult to distinguish from normal red supergiants, it is
presently not clear whether they actually exist. Since this is the
possible fate for the far majority of known HMXBs shown in Fig.~\ref{fig:corbet}, their birth rate in the Galaxy is expected to be 
quite high \citep[$\sim 2\times 10^{-4}\;{\rm yr}^{-1}$,][]{pcr95}. 
Depending on the uncertain lifetime of this phase (which is limited, e.g. by wind mass loss), a few to 10\% of all red supergiants 
with a luminosity comparable to or above the Eddington limit for a NS ($L_{\rm Edd}\sim 10^5\;L_{\odot}$) may harbour NS cores. 
Massive T\.ZOs would be distinguishable from normal red supergiants through their anomalously
large abundances of proton-rich elements, in particular molybdenum \citep{bie91,can93}. 
Despite numerous searches, to date only one candidate T\.ZO has been identified \citep[HV~2112 in the SMC,][]{lmzm14}
and its interpretation as a T\.ZO has remained controversial \citep{md16}. This suggests that the lifetime of the T\.ZO phase (if it exists) 
is much shorter than previously estimated.

\subsubsection{Wide-orbit HMXBs and formation of DNS systems}
If the orbital period of the HMXB hosting a NS is relatively long ($\ga 1\;{\rm yr}$), the binding energy of the donor star envelope at the onset of the CE is
sufficiently small to allow for envelope ejection, depending on the hydrodynamical conversion of the released orbital energy
and the mass of the donor star. \citet{ktl+16} demonstrated that solutions exist for ejecting the CE in such post-HMXB
systems with donor star masses below $\sim\! 22\;M_{\odot}$ and $\sim\! 26\;M_{\odot}$, for metallicities equivalent to
that of the Milky Way average and $Z=Z_{\odot}/50$, respectively. 
From the orbital period distribution of the known Galactic HMXBs we conclude that very few, if any, of these binaries will produce DNS systems 
--- a fact that is not in contradiction with population statistics given that the radio lifetime of DNS systems ($10^8-10^{10}\;{\rm yr}$ 
for the recycled pulsar, cf. Fig.~\ref{fig:PPdot}) is much longer than the X-ray lifetime of HMXBs (typically between some $10^5$ and a few $10^6\;{\rm yr}$).
For recycled pulsars, the characteristic age is an approximate measure of the {\it remaining} lifetime as an active radio pulsar, given that
$\tau = P/(2\dot{P}) = E_{\rm rot}/|\dot{E}_{\rm rot}|$ \citep[for discussions, see][]{tlk12}. 
The requirement of a wide orbit for the HMXB system to survive the CE phase has interesting consequences for the 
magnitude of the NS kick in the first SN explosion, cf. discussion in Section~\ref{subsec:selection_kick}.

In the case of successful CE ejection, the post-CE system will be a much closer binary consisting of the NS in
orbit with the hydrogen-exhausted core of the massive secondary star. If this remaining helium star has a mass  
$\ga 3-4\;M_\odot$, it will have a significant stellar wind \citep[e.g.][and references therein]{ywl10}. If even a small fraction of this wind is accreted by the 
NS (cf. Section~\ref{subsec:WRwind}), the system will again appear as an X-ray binary. Eventually, and in many cases following
an additional phase of mass transfer (Case~BB RLO, Section~\ref{subsec:BBRLO}), the helium star
will explode in an (ultra-)stripped SN of Type~Ib/Ic and will itself produce a NS (Fig.~\ref{fig:vdHcartoon}). Depending on the magnitude of the natal NS 
kick, the system may be disrupted in this second SN event (see Sections~\ref{subsec:kick_mag} and \ref{subsec:selection_kick}). 
If this is the case, both NSs (one a young pulsar, the other a relatively old and mildly recycled NS) will move apart as
runaway NSs \citep{rs85,tt98,lma+04}.
On the other hand, if the system remains bound, the surviving binary finally becomes a DNS system.

\subsection{ULXs: origin and connection to DNS progenitors?}\label{subsec:ULX}
\begin{figure}
  \begin{center}
     \includegraphics[width=0.72\columnwidth,angle=-90]{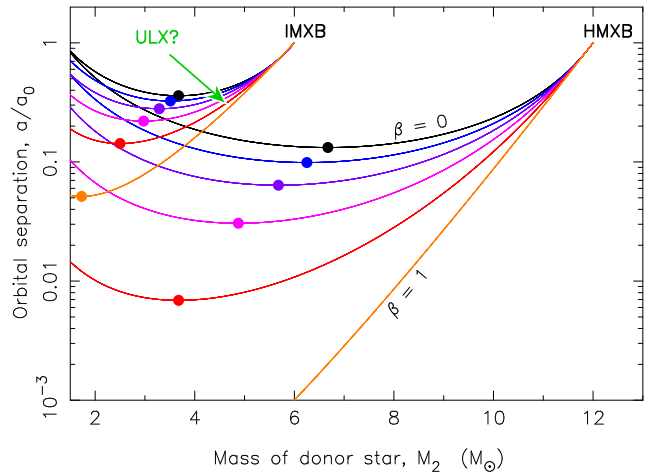}
    \vspace{0.2cm}
    \caption{Orbital evolution of IMXBs and HMXBs based on the isotropic re-emission model. The plot shows the decrease in orbital separation 
             (in units of the pre-RLO orbital separation, $a_0$) as a function of donor star mass, for an IMXB system ($M_2=6\;M_{\odot}$) and a HMXB system ($M_2=12\;M_{\odot}$).
             Minimum values of $(a/a_0)$ are shown with circles for different values of $\beta$ in steps of 0.2.
             NS~ULXs are likely to be IMXBs where the RLO can remain stable for a relatively long time, unlike the situation for HMXBs which evolve with strong orbital decay -- see Section~\ref{subsec:ULX} for discussions.}
  \label{fig:ULX}
  \end{center}
\end{figure}
The recent discovery that a number of ultra-luminous X-ray sources (ULXs: M82~X-2, NGC~7793~P13 and NGC~5907) are pulsating NSs \citep{bhw+14,ipe+17,fwh+16,ibs+17}
has lead to the suggestion that ULXs might be HMXBs \citep{eac+15,kl15,sl15,mstp15}, including BeHMXBs \citep{km16,clk+17}, and which could therefore suggest that some ULXs
might potentially lead to the production DNS systems.

The Chandra discovery of a population of ULXs with lifetimes less than about 10~Myr in the Cartwheel galaxy suggests indeed that ULXs could be related to HMXBs \citep{kin04}.
However, these ULX sources were not detected to be pulsating and might therefore host accreting BHs. If they host BHs (which are more massive than NSs) their
orbital evolution is much more stable.

ULXs have also been suggested to be intermediate-mass X-ray binaries \citep[IMXBs,][]{kl16,kar16,ibs+17} which have typical donor stars of $3-8\;M_{\odot}$.
Theoretically, it has been established that the orbital evolution of IMXBs is dynamically stable for a range of initial donor star masses and orbital periods \citep{tvs00,prp02,sl12}, 
and that these systems therefore avoid formation of a CE \citep{kb99}. During the X-ray phase, IMXBs might mimic systems like SS433 \citep{ktb00} and eventually they may evolve 
into systems like Cyg~X-2 \citep{kr99,pr00}. The final product of such IMXBs, however, is not a DNS system but instead a recycled NS orbited by a (typically CO) WD companion \citep{tvs00,prp02,lrp+11,tlk11}. 

To investigate whether ULXs are indeed IMXBs or HMXBs (and in the latter case, potential progenitors of DNS systems), we have analysed the orbital evolution of such systems. 
In Fig.~\ref{fig:ULX}, we have plotted the relative change in orbital separation ($a/a_0$) as a function of the decreasing donor star mass, $M_2$ for a HMXB (initially $M_2=12\;M_{\odot}$) and
an IMXB (initially $M_2=6\;M_{\odot}$). To calculate these tracks of non-conservative RLO we applied the so-called isotropic re-emission model \citep{bv91,vdh94a}. 
By integrating the orbital angular momentum balance equation, one can derive the analytical expression \citep{tau96}:
\begin{equation}\label{eq:tau96}
  \frac{a}{a_0}= \left( \frac{q_0(1-\beta)+1}{q(1-\beta)+1} \right) ^{\frac{3\beta-5}{1-\beta}}
                 \left( \frac{q_0+1}{q+1} \right) \left( \frac{q_0}{q} \right) ^2,
\end{equation}
where $a_0$ and $a$ refer to the orbital separation before and during RLO, respectively, and where $q_0$ and $q$ represent the
mass ratios at these two epochs. The parameter $\beta$ represents the (assumed constant) fraction of transferred material which is
lost from the vicinity of the NS. 
Here we assume that the wind-mass loss rate is negligible compared to the rate at which material is transfered via L1 during RLO, and we disregard the possibility of formation of a circumbinary disk.

Given that the thermal timescale mass-transfer rate resulting from RLO in HMXB and IMXB systems is much higher than the Eddington limit
for an accreting NS ($\sim 1.8\times 10^{-8}\;M_{\odot}\,{\rm yr}^{-1}$), we expect evolution close to $\beta \approx 1$.
There is evidence from strong spin-up torques that NSs in ULXs are able to accrete substantially above the Eddington limit \citep[even on of the order of $\sim100\;\dot{M}_{\rm Edd}$,][]{ibs+17}.
However, given that IMXBs and HMXBs immediately reach values of $|\dot{M}_2|>10^{-5}\;M_{\odot}\,{\rm yr}^{-1}$ after onset of RLO, the estimate of $\beta >0.9$ will likely still remain valid. 

From Fig.~\ref{fig:ULX}, it can be seen that the orbital shrinkage is substantially larger for a HMXB system compared to an IMXB system by more than an order of magnitude.
Therefore, as discussed previously, the HMXBs are bound to become dynamically unstable upon efficient accretion \citep[delayed-dynamical instability (DDI),][]{hw87} 
--- unless assuming some (most likely unrealistic) very large value of the wind-mass loss rate exceeding the rate at which matter is transfered via L1 during RLO \citep[i.e. $\alpha >0.9$,][]{flks15}, 
or other unconventional types of applied angular momentum losses \citep[however, see also][]{pibv16}. 

The (thermal) timescale on which such a HMXB becomes dynamically unstable after initiating RLO (and presumably forming a CE) is of the order $10^2-10^4\;{\rm yr}$ \citep{sav78}, depending on the orbital period at RLO. 
In comparison, the IMXB systems are typically able to provide mass-transfer rates, $|\dot{M}_2|>10^{-5}\;M_{\odot}\,{\rm yr}^{-1}$ for a few $10^5-10^6\;{\rm yr}$ \citep[e.g.][]{tvs00,tlk11},
followed by RLO at a lower rate for some $10^6-10^7\;{\rm yr}$, depending on the orbital period.
Therefore, based on the above arguments and timescale estimates, we find it much more plausible that any detected NS ULX system is an IMXB source compared to a HMXB source (although the latter case cannot be completely ruled out). 
Hence, we find that the recently observed pulsating NS ULXs are unlikely to be progenitors of DNS systems. The main points of the above discussions are shown in Table~\ref{table:ULX}.

\begin{table}[b]
  \caption[]{Linking IMXBs, HMXBs, NS~ULXs and DNS systems}
  \begin{center}
  \begin{tabular}{cccc}
        \hline
        \hline
	\noalign{\smallskip}
	\noalign{\smallskip}
            $M_2/M_{\odot}$ & Stability & NS~ULX source & Final outcome\\ 
           \noalign{\smallskip}
           \hline
           \noalign{\smallskip}
            $\la 5$   & stable RLO            & yes & NS + He/CO WD \\
            $6-9$     & DDI $\rightarrow$ CE  & ?   & NS + CO/ONeMg WD \\
            $\ga 10$  & CE                    & no  & DNS \\
	\noalign{\smallskip}
        \hline
        \hline
  \end{tabular}
\end{center}
%\begin{flushleft}
       {\bf Notes} --- The division between IMXBs and HMXBs is somewhat unclear but typically at a donor star mass of $M_2\simeq 8-10\;M_{\odot}$. 
       A DNS system is only produced from a HMXB evolving through a CE phase if the initial orbital period is wide enough (see~Section 3.4.2).
%\end{flushleft}
\label{table:ULX}
\end{table}

We have recently been notified that optical monitoring of the NS ULX system NGC~7793~P13 \citep{mps+14} 
has revealed a $\sim 20\;M_{\odot}$ donor star and an orbital period of $\sim 64\;{\rm days}$. If this interpretation is correct, it is a very puzzling
result and it challenges current understanding of long-term stability of such HMXBs (unless this source initiated RLO very
recently and was fortunate to be discovered in the short time span before it becomes dynamically unstable). If the system in
the future evolves through a CE phase, then we expect it to coalesce and produce a T\.ZO, as explained in Section~3.4).

To summarize our analysis on DNS progenitor systems in this section, we conclude that only wide-orbit HMXBs (possibly only a few, if any, 
of the observed Galactic HMXBs) will eventually survive the CE evolution and potentially produce DNS systems. 
NS~ULX systems are possibly IMXBs and, if so, not massive enough to be progenitors of DNS systems.

%%%%%%%%%%%%%%%%%%%%%%%%%%%%%%%%%%%%%%%%%%%%%%%%%%%%%%%%%%%%%%%%%%%%%%%%%%%
\section{Amount of mass accreted by recycled pulsars in DNS systems}\label{sec:Mrecycled}
A first estimate of the amount of matter accreted by recycled pulsars in different types of binary pulsar systems was made by \citet{tv86}. 
Here, to infer the birth masses of the first-born NSs in DNS systems (and allow for direct comparison with the masses of the second-born NSs, or NSs in other binaries like HMXBs), 
we have identified five consecutive phases of evolution where the first-born NS in a DNS system will potentially accrete material (cf. Fig.~\ref{fig:vdHcartoon}): 
\begin{itemize}
  \item[i)] wind accretion in the HMXB stage; 
  \item[ii)] CE and spiral-in evolution; 
  \item[iii)] wind accretion from the helium/Wolf-Rayet star; 
  \item[iv)] Case~BB Roche-lobe overflow; 
  \item[v)] shell impact from the SN of the secondary star. 
\end{itemize} 
We now discuss the expected amount of matter accreted by the NS in each of these five phases.

\subsection{Wind accretion in the HMXB stage}\label{subsec:HMXBwind}
The stellar wind mass-loss rate of early type stars is subject to large uncertainties. 
The observational and theoretical estimated values depend on the modelling of the terminal wind velocity, clumping effects, metallicity 
and absorption properties in the local environment \citep{kp00,vdl01,rph04,clw06,mdv+07,mp08,lan12,sha+13,cha13,fbl+15}.  
In the following, we shall begin by assuming a wind mass-loss rate of a typical OB-star companion in a HMXB of 
$|\dot{M}_{\rm 2,wind}|\simeq 10^{-6}-10^{-5}\;M_{\odot}\,{\rm yr}^{-1}$ (assuming that the effect of X-ray irradiation on the
wind mass-loss rate of the companion star is negligible). 
The wind velocity can be approximated by the escape velocity at the stellar surface: $v_{\rm wind}\simeq v_{\rm esc}=\sqrt{2GM_2/R_2}$,
where $M_2$ and $R_2$ are the companion star mass and radius, and $G$ is the constant of gravity.
Moreover, we assume Bondi-Hoyle-like accretion to be valid, given the highly supersonic flow of the stellar wind from the companion star.
The gravitational capture radius of the accreting NS is then given by: $R_{\rm acc}=2GM_{\rm NS}/(v_{\rm rel}^2+c_s^2)$,
where $c_s=\sqrt{\gamma P/\rho}\approx 11\,\sqrt{T/(10^4\,K)}\;{\rm km\,s}^{-1}$ is the local sound speed of the 
ambient medium ($\gamma$ is the adiabatic index, $P$ is the pressure, $\rho$ is the mass density and
$T$ is the temperature).
The relative velocity between the wind and the NS ($\vec{v}_{\rm rel}=\vec{v}_{\rm wind}+\vec{v}_{\rm orb}$) is usually dominated
by the wind velocity compared to the orbital velocity, i.e. $v_{\rm wind}\gg v_{\rm orb}$ (except for very tight, and/or highly eccentric binaries).  
Similarly, we have $v_{\rm wind}\gg c_s$. 
The resulting accretion rate onto the NS, $\dot{M}_{\rm NS}\approx \pi R_{\rm acc}^2\rho\, v_{\rm wind}$
can be combined with the continuity equation: $\rho=|\dot{M}_{\rm 2,wind}|/(4\pi a^2\,v_{\rm wind})$
to yield:
\begin{equation}
  \dot{M}_{\rm NS}\approx \frac{(GM_{\rm NS})^2\,|\dot{M}_{\rm 2,wind}|}{a^2 v_{\rm wind}^4},
\end{equation}
where $a$ is the orbital separation.
An estimate from this equation indicates that, in wind-driven scenarios, $\dot{M}_{\rm NS}$ is typically 
$10^3-10^4$ times lower than $|\dot{M}_{\rm 2,wind}|$ (for helium-star winds the ratio is even lower
by an order of magnitude, cf. Section~\ref{subsec:WRwind}).
Hence, we adopt a typical NS accretion rate of $\dot{M}_{\rm NS}\approx 10^{-9}\;M_{\odot}\,{\rm yr}^{-1}$,
corresponding to an X-ray luminosity of the order, $L_x\approx 10^{37}\;{\rm erg\,s^{-1}}$.

To estimate the amount of material accreted by the NS during wind accretion in the HMXB stage,
we then simply multiply this rate with the expected lifetime of the HMXB companion star during
which NS wind accretion is active: ${\Delta M}_{\rm NS}=\dot{M}_{\rm NS} \cdot \Delta t_2$.
However, an analysis of the X-ray luminosity function of Galactic HMXBs \citep{ggs02} shows
that only $\sim\!$10\% of the sources have $L_x \ge 10^{37}\;{\rm erg\,s^{-1}}$.
This fact probably reflects a limited amount of time in which a HMXB companion star is able
to deliver a sufficiently high wind mass-loss rate and/or our initially assumed (average) wind mass-loss rate, $|\dot{M}_{\rm 2,wind}|$ is too large. 
During most of the (main sequence) lifetime of the HMXB donor star, the value of $|\dot{M}_{\rm 2,wind}|$ is often lower
than $10^{-6}\;M_{\odot}\,{\rm yr}^{-1}$. 
Indeed, from a study of 10 eclipsing HMXBs, \citet{fbl+15} list typical values of $|\dot{M}_{\rm 2,wind}|\approx 10^{-7}-10^{-6}\;M_{\odot}\,{\rm yr}^{-1}$,
based on observations of their average NS accretion rates.  
Hence, by adopting a value of $\Delta t_2 \sim 10^7\;{\rm yr}$ and slightly smaller values of $|\dot{M}_{\rm 2,wind}|$ and $\dot{M}_{\rm NS}$,  we finally obtain
${\Delta M}_{\rm NS}\approx {\rm a~few}\,\times 10^{-3}\;M_{\odot}$ for wind accretion during the HMXB phase.

The structured nature of the stellar wind of a supergiant star (i.e. a wind where cool dense clumps are embedded in a photoionized gas) 
makes it difficult to derive an estimate of its mass-loss rate that is more accurate than a factor of a few \citep[e.g.][]{mkb+17}.
Another caveat is that the presence of a NS causes hydrodynamical interactions of the wind flow in close binaries \citep{mwb12}.
However, it is also possible that a strong reduction of the long-term mass-accretion rate is produced by the effect of
a strong magnetic field and/or a fast spin of the NS, which could lead to magnetic and/or centrifugal barriers, respectively \citep{bfs08,boff16}.
Such complications are not taken into account in our simplified Bondi-Hoyle-like treatment of the accretion process (see also Section~\ref{subsec:NSefficiency}).

\subsection{CE and spiral-in evolution}\label{subsec:CE}
The question of accretion onto a NS undergoing spiral-in during a CE
event has been a subject of long debate. From a theoretical point of view, 
it has been suggested that a NS in a CE might suffer from hypercritical accretion \citep{che93}
leading to its collapse into a BH. Thus to explain the observed DNS systems, an alternative
scenario without CE evolution was invented: the {\it double core scenario} \citep{bro95,dps06}.
In this scenario, two stars with an initial mass ratio close to unity evolve in parallel and reach 
the giant stage roughly at the same time. Therefore, when the CE forms it will embed both stars in their giant stages (or as a giant star and a helium core)
thereby avoiding the formation of a CE with a NS. 
This scenario, however, only works for the evolution of two stars with a mass ratio close to 1
and thus cannot explain the observations of tight binaries with, for example, a NS orbited by a WD companion, e.g. PSR~B0655+64 \citep{dbtb82,kul86}, 
PSR~J1802$-$2124 \citep{fsk+10} and PSR~J1952+2630 \citep{ltk+14}. 
Such systems show clear evidence of having evolved through a stage with a NS embedded in a CE \citep{vt84,tlk12}.

Recent theoretical works by \citet{mr15a,mr15b} seem to resolve this paradox. From their hydrodynamical simulations they find that
a NS embedded in a CE only accretes a very modest amount of material during its spiral-in as a result of a density gradient across its
accretion radius, which strongly limits accretion by imposing a net angular momentum to the flow around the NS.
\citet{mr15b} find an upper limit of $\Delta M_{\rm NS}<0.1\;M_{\odot}$. However, they remark in their paper that they
{\it make several approximations that likely lead their calculations of accreted mass in a CE to be an overestimate},
including an assumption of a static structure for the CE, effective hypercritical accretion and cooling by neutrinos, as well as 
propagation of accreted mass all the way down to the shock radius. 

Observational support for such a small amount of NS accretion during the CE phase is
given by the four DNS systems (Table~\ref{table:DNS}) in which the first-born NS (the recycled component) has a mass of
less than $1.38\;M_{\odot}$, and also PSR~J1802$-$2124 where the NS mass is estimated to be $\sim\!1.24\pm0.11\;M_{\odot}$ \citep{fsk+10}. 
If all these NSs had accreted of the order $0.1\;M_{\odot}$ they would need to have been born with a mass of
$M_{\rm NS}<1.28\;M_{\odot}$.
While this is still possible, it would be unexpected for the mass of the first-born NS in so many systems.
Furthermore, the NS is also expected to accrete material in the subsequent phases discussed below, which would require even
smaller birth masses in these cases.
In the following, we take ${\Delta M}_{\rm NS}=0.01\;M_{\odot}$ as a reasonable estimate for the upper limit of the amount of mass accreted by
a NS during a CE phase, although we cannot fully rule out the possibility for further accretion.

\subsection{Wind accretion from the helium/Wolf-Rayet star }\label{subsec:WRwind}
The recipe for calculating the amount of mass accreted by the NS from the fast wind of its naked helium (or Wolf-Rayet, WR) star companion
is similar to that given in Section~\ref{subsec:HMXBwind}, i.e. $\Delta M_{\rm NS}\simeq |\dot{M}_{\rm He,wind}|\cdot f_{\rm acc} \cdot \Delta t_{\rm He}$,
where $\Delta t_{\rm He}$ is the lifetime of the helium star, $|\dot{M}_{\rm He,wind}|$ is its mass-loss rate, and
$f_{\rm acc}=\dot{M}_{\rm NS}/|\dot{M}_{\rm He,wind}|$ is the wind accretion efficiency.
It should be noted that helium stars with mass $\la 8\;M_{\odot}$ are never seen as WR~stars, meaning that the strong WR winds
occur only in helium stars $\ga 8\;M_{\odot}$ \citep{cro07}. Such massive helium stars, however, rarely end up as progenitors of DNS systems. 
From stellar evolution modelling of NS--helium star binaries \citep{tlp15}, we find that typically
$f_{\rm acc}\simeq {\rm a~few~}\times\;10^{-4}$ \citep[see also][]{isr96},
and we obtain values of $\Delta M_{\rm NS}<4\times 10^{-4}\;M_{\odot}$ when integrating throughout the 
wind-accretion phase of the NS--helium star binaries. 

\subsection{Case~BB Roche-lobe overflow}\label{subsec:BBRLO}
The amount of material which can be accreted by the NS as a result of Case~BB RLO is limited
by the short duration of this mass-transfer phase ($<10^5\;{\rm yr}$) combined
with an Eddington-limited accretion rate onto the NS, which is roughly given by (cgs units in the central term): 
\begin{equation}
 \dot{M}_{\rm Edd}=\frac{4\pi c\,R_{\rm NS}}{0.2\,(1+X)}\;\simeq \;3.0\times 10^{-8}\;M_{\odot}\,{\rm yr}^{-1} \,,
\end{equation}
for accretion of helium ($X=0$).
Applying detailed binary stellar evolution modelling, \citet{tlp15} systematically investigated a wide parameter space 
of initial orbital periods and helium-star masses to calculate the amount of mass accreted by the NS. 
For binaries leading to formation of DNS systems they found: $\Delta M_{\rm NS}=5\times10^{-5}-3\times 10^{-3}\;M_{\odot}$,
where $\Delta M_{\rm NS}$ is systematically decreasing with increasing values of $P_{\rm orb}$. 
These values are in agreement with previous studies of Case~BB RLO \citep{dpsv02,dp03,ibk+03,tlk12,ltk+14}.

Besides increasing the mass of the accreting NS star, Case~BB RLO also serves to transfer
angular momentum and thus spin up this first-born NS to become observable as a (mildly) recycled radio pulsar. 
Based on standard assumptions of the torque acting on the NS magnetosphere at the inner edge of the
accretion disk, combined with observational data of recycled pulsars in DNS systems,
\citet{tlk12} and \citet{ltk+14} concluded that Case~BB RLO should allow for NS accretion rates above the standard
value of $\dot{M}_{\rm Edd}$ (see above) by a factor of at least $2-3$. This result seems to be
necessary to explain the fastest spins of recycled pulsars with NS or massive WD companions
(which are also believed to have evolved via post-CE Case~BB RLO).  
An X-ray binary can circumvent the value of $\dot{M}_{\rm Edd}$, for example, due to a strong B-field of the NS (see discussions in Section~\ref{subsec:theoPorbPspin}).
Hence, we conclude that Case~BB RLO may, in a few cases, result in accretion of up to $(6-9)\times10^{-3}\;M_{\odot}$.

\subsection{Shell impact from the SN of the secondary star}\label{subsec:SN_shell_impact}
In case the secondary star explodes at a very close distance to the first-born NS, the latter
may, in principle, accrete material from the SN ejecta \citep{frr14}.
However, in order for this to happen the evolution leading to the second SN event has to
be extremely finetuned with respect to both the preceding CE phase and the evolutionary stage
of the secondary star. The system would have to almost, but not quite, merge during the CE
and at the same time secure that the secondary star explodes before the system merges
as a result of GW radiation. Although one cannot rule out that such a finetuning might
be possible in a few extremely rare cases --- leading to orbital periods of $2-10\;{\rm mins}$ at the
moment of the explosion --- such a scenario can safely be ignored for DNS progenitors in general
and thus $\Delta M_{\rm NS}\ll 10^{-3}\;M_{\odot}$.

\subsection{The NS accretion efficiency}\label{subsec:NSefficiency}
Although the HMXB donor star (the progenitor of the second-born NS) provides the material for
potential accretion onto the first-born NS in all above-mentioned phases, it is not certain
how much of that material transfered will actually be accreted. For example, some of it may be lost
from the system due to ejector or propeller effects, accretion disk instabilities and direct irradiation
of the donor’s atmosphere from the pulsar \citep[e.g.][]{is75,vpa96,dlhc99,rukl09}. 
Evidence of low NS accretion efficiencies in LMXB binaries is seen in several binary pulsar
systems with fully recycled MSPs and yet relative small NS masses \citep{ts99,avk+12,ato+17}.
An extreme example is PSR~J1918$-$0642, which is an MSP with $P=7.6\;{\rm ms}$, in orbit with a low-mass He~WD and $P_{\rm orb}=10.9\;{\rm days}$,
and yet it has a mass of only $1.18^{+0.10}_{-0.09}\;M_{\odot}$ \citep{fpe+16}. 

As discussed in Section~\ref{subsec:HMXBevol}, only wide-orbit HMXBs are expected to survive CE evolution
and produce DNS systems \citep[e.g.][]{taa96}. 
For such HMXBs, both $\Delta M_{\rm NS}$ and the efficiency of angular-momentum transfer is even more uncertain. 
Moreover, there is evidence of quite a variety in NS accretion efficiencies from the location
of the different populations of HMXBs in the Corbet diagram \citep{wvk89,cha13,lsl16}.
It is therefore likely that the estimated values of $\Delta M_{\rm NS}$ at various stages discussed above
may represent overestimates due to magnetospheric and accretion disk effects which were not taken into account.  

\subsection{Total amount of mass gained by the first-born NS}
Adding the expected amounts of material accreted by the first-born NS from each of the above-mentioned stages 
leads to a total accumulated amount of $\Delta M_{\rm NS}< 0.02\;M_{\odot}$. The exact amount depends on CE physics and 
magnetospheric conditions, i.e. the B-field and the spin evolution of the NS.
The spin-up, on the other hand, is expected to be most efficient during Case~BB RLO where an accretion disk
is always present \citep[unlike the situation in a CE,][]{mmr+17}, leading to a strong and stable accretion torque acting on the NS.
Hence, the spin period of the recycled pulsar is primarily determined by the amount of accreted material during Case~BB RLO.

\subsection{Spin periods of recycled NSs in DNS systems}\label{subsec:DNS_spins}
The minimum amount of mass accreted to recycle a pulsar to reach a certain equilibrium
spin period is estimated to be roughly given by \citep[see discussions in Sections~4.3 and 4.4 in][]{tlk12}\footnote{Under the assumption that the
location of the NS magnetosphere is approximately kept fixed and the accreting NS spins near/at equilibrium while accumulating most of its material.}:
\begin{equation}
     \Delta M_{\rm eq} \simeq 0.22\,M_{\odot}\; \frac{(M_{\rm NS}/M_{\odot})^{1/3}}{P_{\rm ms}^{4/3}} \,.
  \label{eq:deltaM}
\end{equation}
From the fact that the observed spin periods of the (mildly) recycled NSs in DNS systems are in the interval 
$23-185\;{\rm ms}$ (cf. Table~\ref{table:DNS}), we notice that only a very modest amount,  
$\Delta M_{\rm NS}=(0.2-4)\times 10^{-3}\;M_{\odot}$ is in principle needed to explain these spin periods.
Interestingly enough, these values of $\Delta M_{\rm NS}$ are in fine agreement with the
amount of accreted material derived above. 

According to the mass-transfer modelling of \citet{tlp15}, the fastest spinning recycled DNS pulsars (outside globular clusters) are expected to possess spin periods 
down to $\sim\! 11\;{\rm ms}$. Hence, no DNS systems are expected to be discovered with a spin period below this value.
This argument strengthens the hypothesis that the globular cluster DNS J1807$-$2500B, which has a spin period of 
only 4.2~ms, is the outcome of a dynamical encounter event \citep{lfrj12}. This NS was most likely spun-up in a LMXB system and where the low-mass companion (WD)
was subsequently exchanged with a more massive ($1.21\;M_{\odot}$) compact object to produce the presently observed DNS system. 

A caveat in the above discussion is the possibility of long-term accretion at a rate $\dot{M}_{\rm NS}\gg\dot{M}_{\rm Edd}$. Such a high accretion rate is apparently evident in a 
ULX system with a pulsating NS accretor \citep{ibs+17}. If such a ULX system is indeed a HMXB (rather than an IMXB, as we argue in favor of in Section~\ref{subsec:ULX}),
and {\it if} such a system could avoid a delayed dynamical instability and actually form a DNS system, then it might be possible to produce a DNS system
in the Galactic disk with a NS being fully recycled to a spin period of $1-2\;{\rm ms}$ (or even a sub-ms pulsar). A remaining puzzle, however, is how the strong B-field (if this is 
the reason behind a highly super-Eddington accreting NS) would remain strong without decaying \citep{bha02}, given the vast amount of material accreted by such a NS.  
If a hypothetical post-ULX DNS system hosted a recycled pulsar with a large B-field it would only be detectable for a short time due to rapid loss of rotational energy. 
However, the second-born NS would be detectable as a normal pulsar.

\subsection{Birth masses of NSs in DNS systems}\label{subsec:NSbirth}
The above analysis demonstrates very limited accretion ($< 0.02\;M_{\odot}$) by the first-born NS during/after the HMXB phase, and 
we can hereby conclude that the distribution of NS masses in DNS systems (Fig.~\ref{fig:NSmass}), being either the first-born (mildly recycled) NS or the second-born (young) NS,
to a large degree reflects the birth distribution of NS masses in such observed systems. 
Two interesting questions are therefore: 
  i) whether the mass difference (roughly $\sim\!0.1\;M_{\odot}$) between the first- and the second-born NSs can be understood from a stellar evolution (and/or a SN explosion) point of view; and 
 ii) how the birth masses of the NSs in DNS systems compare to the measured masses of NSs observed in X-ray binaries or in binary radio pulsar systems with a WD companion. 

To answer the first question, it is of interest to consider 
studies of the pre-SN stripping effects of the exploding stars in close orbits (Type~Ib/c SNe). Due to the presence of a non-degenerate companion star in the first SN, it is clear that the progenitor 
of the first-born NS \citep{ywl10} is usually stripped significantly less than the progenitor of the second-born NS \citep{tlp15}. 
The reason being that in the second SN, the progenitor of the exploding star is being stripped significantly deeper when the companion is a NS. 
This might give a natural explanation for potential differences in the resulting NS masses.
Note that we must be aware of potential selection effects. 
If a small NS mass is connected with a small kick (see Section~\ref{subsec:kick_mass}), the systems with the smallest masses of the second-born NSs 
have a larger probability of surviving the second SN. This provides a selection effect favouring low secondary NS masses to be overrepresented among the DNS systems.
In Section~\ref{subsec:selection_kick} we demonstrate, however, that this selection effect is rather limited in practice. 

For the second question, we see in Fig.~\ref{fig:NSmass} that while the lower end of the NS mass distribution matches fairly well between
NSs in DNS and NS+WD systems, so far no massive NS ($\ge 1.70\;M_{\odot}$) has been found in a DNS system. 
However, we are dealing with relatively small number statistics and we know of NS masses in HMXBs (the progenitors of DNS systems)
which may be as high as $\sim\!2.0\;M_{\odot}$, e.g. Vela~X-1 \citep{bkv+01,fbl+15}.
Given that wide-orbit HMXBs are progenitors of DNS systems (Section~\ref{subsec:HMXBevol}), one would indeed expect the NS masses in these HMXBs to resemble the masses of 
the first-born NSs in DNS systems based on our presented evidence for very limited accretion during/after the HMXB phase. 
Unfortunately, the masses of NSs in HMXBs \citep{fbl+15} are not measured to the same precision as radio pulsar masses, which limits the
possibility for a useful direct comparison.

%%%%%%%%%%%%%%%%%%%%%%%%%%%%%%%%%%%%%%%%%%%%%%%%%%%%%%%%%%%%%%%%%%%%%%%%%%%
%\section{The $(P_{\rm \MakeLowercase{orb}},\,P,\,\MakeLowercase{e})$--correlations}\label{sec:PorbPspinEcc}
\section{The $(P_{\rm orb},\,P,\,e)$--correlations}\label{sec:PorbPspinEcc}
An important diagnostic tool for probing formation of DNS systems is an investigation of correlations between orbital parameters
(orbital period and eccentricity) and spin period of the recycled pulsars. Any such correlation will yield valuable information
about the previous mass-transfer phase and the SN explosion. The first such investigations were carried out by \citet{dpsv02,dp03,fkl+05,dpp05}. 
A similar method is applied when investigating the formation and evolution of Be-HMXBs (Sections~\ref{subsec:HMXBpop} and \ref{subsec:HMXBevol}) 
following the first SN, see e.g. \citet{prps02} and \citet{kcp11}.

In the following, we first demonstrate an empirical correlation between $P_{\rm orb}$ and $P$ (Section~\ref{subsec:PorbPspin})
followed by theoretical modelling of such a correlation (Section~\ref{subsec:theoPorbPspin}). We then discuss correlations between $P_{\rm orb}$ and
eccentricity (Section~\ref{subsec:PorbEcc}) and $P$ and eccentricity (Section~\ref{subsec:PEcc}), with a particular focus
on the NS kick from the second SN. More detailed analyses and discussions on NS kicks follow thereafter in Section~\ref{sec:kicks}.

\subsection{The $(P_{\rm orb},\,P)$--correlation}\label{subsec:PorbPspin}
\begin{figure}[t]
 \centering
 \includegraphics[width=0.72\columnwidth,angle=-90]{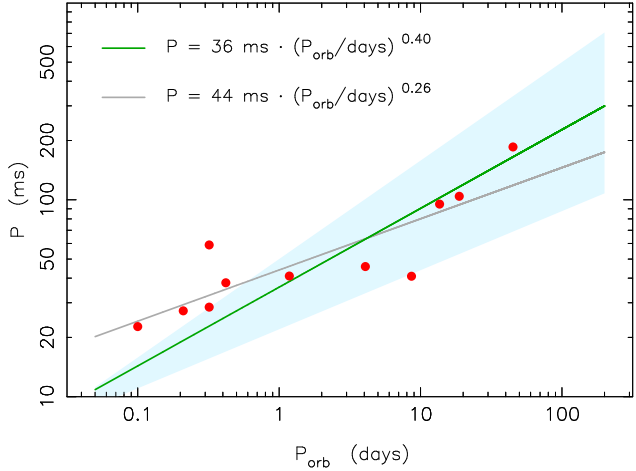}
 \caption{Spin period as a function of orbital period for all first-born (recycled) NSs in DNS systems.
          The observational data are plotted with red points (the error bars are too small to be seen). 
          The grey line (Eq.~\ref{eq:PspinPorb0}) is a fit to the raw data.
          The green line (Eq.~\ref{eq:PspinPorb}) is our estimated empirical relation at birth (i.e. right after the second SN),  
          and the shaded region is its uncertainty. 
          }
 \label{fig:PorbP}
\end{figure}

In Fig.~\ref{fig:PorbP}, we have plotted the spin period of the recycled pulsars as a function of $P_{\rm orb}$ for all observed Galactic disk DNS systems. 
A fit to the raw data results in the following correlation (grey line): 
\begin{equation}
  P = 44\;{\rm ms}\;(P_{\rm orb}{\rm /days})^{0.26} \,,
  \label{eq:PspinPorb0}
\end{equation}
based on a linear regression in the ($\log P_{\rm orb},\log P$)--plane, and with a regression coefficient of $R^2=0.73$.
However, as we shall demonstrate below, differences in the progenitor stars and their SN explosions result in a scatter in this diagram.
Furthermore, the observed data of DNS systems does not necessarily reflect their birth properties of $P_{\rm orb}$ and $P$, 
since pulsars lose rotational energy with time, and for tight DNS systems the orbits decay due to GW radiation 
(i.e. $P$ increases and $P_{\rm orb}$ decreases, cf. Section~\ref{sec:mapping}).
By taking these effects into account in a qualitative manner, and placing particular weight on the very wide-orbit DNS 1930$-$1852 \citep{srm+15}, 
we obtain the following simple empirical birth correlation (green line): 
\begin{equation}
  P\approx 36\pm14\;{\rm ms}\;(P_{\rm orb}{\rm /days})^{0.40\pm0.10}.
  \label{eq:PspinPorb}
\end{equation}

\begin{figure}[t]
\hspace{-0.5cm}
\includegraphics[width=1.05\columnwidth,angle=0]{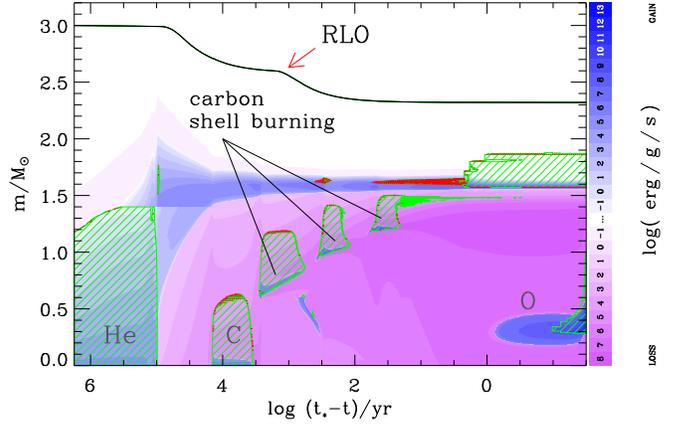}
\caption{Kippenhahn diagram of a $3.0\;M_{\odot}$ helium star undergoing Case~BB RLO to a NS
in a wide binary with $P_{\rm orb,i}=50\;{\rm days}$ (cf. data in Table~\ref{table:He3.0}).
The plot shows the evolving cross-section of the star, in mass coordinate on the y-axis, as a function of remaining
calculated lifetime on the x-axis. The green hatched areas denote zones with convection and the
intensity of the blue/purple colour indicates the net energy-production rate.
The total mass of the star is shown by the solid black line. For this wide-orbit binary 
the onset of the RLO (marked by a red arrow) will occur at a late evolutionary stage (carbon shell burning) 
such that the star will undergo core collapse within $\sim1000\;{\rm yr}$, i.e. before the end of the RLO.
Mass loss prior to RLO is due to a stellar wind.}
\label{fig:kipp_50days}
\end{figure}

A corresponding theoretical $(P_{\rm orb},\,P)$--correlation can also be derived for DNS systems 
and which can easily be understood qualitatively from binary stellar evolution arguments \citep{tlp15}.
The wider the initial orbit of the DNS progenitor system (i.e. the NS--helium~star system following the CE phase), the more evolved is the
helium star by the time it fills its Roche~lobe and initiates mass transfer. Hence, in a wide system the helium star has less time to transfer material toward the accreting NS
before it terminates its nuclear evolution and explodes in a SN. As a result, the wider the progenitor system is, the
less efficient is the recycling of the first-born NS and the larger (slower) is its rejuvenated spin period.
For example, helium star donors in wide-orbit systems may already be undergoing carbon shell burning at the onset of Case~BB~RLO (strictly speaking Case~BC~RLO) 
whereby their remaining lifetime is less than a few thousand years before they collapse.
Such a case is demonstrated in Fig.~\ref{fig:kipp_50days}. 
In the widest DNS systems, one would therefore expect to find NSs with $P>100\;{\rm ms}$ and relatively large B-fields,
i.e. {\it marginally recycled pulsars} \citep{tlp15}.
The recently discovered DNS system PSR~J1930$-$1852 \citep{srm+15} is an excellent example of such a case  
($P=185\;{\rm ms}$, $\dot{P}=1.8\times 10^{-17}\;{\rm s}\,{\rm s}^{-1}$, $P_{\rm orb}=45\;{\rm days}$, and eccentricity $e=0.40$, see Fig.~\ref{fig:PPdot}).
Its discovery is important for testing formation scenarios of such wide-orbit marginally recycled pulsars.

\subsection{Calculating a theoretical $(P_{\rm orb},\,P)$--correlation}\label{subsec:theoPorbPspin}
Based on a few assumptions, one can derive a theoretical $(P_{\rm orb},\,P)$--correlation as follows.
From the detailed grid of Case~BB RLO models by \citet{tlp15} we obtain the amount of material accreted by the NS, $\Delta M_{\rm NS}$.
Assuming the NS reaches equilibrium spin during Case~BB RLO, we can use Eq.~(\ref{eq:deltaM})
to estimate the spin period of the recycled pulsar, $P$ at the end of the mass-transfer phase \citep[see also][]{dpp05}. 
The orbital period at the end of Case~BB~RLO, just prior to the (ultra-stripped) SN explosion, is also provided by the models in \citet{tlp15}.
Assuming, as a first-order approximation, that this second SN is symmetric (without any kick imparted on the newborn NS, i.e. $w=0$, cf. Section~\ref{sec:kicks} for discussions)
we can use simple analytic prescriptions to calculate the post-SN orbital period, $P_{\rm orb}$ and the eccentricity (cf. Section~\ref{subsec:symmetric}).

\begin{figure}[t]
 \centering
 \includegraphics[width=0.72\columnwidth,angle=-90]{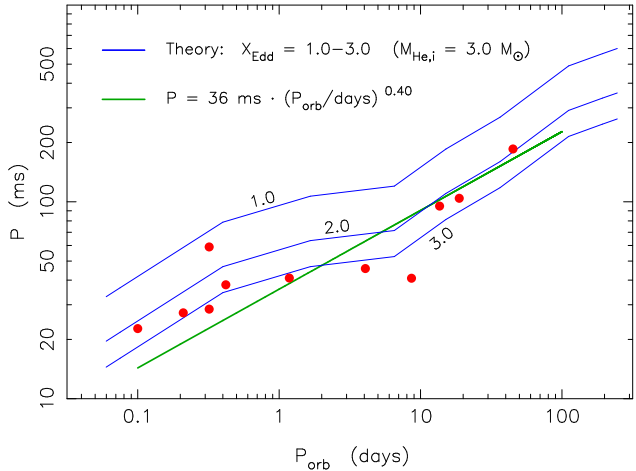}
 \caption{Theoretical correlations between orbital period and spin period of the first-born NSs in DNS systems (blue lines),
          calculated for three different values of the Eddington accretion efficiency parameter, $X_{\rm Edd}$,
          and assuming $M_{\rm He,i}=3.0\;M_{\odot}$ and no kick during the SN. The observational data are plotted with red points, 
          and the green line is our birth relation based on this data (cf. Fig.~\ref{fig:PorbP}).} 
 \label{fig:PorbP-theory1}
\end{figure}

Fig.~\ref{fig:PorbP-theory1} shows examples of such theoretical $(P_{\rm orb},\,P)$--correlations
based on purely symmetric SNe.
In this plot, we assumed in all cases an exploding star which is a stripped helium star with an
initial mass of $3.0\;M_{\odot}$. 
The parameters of the applied models from \citet{tlp15} are shown in Table~\ref{table:He3.0}.
Three different blue curves are shown in Fig.~\ref{fig:PorbP-theory1}, depending on the assumed value for the Eddington accretion efficiency parameter, $X_{\rm Edd}$.
This parameter is a measure of the maximum NS accretion rate in units of the classical Eddington limit, $X_{\rm Edd}\equiv\dot{M}_{\rm NS}^{\rm max}/\dot{M}_{\rm Edd}$.
The larger the value of $X_{\rm Edd}$, the more efficient is the recycling (more material is accreted by the NS),
and the smaller is the final spin period of the recycled pulsar.
The quoted values of $\Delta M_{\rm NS}$ in Tables~\ref{table:He3.0} and \ref{table:He_mix} are based on an assumed accretion efficiency value of $X_{\rm Edd}=1$. 
However, as discussed in Section~\ref{subsec:BBRLO}, it is possible that the amount of material accreted by the NS, $\Delta M_{\rm NS}$ is larger by 
a factor of three, based on evidence for a high accretion efficiency in PSR~J1952+2630 \citep{ltk+14}.

Super-Eddington accretion by a factor of a few is easily obtained by a thick accretion disk with a central funnel \citep{acn80}, or a
reduction in the electron scattering cross-section caused by a strong B-field \citep{bs76,pac92}.
In fact, the latter effect might even be responsible for explaining the highly super-Eddington nature (factor $\sim\!100$) of accreting
NSs in some ULXs, see Section~\ref{subsec:ULX}. However, as mentioned in Section~\ref{subsec:DNS_spins}, the current spin distribution
of observed DNS systems is fully consistent without the need for long-term accretion at rates exceeding $\dot{M}_{\rm Edd}$ by a large factor. 

Applying a SN mass cut at the edge of the metal core, and using the release of gravitational binding energy following \citet{ly89}, 
give the mass of the newborn NS: $M_{\rm NS}=M_{\rm core,f}-M_{\rm NS}^{\rm bind}$, where the
released binding energy of the NS corresponds to a mass decrease of $M_{\rm NS}^{\rm bind}=0.084\;M_{\odot}\cdot (M_{\rm NS}/M_{\odot})^2$.
The resulting masses of the newborn NSs are in this specific case between 1.31 and $1.41\;M_{\odot}$
(cf. $M_{\rm NS,2}$ in Table~\ref{table:He3.0}), given the relatively small initial helium star masses, $M_{\rm He,i}=3.0\;M_{\odot}$ assumed prior to Case~BB RLO.
Finally, knowing the mass of the newborn NS and all relevant pre-SN quantities, we are able to calculate the post-SN values of $P_{\rm orb}$.

\begin{table}[b]
  \caption[]{Properties of Case~BB RLO systems with $M_{\rm He,i}=3.0\;M_{\odot}$.}
  \begin{center}
  \begin{tabular}{ccccccc}
        \hline
        \hline
	\noalign{\smallskip}
	\noalign{\smallskip}
           $M_{\rm He,i}$ & $P_{\rm orb,i}$ & $M_{\rm He,f}$ & $M_{\rm core,f}$ & $P_{\rm orb,f}$  & $\Delta M_{\rm NS}$ & $M_{\rm NS,2}$\\ 
           ($M_{\odot}$)  & (days)          & ($M_{\odot}$)  & ($M_{\odot}$)    & (days)           & ($M_{\odot}$)       & ($M_{\odot}$)\\
           \noalign{\smallskip}
           \hline
           \noalign{\smallskip}
           3.0            & 0.08            & 1.49           & 1.45             & 0.052            & 2.3e-3              & 1.31 \\
           3.0            & 0.10            & 1.58           & 1.50             & 0.063            & 2.0e-3              & 1.35 \\
           3.0            & 0.50            & 1.73           & 1.56             & 0.311            & 7.2e-4              & 1.40 \\
           3.0            & 2.00            & 1.80           & 1.57             & 1.23             & 4.8e-4              & 1.40 \\
           3.0            & 5.00            & 1.86           & 1.57             & 4.61             & 4.1e-4              & 1.40 \\
           3.0            & 10.0            & 2.00           & 1.57             & 9.51             & 2.3e-4              & 1.40 \\
           3.0            & 20.0            & 2.17           & 1.58             & 20.0             & 1.4e-4              & 1.41 \\
           3.0            & 50.0            & 2.32           & 1.58             & 53.2             & 6.3e-5              & 1.41 \\
           3.0            & 80.0            & 2.37           & 1.57             & 87.1             & 5.2e-5              & 1.40 \\
           3.0            & 100.0           & 2.38           & 1.57             & 109.7            & 4.8e-5              & 1.40 \\
	\noalign{\smallskip}
        \hline
        \hline
  \end{tabular}
\end{center}
%\begin{flushleft}
       {\bf Notes} --- Data taken from the calculations of \citet{tlp15} based on 
       an initial NS accretor of mass $M_{\rm NS}=1.35\;M_{\odot}$ and initial orbital periods in the interval $P_{\rm orb,i}=0.08-100\;{\rm days}$.
       The quantities in the columns (left to right) are: the initial helium star mass ($M_{\rm He,i}$) and orbital period ($P_{\rm orb,i}$),
       the final mass (and metal core mass) of the exploding helium star ($M_{\rm He,f}$ and $M_{\rm core,f}$), the final orbital period prior to
       the SN explosion ($P_{\rm orb,f}$), the amount of mass accreted by the NS ($\Delta M_{\rm NS}$) assuming $X_{\rm Edd}=1.0$,
       and the estimated mass of the newborn secondary NS ($M_{\rm NS,2}$) --- see Section~\ref{subsec:theoPorbPspin}.
%\end{flushleft}
\label{table:He3.0}
\end{table}

At first look from Fig.~\ref{fig:PorbP-theory1}, one could be tempted to conclude that theoretical modelling with $X_{\rm Edd}\simeq 2.0$ can best reproduce
the observational data. However, as we shall now demonstrate, the location of each observed DNS system in any $(P_{\rm orb},\,P,\,e)$--plane
is strongly dependent on both the mass of the progenitor of the exploding star (i.e. the helium star) and the magnitude and direction of the
kick during the second SN.

\begin{figure}[t]
 \centering
 \includegraphics[width=0.72\columnwidth,angle=-90]{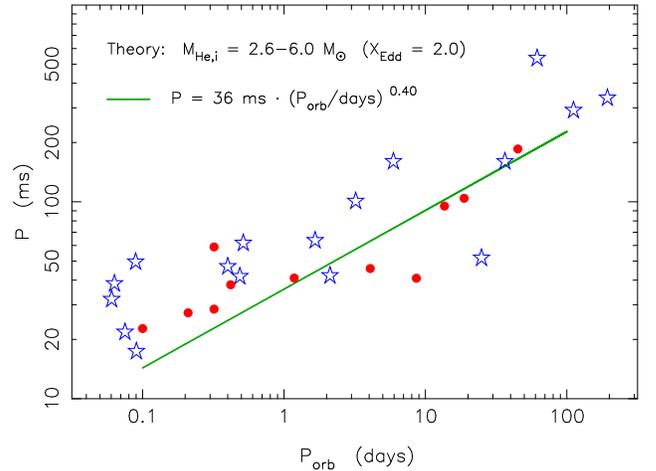}
 \caption{Spread in the $(P_{\rm orb},\,P)$--correlation is illustrated here (blue stars) when 
          applying a wide range of initial binary parameters ($M_{\rm He,i}$ and $P_{\rm orb,i}$) 
          which lead to quite different masses and orbital periods of the exploding helium stars.
          Calculations are based on data in Table~\ref{table:He_mix} and assuming a purely symmetric SN ($w=0$).
          The observational data are plotted with red points, 
          and the green line is our birth relation based on this data (cf. Fig.~\ref{fig:PorbP}).} 
 \label{fig:PorbP-theory2}
\end{figure}

To illustrate the spread in theoretically estimated points in the $(P_{\rm orb},\,P)$--plane using different initial masses of the 
progenitor of the exploding star, we show in Fig.~\ref{fig:PorbP-theory2} a number of points based on sample models from \citet{tlp15} for different
initial helium star masses in the interval $M_{\rm He,i}=2.6-6.0\;M_{\odot}$ and with different initial orbital periods (cf. Table~\ref{table:He_mix}). 
In these models, the masses of the helium stars at the moment of the (electron capture or iron core collapse, see Section~\ref{sec:kicks}) SN explosions are $1.43-2.37\;M_{\odot}$, 
with metal cores of $1.37-2.15\;M_{\odot}$.
The resulting NS masses cover a range between $1.2-1.9\;M_{\odot}$. This range is possibly wider by $\pm 0.1\;M_{\odot}$, 
if accounting for the uncertainty in the location of the mass cut and the NS EoS.
For the theoretical values, shown as blue stars in Fig.~\ref{fig:PorbP-theory2}, we assumed a fixed value of $X_{\rm Edd}=2.0$.
The fact that the scatter of observed data points is smaller than that obtained from our applied range of theoretical models 
suggests some uniformity of the parent (pre-SN) systems.  
Note, following the method described above, we also find evidence for the formation of a few DNS systems outside the range of the plane shown in Fig.~\ref{fig:PorbP-theory2}, 
i.e. DNS systems with larger values of either $P_{\rm orb}$ or $P$.

The presence of (even small) kicks also contributes significantly to smearing out the expected theoretical correlation
between $P_{\rm orb}$ and $P$ obtained for symmetric SNe. In the following sections we illustrate the effect of an asymmetric SN. 

\begin{table}[b]
  \caption[]{Properties of various Case~BB RLO systems.}
  \begin{center}
  \begin{tabular}{ccccccc}
        \hline
        \hline
	\noalign{\smallskip}
	\noalign{\smallskip}
           $M_{\rm He,i}$ & $P_{\rm orb,i}$ & $M_{\rm He,f}$ & $M_{\rm core,f}$ & $P_{\rm orb,f}$  & $\Delta M_{\rm NS}$ & $M_{\rm NS,2}$\\ 
           ($M_{\odot}$)  & (days)          & ($M_{\odot}$)  & ($M_{\odot}$)    & (days)           & ($M_{\odot}$)       & ($M_{\odot}$)\\
           \noalign{\smallskip}
           \hline
           \noalign{\smallskip}
           2.8            & 0.10            & 1.43           & 1.39             & 0.079            & 2.7e-3              & 1.26 \\
           3.0            & 0.10            & 1.58           & 1.50             & 0.063            & 2.0e-3              & 1.35 \\
           3.5            & 0.10            & 1.88           & 1.76             & 0.048            & 1.2e-3              & 1.56 \\
           4.0            & 0.10            & 2.12           & 1.96             & 0.048            & 9.4e-4              & 1.71 \\
           6.0            & 0.10            & 2.37           & 2.15             & 0.064            & 6.7e-4              & 1.86 \\
           2.7            & 0.50            & 1.48           & 1.41             & 0.415            & 8.4e-4              & 1.27 \\
           3.0            & 0.50            & 1.73           & 1.56             & 0.311            & 7.2e-4              & 1.40 \\
           3.5            & 0.50            & 2.07           & 1.80             & 0.365            & 5.0e-4              & 1.59 \\
           2.6            & 2.00            & 1.46           & 1.37             & 1.78             & 8.3e-4              & 1.24 \\
           3.0            & 2.00            & 1.80           & 1.57             & 1.23             & 4.8e-4              & 1.40 \\
           3.5            & 2.00            & 2.39           & 1.81             & 1.78             & 2.6e-4              & 1.60 \\
           4.0            & 2.00            & 2.95           & 2.02             & 2.45             & 1.4e-4              & 1.76 \\
           2.6            & 20.0            & 1.56           & 1.37             & 19.3             & 6.3e-4              & 1.24 \\
           3.0            & 20.0            & 2.17           & 1.58             & 20.0             & 1.4e-4              & 1.41 \\
           3.2            & 20.0            & 2.67           & 1.70             & 23.8             & 2.8e-5              & 1.51 \\
           3.0            & 50.0            & 2.32           & 1.58             & 53.2             & 6.3e-5              & 1.41 \\
           3.0            & 80.0            & 2.37           & 1.57             & 87.1             & 5.2e-5              & 1.40 \\
	\noalign{\smallskip}
        \hline
        \hline
  \end{tabular}
\end{center}
%\begin{flushleft}
       {\bf Notes} --- Data from \citet{tlp15}. See notes for Table~\ref{table:He3.0}.
%\end{flushleft}
\label{table:He_mix}
\end{table}

\subsection{The $(P_{\rm orb},\,e)$--correlation}\label{subsec:PorbEcc}
\begin{figure}[t]
 \centering
 \includegraphics[width=0.72\columnwidth,angle=-90]{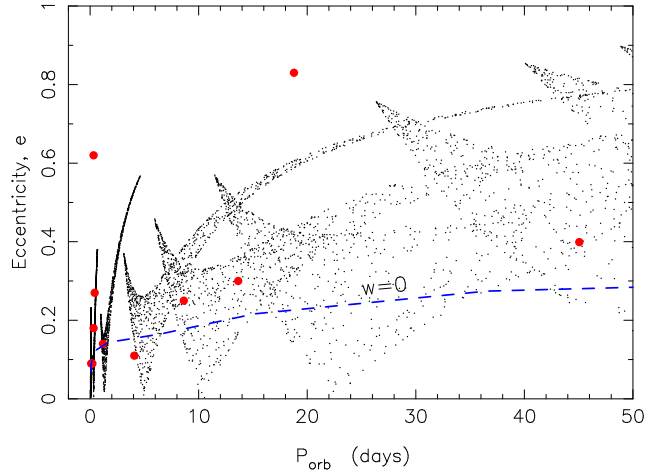}
 \caption{Eccentricity as a function of orbital period for all confirmed DNS systems observed in the Galactic disk (red points).
          Black dots are simulated systems based on
          models from \citet{tlp15}, using $M_{\rm He,i}=3.0\;M_{\odot}$ and $P_{\rm orb,i}=0.08-100\;{\rm days}$.
          For each model, 1000~resulting bound DNS systems were simulated based on SNe with a fixed kick of $w=50\;{\rm km\,s}^{-1}$ 
          and applied in a random (isotropic) direction. Unbound systems are not included. The dashed blue line
          is connecting discrete data point solutions for the resulting correlation in case of symmetric SNe ($w=0$).}
 \label{fig:ecc}
\end{figure}

In Fig.~\ref{fig:ecc}, we have plotted the eccentricities of observed DNS systems (red points)
as a function of their orbital periods. The measured error bars are too small to be seen. Also, a (weak) correlation between $P_{\rm orb}$ and eccentricity  
is expected from binary stellar evolution during Case~BB~RLO \citep{tlp15}.
Again, the argument relates to the decreasing amount of material transfered from 
helium star donors with increasing values of $P_{\rm orb}$ because of their shorter lifetimes prior to collapse
--- see fig.~16 in \citet{tlp15}. 
As a result, these wide-orbit stars possess a large envelope mass by the time they explode which increases
the instantaneous mass loss during the SN, thereby increasing the post-SN eccentricity (see dashed blue line in Fig.~\ref{fig:ecc}
as an illustrative example for symmetric SNe without a kick).
However, this correlation also depends on the initial values of $M_{\rm He,i}$. 
Furthermore, it is more sensitive to even small kicks compared to the $(P_{\rm orb},\,P)$--correlation. 

Indeed, the observed spread in eccentricities for a given value of $P_{\rm orb}$ can be understood if the second SN
in these binaries (producing the second-born, and non-recycled, pulsar) is even slightly asymmetric, e.g. $w\la50\;{\rm km\,s}^{-1}$.
In Fig.~\ref{fig:ecc}, we simulated the dynamical effects of applying kicks of $w=50\;{\rm km\,s}^{-1}$ to the stellar
models presented in Table~\ref{table:He3.0} with circular orbits prior to the SN. We performed 1000 trials with randomly oriented (isotropic) kicks.
As can be seen from the figure, the resulting post-SN systems are spread out over a large area in the $(P_{\rm orb},\,e)$-- plane.
An immediate result, however, is that with such small kicks we cannot explain two of the DNS systems. 
These are the well-known PSR~B1913+16 (the Hulse-Taylor pulsar, $P_{\rm orb}=0.323\;{\rm days}$ and $e=0.617$) and
PSR~J1811$-$1736 ($P_{\rm orb}=18.8\;{\rm days}$ and $e=0.828$). As we shall discuss in more detail in Section~\ref{sec:sim},
the latter DNS system can actually be reproduced from a small kick by increasing the value of $M_{\rm He,i}$, 
whereas PSR~B1913+16 must have had $w\ga 200\;{\rm km\,s}^{-1}$ for any mass of the exploding star to match observations. 

\begin{figure}[t]
% \centering
%\includegraphics[width=0.70\columnwidth,angle=-90]{Figures/DNS_paper_6.ps}
\includegraphics[width=0.76\columnwidth,angle=-90]{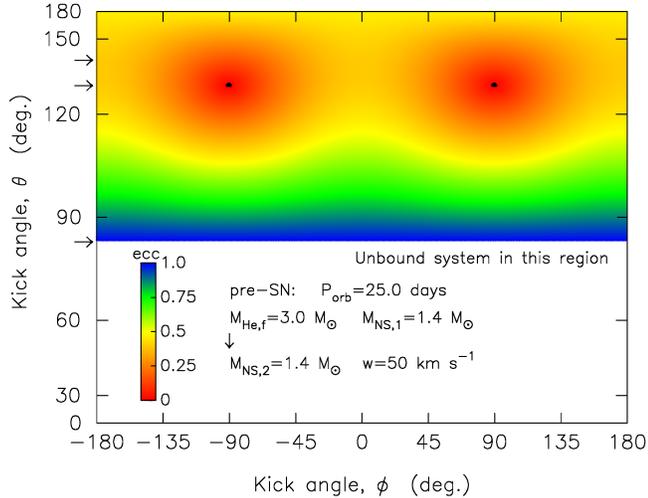}
 \caption{Dependence of the post-SN eccentricity on the direction of the kick (see Fig.~\ref{fig:kick_sphere}).
          About half (56\%) of the systems remain bound, with an average 3D systemic velocity of $\langle v_{\rm sys}\rangle =40\;{\rm km\,s}^{-1}$.
          The value of their eccentricities ($0<e<1$) are color coded, cf. scale. 
          All systems with $\theta < \theta_{\rm crit}$, however, will not survive the kick.
          The small black areas in centre of the red regions mark almost circular post-SN systems with $e<0.015$.
          The three arrows on the left side mark (bottom to top): $\theta_{\rm crit}$, $\theta_{\rm a}$ and $\theta_{\rm P}$ --- see Section~\ref{sec:kicks}.
          For another version of this plot, see Fig.~\ref{fig:ecc_angle2}.}
 \label{fig:ecc_angle}
\end{figure}

In relatively wide-orbit pre-SN binaries, even small kicks are able to produce a full range of post-SN eccentricities
between $0<e<1$, or even disrupt the system. This is demonstrated in Fig.~\ref{fig:ecc_angle}, where we modelled the
explosions of $M_{\rm He,f}=3.0\;M_{\odot}$ stars with a fixed kick magnitude of $w=50\;{\rm km\,s}^{-1}$, and varied the direction
of the applied kicks by systematically changing the two kick angles, $\theta$ and $\phi$ (see Section~\ref{sec:kicks} for a definition). 
The pre-SN orbital period was assumed to be $P_{\rm orb,f}=25\;{\rm days}$ in all cases. 
It is seen that systems with $\theta$ less than some critical value, $\theta_{\rm crit}\simeq 83^{\circ}$ will always be disrupted
--- cf. discussions in Section~\ref{sec:kicks}. Another version of this plot, for an assumed ultra-stripped SN with a large kick
in a tight orbit, is shown in the Appendix (Fig.~\ref{fig:ecc_angle2}).

\subsection{The $(P,\,e)$--correlation}\label{subsec:PEcc}
In the previous sections, we have investigated the correlations for DNS systems between $P_{\rm orb}$ and $P$, and between $P_{\rm orb}$ and eccentricity,
based on binary stellar evolution calculations starting from NS--helium star binaries. As a consequence, one might also expect a correlation between $P$ and eccentricity.
Indeed, such a correlation for DNS systems was suggested by \citet{mlc+05,fkl+05}, and investigated theoretically by \citet{dpp05}, based on the discovery
of the first seven DNS systems which could be fitted reasonably well onto a straight line in the $(P,\,e)$--plane.

\citet{fkl+05} suggested that in NS--helium star binaries, where a large amount of material
is transfered from the helium star to the NS, leading to efficient spin up and thus a small value of $P$, 
less material will be ejected during the SN explosion which results in smaller eccentricities.
This qualitative argument is in good agreement with the more detailed analysis presented here and in \citet{tlp15}.

\begin{figure}[t]
 \centering
 \includegraphics[width=0.72\columnwidth,angle=-90]{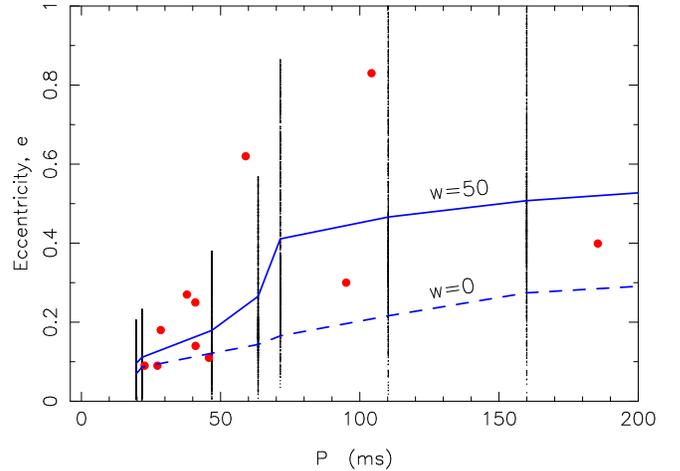}
 \caption{Eccentricity as a function of spin period for all first-born (recycled) NSs in DNS systems.
          The observational data are shown with red dots. The error bars are too small to be seen.
          Black dots along discrete vertical lines are simulated systems based on models from \citet{tlp15}, see Fig.~\ref{fig:ecc} for details.
          The dashed and solid blue lines show the resulting correlation for $w=0$ and the average values of eccentricity for bound systems using $w=50\;{\rm km\,s}^{-1}$, respectively.}
 \label{fig:ecc_spin}
\end{figure}

In Fig.~\ref{fig:ecc_spin}, we show such an expected correlation between $P$ and eccentricity for the same systems 
displayed in Fig.~\ref{fig:ecc} (cf. Table~\ref{table:He3.0}), 
and which subsequently exploded in either symmetric SNe ($w=0$, dashed blue line) or in a SNe with a fixed kick of $w=50\;{\rm km\,s}^{-1}$.
The solid blue line represents the outcome of our theoretical models with applied kicks of $w=50\;{\rm km\,s}^{-1}$ averaged over all kick directions
which resulted in bound systems. 
We assumed again an Eddington accretion efficiency parameter $X_{\rm Edd}=2.0$ for the NS.
It is clear that the observed systems (red points) do not follow the theoretical curve for a symmetric SN ($w=0$). 
The reason is again, that even a small asymmetry in the SNe will affect the eccentricities \citep[see also simulations and discussion in][]{dpp05}.
This is also seen in Fig.~\ref{fig:ecc_spin}, where the black dots clustered on discrete vertical lines represent MC simulations 
for each of the 10~values of $P_{\rm orb,i}$, assuming a random (isotropic) kick direction for a fixed kick magnitude of $w=50\;{\rm km\,s}^{-1}$ (see also Fig.~\ref{fig:ecc}).
Note, considering only the eccentricities of the DNS systems as a function of $P$ we can explain (almost) all of
the observed systems using $w\le 50\;{\rm km\,s}^{-1}$. However, there are further observational constraints than
$P$, $P_{\rm orb}$ and eccentricity, such as proper motion, which in some cases require larger values of $w$ (see discussions in Sections~\ref{sec:kicks} and \ref{sec:sim}).

In the previous four subsections (Sections~\ref{subsec:PorbPspin}--\ref{subsec:PEcc}), we have discussed the correlations
between $P$, $P_{\rm orb}$ and eccentricity for DNS systems, as expected from theoretical modelling of binary pre-SN stars
and mainly under the assumption of purely symmetric SNe. However, we have also demonstrated (cf. Figs.~\ref{fig:PorbP-theory2}, \ref{fig:ecc} and 
\ref{fig:ecc_spin}) that applying different masses of the exploding star and/or taking into account (even small) kicks will camouflage 
any such correlation. We now continue discussing kicks in the context of a $(P,\,e)$--correlation.

\subsubsection{Kick magnitudes and selection effects}\label{subsubsec:kick_selections}
The importance of the kick magnitude during the second SN in DNS systems for the suggested correlation between $P$ and eccentricity 
was investigated in detail by \citet{dpp05} using a population synthesis method, also taking into account orbital and spin evolution
since the birth of a given DNS system. They found that large kicks (using a Maxwellian kick velocity distribution with a 1D dispersion of $\sigma =190\;{\rm km\,s}^{-1}$) 
would completely randomize any correlation, whereas small kicks with typically $w<50\;{\rm km\,s}^{-1}$ ($\sigma =20\;{\rm km\,s}^{-1}$) preserve a 
correlation.

In addition, \citet{dpp05} found that the $(P,\,e)$--correlation cannot be explained simply by selection effects. Such a hypothesis
had been suggested by \citet{cb05}, who argued that the correlation is, at least partly, caused by close-orbit (thus small $P$) systems with high eccentricities 
merging on a short timescale due to GWs, thereby removing them from the observed sample. 
This hypothesis would then explain an observed DNS population with some correlation between
$P$ and eccentricity. However, using population synthesis, \citet{dpp05} showed that although close-orbit DNS systems are removed
from the expected observed sample due to merger events, the remaining systems still show a random distribution of eccentricities for 
a given value of $P$ if the applied kicks are large ($\sigma =190\;{\rm km\,s}^{-1}$, corresponding to average kicks of $\langle w \rangle = 450\;{\rm km\,s}^{-1}$). Only by applying a small kick distribution
($\sigma =20\;{\rm km\,s}^{-1}$, $\langle w \rangle = 47\;{\rm km\,s}^{-1}$), does a correlation between $P$ and eccentricity remain --- in agreement with our findings.
As we shall discuss further below, a small kick imparted onto the newborn NS in the second SN is indeed, in general, providing a better fit to observations \citep[see also][]{be16}.  
Further discussions on selection effects of the observed population due to kicks are discussed in Section~\ref{subsec:selection_kick}.

%%%%%%%%%%%%%%%%%%%%%%%%%%%%%%%%%%%%%%%%%%%%%%%%%%%%%%%%%%%%%%%%%%%%%%%%%%%%%%%%%%%%%%%%%%%%%%%%%
\section{General dynamical effects of SN{\footnotesize e}}\label{sec:kicks}
The dynamical effects of SNe in close binaries (and hierarchical triples) have been studied both analytically and numerically in a number of papers
over the last four decades, since the discovery of the Hulse-Taylor pulsar,
e.g. \citet{fv75,sut78,hil83,bp95,tb96,kal96,tt98,wkk00,pcp12}.

\begin{figure}[h]
 \centering
 \includegraphics[width=0.95\columnwidth,angle=0]{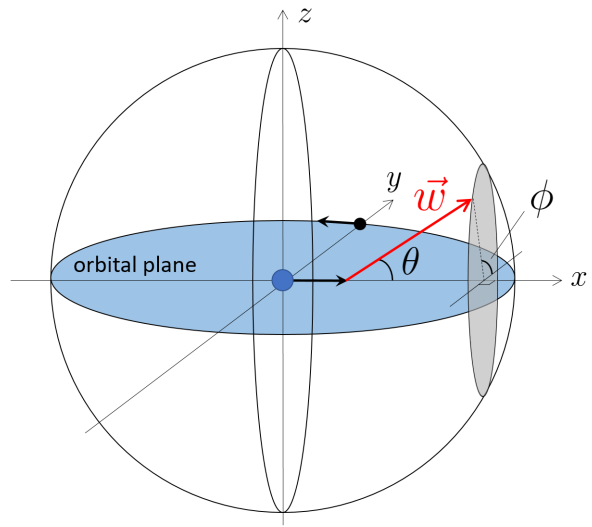}
 \caption{Illustration of the geometry of the two kick angles ($\theta,\phi$) and the kick sphere surrounding the exploding star (blue). 
          The pre-SN orbital angular momentum vector is along the z-axis.
          The kick velocity vector ($\vec{w}$) is shown in red.}
 \label{fig:kick_sphere}
\end{figure}

To solve for the dynamical effects of the SN explosion, the SN event can be assumed to be instantaneous given that
the SN ejecta velocity is much greater than the binary orbital velocity\footnote{We ignore the extremely rare
cases discussed in Section~\ref{subsec:SN_shell_impact}.}.
The mass loss reduces the absolute value of the potential energy and affects the orbital kinetic energy
by decreasing the reduced mass of the system. In addition, a kick is imparted on the newborn NS. Considering the change in
total energy of the system, the change in the orbital semi-major axis (ratio of final to initial value) 
can be expressed by \citep{hil83}:
\begin{equation}\label{eq:a_ratio_SN}
  \frac{a_{\rm f}}{a_{\rm i}}= \left[\frac{1-\Delta M/M}{1-2\Delta M/M -(w/v_{\rm rel})^2 -2\cos\theta \,(w/v_{\rm rel})} \right] \;,
\end{equation}
where $\Delta M$ is the amount of instantaneous mass loss from the exploding star (in our notation for the second SNe applied here,
$\Delta M = M_{\rm He,f}-M_{\rm NS,2}$), $M=M_{\rm He,f}+M_{\rm NS,1}$ is the total mass of the pre-SN system and
$v_{\rm rel}$ is the relative velocity between the two stars ($\sqrt{G(M_{\rm He,f}+M_{\rm NS,1})/a_{\rm i}}$).
The kick angle, $\theta$ is defined as the angle between the kick velocity vector, $\vec{w}$ and the
pre-SN orbital velocity vector of the exploding star, $\vec{v}_{\rm He}$ in the pre-SN centre-of-mass rest frame (cf. Fig.~\ref{fig:kick_sphere}). 
Using Kepler's third law, the change in $P_{\rm orb}$ can be obtained.

Eq.~(\ref{eq:a_ratio_SN}) applies to a circular pre-SN binary. This is most likely a good approximation here given the tidal interactions during Case~BB RLO.
For the more general case, see \citet{hil83}.
While the shell-impact effects on the companion star from exploding helium stars in tight binaries can potentially
be important for the first SN explosion \citep[e.g.][]{wlm75,tau15,ltr+15}, we disregard such effects in the second SN where the companion star is a NS
(essentially a point mass).

Solving for the denominator being equal to zero in Eq.~(\ref{eq:a_ratio_SN}) yields the critical angle, $\theta _{\rm crit}$,
so that $\theta < \theta _{\rm crit}$ will result in the disruption of the orbit, cf. Fig.~\ref{fig:ecc_angle}. Thus, the probability of a binary system 
surviving a SN with a kick in a random (isotropic) orientation can be found by integration and yields
$P_{\rm bound}=1-(1-\cos \theta _{\rm crit})/2$, or equivalently \citep{sut78,hil83}:
\begin{equation}
  P_{\rm bound}=\frac{1}{2}\, \bigg\{1+\left[\frac{1-2\Delta M/M -(w/v_{\rm rel})^2}{2\,(w/v_{\rm rel})}\right]\bigg\} \;,
\label{eq:bound}
\end{equation}
where $P_{\rm bound}$ is restricted to [0,\,1], and takes the value of 0 or 1, below and above this interval, respectively.  

The eccentricity of the post-SN system can be evaluated directly from the post-SN orbital angular momentum, $L_{\rm orb,f}$ and  is given by:
\begin{equation}\label{eq:ecc} 
  e=\sqrt{1+\frac{2\,E_{\rm orb,f}\,L_{\rm orb,f}^2}{\mu_{\rm f}\,G^2 M_{\rm NS,1}^2 M_{\rm NS,2}^2}} \;,
\end{equation} 
where
\begin{equation} 
  L_{\rm orb,f}=a_{\rm i}\,\mu_{\rm f}\,\sqrt{\left(v_{\rm rel}+w\cos\theta\right)^2+\left(w\sin\theta\sin\phi\right)^2} \;.
\end{equation} 
Here $\mu_{\rm f}$ and $E_{\rm orb,f}=-GM_{\rm NS,1}M_{\rm NS,2}/2a_{\rm f}$ are the post-SN reduced mass and orbital energy, respectively. 
The kick angle, $\phi$ is measured in the plane perpendicular to the pre-SN velocity vector of the exploding star, $\vec{v}_{\rm He}$ \citep{hil83,tt98},
such that the component of the kick velocity pointing directly toward to companion star is given by $w_y=w\,\sin\theta\,\cos\phi$ (Fig.~\ref{fig:kick_sphere}). 

Solving for the right-hand side in Eq.~(\ref{eq:a_ratio_SN}) being equal to 1 yields another interesting angle, $\theta _{\rm a}$: 
\begin{equation}\label{eq:theta_crit_a}
  \theta_{\rm a} = \cos^{-1}\,\left(- \,\frac{\Delta M/M + (w/v_{\rm rel})^2}{2\,(w/v_{\rm rel})}\right) \;,
\end{equation}
which determines whether the post-SN semi-major axis, $a_{\rm f}$ is larger or smaller than the pre-SN orbital radius, $a_{\rm i}$.
Thus, as a consequence of the SN explosion the orbit widens\footnote{It should be noted that in case the term in the parenthesis in Eq.~(\ref{eq:theta_crit_a}) takes a value less than $-1$, it follows that  
$\theta_{\rm a}=180^{\circ}$, meaning that all post-SN systems will widen.} if the kick angle $\theta < \theta_{\rm a}$ and it shrinks if $\theta > \theta_{\rm a}$.
The ratio ($a_{\rm f}/a_{\rm i}$) is a simple monotonically decreasing function of $\theta$ (cf. Eq.~\ref{eq:a_ratio_SN}) such that minimum
size post-SN orbits are achieved for $\theta \rightarrow 180^{\circ}$. 
For random kick directions, we find that the probability that the post-SN orbit shrinks ($a_{\rm f}<a_{\rm i}$) is given by\footnote{$P^{-}_{\rm a}$ is restricted to [0,\,$\frac{1}{2}$] 
and is always 0 below this interval.}:
\begin{equation}\label{eq:prob_crit_a}
  P^{-}_{\rm a} = \frac{1}{2} - \left[\frac{\Delta M/M + (w/v_{\rm rel})^2}{4\,(w/v_{\rm rel})}\right] \; .
\end{equation}
This expression is convenient to use as an indicator for estimating the distribution of merger timescales of DNS systems produced by the second SN. 

Combining Eqs.~(\ref{eq:ecc}) and (\ref{eq:theta_crit_a}) we can thus derive the two finetuned solutions for the kick angles 
for which the post-SN system is circular (i.e. $e=0$). The result is simply: $(\theta,\,\phi)=(\theta_{\rm a},\,\pm 90^{\circ})$,
see Fig.~\ref{fig:ecc_angle} for an example.

As a final twist in the tail, using Kepler's third law we can also work out a critical angle, $\theta _{\rm P}$: 
\begin{equation}\label{eq:theta_crit_P}
  \theta_{\rm P} = \cos^{-1}\,\left( \frac{1-2\Delta M/M - (1-\Delta M/M)^{2/3} - (w/v_{\rm rel})^2}{2\,(w/v_{\rm rel})}\right) \;,
\end{equation}
above which the post-SN orbital period is smaller than the pre-SN orbital period. The probability that $\theta > \theta_{\rm P}$ (and therefore that $P_{\rm orb}$ decreases)
is thus given by:
\begin{equation}\label{eq:prob_crit_P}
  P^{-}_{\rm P} = \frac{1}{2} - \left[ \frac{2\Delta M/M + (1-\Delta M/M)^{2/3} + (w/v_{\rm rel})^2 - 1}{4\,(w/v_{\rm rel})}\right] \;.
\end{equation}
It is always the case that $\theta_{\rm crit}<\theta_{\rm a}<\theta_{\rm P}$ for any value of $w$ and $\Delta M>0$.
Thus, for systems with a kick angle $\theta_{\rm a}<\theta<\theta_{\rm P}$, the orbital semi-major axis shrinks while at the same time
the orbital period increases as a result of the SN. Examples of $\theta_{\rm crit}$, $\theta_{\rm a}$ and $\theta_{\rm P}$
for a given SN explosion are shown in Fig.~\ref{fig:ecc_angle}. Finally, constraints on retrograde versus prograde post-SN spin--orbits are discussed
in \citet{hil83,bp95}.

\subsection{Misalignment angles}\label{subsec:misalignment}
If the kick applied to the second-born NS is directed out of the orbital plane of the pre-SN system (i.e. if the kick angle
$\phi \ne 0^{\circ}$ and $\phi \ne \pm 180^{\circ}$), then the spin axis of the recycled pulsar will be tilted with respect to 
the post-SN orbital angular momentum vector. This misalignment angle can be calculated as \citep{hil83}:
\begin{equation}\label{eq:misalign} 
  \delta = \tan ^{-1} \left( \frac{w\sin\theta\sin\phi}{\sqrt{(v_{\rm rel}+w\cos\theta)^2+(w\sin\theta\cos\phi)^2}} \right)
\end{equation} 
and leads to geodetic precession of the recycled pulsar \citep[e.g.][]{kra98}. 
To use the observed misalignment angle to constrain kick properties, we must rely on the assumption that accretion torques align 
the spin axis of the first-born NS with the orbital angular momentum vector during the recycling process \citep[e.g.][]{hil83,bv91}.
The fact that no observable effects of geodetic precession have been measured for pulsar~A in the double pulsar system J0737$-$3039, implies that
$\delta < 3.2^{\circ}$ \citep{fsk+13} and thus supports our assumption that $\delta_i=0$ prior to the second SN explosion in all DNS systems. 
Further observational evidence that such a spin-orbit alignment actually occurs during accretion in general was demonstrated for LMXBs by \citet{gt14}, 
who found agreement between the viewing angles of binary MSPs (as inferred from $\gamma$-ray light-curve modelling) 
and their orbital inclination angles. Although the timescale of accretion during Case~BB RLO in DNS progenitor systems is substantially
shorter (by two to four orders of magnitude) than that in LMXBs, the torques at work will be larger due to the 
much higher mass-transfer rates during Case~BB RLO (while the size of the magnetospheres remain roughly the same as a result of 
larger NS B-fields in DNS systems compared to fully recycled MSPs). Therefore, we find that it is reasonable to assume $\delta_i=0$ prior to
the second SN explosion and thus legitimate to use the post-SN measurements of $\delta$ of the recycled NSs to constrain kicks
in the second SN event.

\subsection{Systemic velocities}\label{subsec:velocities}
\begin{figure}[t]
 \centering
 \includegraphics[width=0.72\columnwidth,angle=-90]{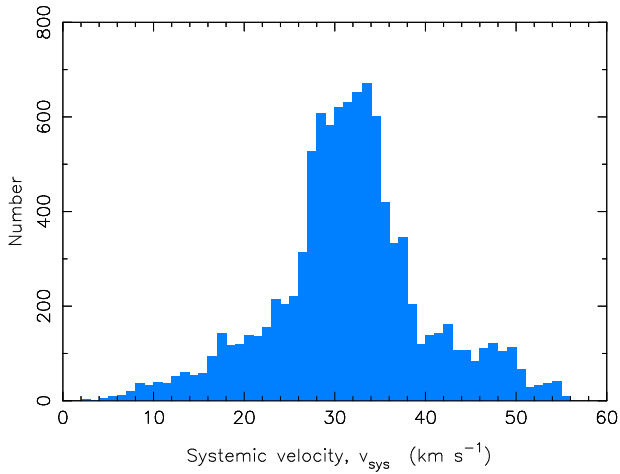}
 \caption{Distribution of $v_{\rm sys}$ for the simulated DNS systems shown in Fig.~\ref{fig:ecc} (using $w=50\;{\rm km\,s}^{-1}$); 
          see text for discussions.}
 \label{fig:velocities}
\end{figure}

Another important diagnostic quantity (besides from $P_{\rm orb}$, $P$, eccentricity, $M_{\rm NS}$ and $\delta$) for understanding DNS formation is  
their post-SN systemic velocity, $v_{\rm sys}$. 
Any DNS system receives a recoil velocity relative to the centre-of-mass rest frame of the pre-SN system.
This is due to the combined effects of sudden mass loss and a kick velocity imparted on the newborn NS.
From simple conservation of momentum considerations \citep[e.g. following][]{tb96} we can write this 3D velocity as: 
\begin{equation} 
     v_{\rm sys} = \sqrt{(\Delta p_x)^2 + (\Delta p_y)^2 + (\Delta p_z)^2} \;/ \;(M_{\rm NS,1}+M_{\rm NS,2}) \;,
  \label{eq:recoil1}
\end{equation}
where the change in momentum is given by:
\begin{equation}
   \begin{array}{lll}
   \vspace{0.2cm}
   \quad \Delta p_x   = M_{\rm NS,2}\,w\cos\theta - \Delta MM_{\rm NS,1}\sqrt{G/(M a_{\rm i})} \;,\\
   \vspace{0.2cm}
   \quad \Delta p_y   = M_{\rm NS,2}\,w\sin\theta\cos\phi \;,\\ 
   \quad \Delta p_z   = M_{\rm NS,2}\,w\sin\theta\sin\phi \;.\\
   \end{array}
\end{equation}
As an example, Fig.~\ref{fig:velocities} shows the calculated 3D systemic velocities from the simulations shown in Fig.~\ref{fig:ecc}
where randomly oriented kicks with $w=50\;{\rm km\,s}^{-1}$ were applied. 
For these resulting systems, the average systemic velocity is $\langle v_{\rm sys}\rangle \simeq 30\;{\rm km\,s}^{-1}$, which is still larger than
the values observed for some of the DNS systems (Table~\ref{table:DNS_v_LSR}) and thus indicating again that, at least in some
cases, the kicks during the second SN have magnitudes of $w<50\;{\rm km\,s}^{-1}$ 
(see detailed discussions further below). 

Finally, in addition to the kinematic effects from the second SN, DNS systems also have a relic systemic velocity component from
the first SN. This value, however, is expected to be small. Based on 
the distribution and the peculiar velocities of HMXBs, \citet{cc13} find that their resulting systemic velocities
are often only $\sim\!10-20 \; {\rm km\,s}^{-1}$. 
Even after taking into account the momentum absorbed by the massive companion star in HMXBs we find that this result 
supports the hypothesis that exploding stars generally produce smaller kicks compared to the distribution of \citet{hllk05}, if they have lost (at least) their 
hydrogen-rich envelope via mass transfer prior to the SN (see Section~\ref{subsec:NS_kicks} for discussions). 

\subsection{Symmetric SNe}\label{subsec:symmetric}
For purely symmetric SNe ($w=0$), the equations governing the dynamical effects of the SN explosion (Eqs.~\ref{eq:a_ratio_SN}--\ref{eq:ecc}) simplify to \citep{fv75}:
\begin{equation}\label{eq:aratio_symm} 
  \frac{a_{\rm f}}{a_{\rm i}}= \left[\frac{1-\Delta M/M}{1-2\Delta M/M} \right] \,,
\end{equation}
and
\begin{equation}\label{eq:ecc_symm} 
  e=\Delta M/(M_{\rm NS,1}+M_{\rm NS,2}) \,,
\end{equation}
and where the probability of remaining bound is always $P_{\rm bound}=1$ for $\Delta M/M<0.5$, whereas 
(following the virial theorem) all systems are disrupted if more than half of the total mass is lost, i.e. if $\Delta M/M>0.5$.

\subsubsection{The eccentricity floor and the NS EoS}\label{subsubsec:ecc_floor}
If the second SN in DNS systems were always symmetric ($w=0$) and the amount of SN ejecta mass negligible
(i.e. assuming extremely ultra-stripped exploding cores), then the only contribution to $\Delta M$ would be the released 
gravitational binding energy ($M_{\rm NS}^{\rm bind}$). This is released when baryonic mass is converted 
into gravitational mass during the collapse (see e.g. Section~\ref{subsec:theoPorbPspin}). Under such circumstances
one would expect to observe an ``eccentricity floor" at approximately $e=\Delta M/M\simeq 0.06$. 
From observations of many DNS systems, the exact inferred value of this ``eccentricity floor" would 
provide a measure of the released gravitational binding energy, $E_{\rm bind}$ and thereby constrain the NS EoS \citep{lfa+16}.

Unfortunately, this method is difficult in practice since even ultra-stripped NS progenitors have small amounts ($<0.20\;M_{\odot}$) of baryonic SN ejecta mass. 
For symmetric explosions with such small amounts of ejecta mass, the post-SN eccentricities of DNS systems are expected to be $e\simeq 0.07-0.13$, 
in agreement with constraints obtained from some of the known DNS systems (cf. Section~\ref{sec:sim}).
Such low eccentricities are, however, also possible for asymmetric SNe, depending on the kick direction. 
Furthermore, even small kicks of $w\sim 50\;{\rm km\,s}^{-1}$ will generally lead to a substantial spread in eccentricities (also to cases with $e<0.06$), 
as we have demonstrated here in Fig.~\ref{fig:ecc},
thereby making the idea of constraining the NS EoS impossible in practice using this method. 
Finally, in sufficiently tight binaries gravitational damping may have decreased the values of the eccentricity
significantly since the second SN explosion \citep{lfa+16}.

\subsection{NS kick magnitudes in binaries}\label{subsec:NS_kicks} 
Whereas average kicks of $400-500\;{\rm km\,s}^{-1}$ have been demonstrated for young isolated radio pulsars \citep{ll94,hllk05}, 
it has been suggested for a couple of decades that exploding stars which are stripped in close binaries (i.e. Type~Ib/c SNe) may produce substantially
smaller NS kicks compared to Type~II SN explosions of isolated, or very wide-orbit, stars \citep{tb96,prps02,vt03,plp+04,vdh04,dpp05,be16}.
This awareness came about from both theoretical and observational arguments ---
the former along the lines of stripped stars often leading to relatively fast explosions with small kicks (see Section~\ref{subsec:kick_mag}), and the latter  
from comparison of the observed radio pulsar velocity distribution, or population synthesis studies based on this distribution, 
with the space velocities, orbital periods and eccentricities of both X-ray binaries, MSPs and DNS systems. 
It is simply not possible to reproduce the observed data if exploding stars in close binaries, in general, would receive kicks of $400-500\;{\rm km\,s}^{-1}$.
The conclusion also holds when selection effects are taken into account (Section~\ref{subsec:selection_kick}). 

Indeed, the evidence for small kicks has already been pointed out in several studies \citep[e.g.][]{py98,prps02,plp+04,vdh04,ps05,spr10,fsk+13,fsk+14,bp16}.
Moreover, for PSR~J0737$-$3039 (the double pulsar) and PSR~J1756$-$2251, \citet{fsk+13,fsk+14} derived small
misalignment angles from observations of these systems, giving further support for small kicks. 

On the other hand, we have clear evidence that even relatively large kicks {\it can} happen in close-orbit DNS progenitor systems.
For example, to explain the characteristics (i.e. large proper motions) of the Hulse-Taylor pulsar (PSR~B1913+16) and PSR~B1534+12, a kick of
at least $w\simeq 200\;{\rm km\,s}^{-1}$ is needed \citep[see Sections~\ref{subsec:1534} and \ref{subsec:1913+16}, and also e.g.][]{wkk00,wwk10}. 
Furthermore, to explain the significant misalignment of the spin-axis of the B-star companion from the orbital
angular momentum vector in the pulsar binary system PSR~J0045$-$7319, a large kick is needed too \citep{kbm+96,wex98}.
Finally, there is some evidence for binaries being disrupted in the second SN, thereby explaining the
observations of isolated mildly recycled radio pulsars with similar properties to the first-born NS in DNS systems \citep{lma+04}. 
The velocities of such ejected NSs can be large even if the kick is small in case the former binary is
tight and disrupted due to a large amount of mass loss during the SN event \citep{tt98}. However, ultra-stripping prior to the second SN event
often prevents disruption due to mass loss and in these cases a large kick is needed to break up the system. 

\subsection{Theoretical estimates of NS kick magnitudes}\label{subsec:kick_mag}
Kicks are associated with explosion asymmetries and may arise from non-radial hydrodynamic instabilities in the collapsing stellar core \citep[i.e. neutrino-driven
convection and the standing accretion-shock instability,][]{bm06,skjm06,fgsj07,mj09,jan12,jan13,blh+15,fkg+15}. These instabilities are thought to lead to large-scale anisotropies
of the innermost SN ejecta, which interact gravitationally with the nascent NS and accelerate it on a timescale of several seconds \citep[e.g.][]{jan12,wjm13}.

The magnitude of the kick velocity imparted on a newborn NS is difficult to calculate, even for a given progenitor star modelled
until the onset of the core collapse.
To predict the outcome of a stellar core collapse, \citet{oo11} defined the {\it compactness} parameter of a stellar core at bounce by: 
\begin{equation}
  \xi_{\rm M} = \frac{M/M_{\odot}}{R(M)/1000\;{\rm km}}\,,
\end{equation}
where $R(M)$ is the radius of the coordinate enclosing mass, $M$. 
\citet{sw14} argued that $\xi$ at bounce is correlated with $\xi$ at the onset of core collapse, and hence the calculated 
pre-SN density structures can help in forecasting the mapping between pre-SN (ideally even ZAMS for single stars) parameters and the outcome of a core collapse.
Therefore, to enable an estimate of the expected kicks, it is necessary to calculate stellar density structures at the end of 
Case~BB RLO evolutionary models \citep{tlm+13,tlp15}. So far, however, these progenitor models have not been calculated self-consistently to 
the very onset of core collapse \citep[although see][for a composite calculation model integrating two stellar evolution codes]{mmt+17}. 

In \citet{wjm13}, a formula is presented (see their eq.~10) for estimating {\it maximum} kick magnitudes.
This formula is based on the assumption of a dipolar mass asymmetry, $\Delta m$ between the two hemispheres 
which is a fraction (maybe at most 20--30\%) of the total mass enclosed between the SN shock radius and the
proto-NS at the time the explosion sets in. 
This mass can be determined numerically and scales with the mass in the postshock layer and thus, roughly, with the mass-infall rate 
of the collapsing stellar core as a progenitor-core dependent quantity.
Moreover, the resulting kick depends on the shock velocity and thus on the explosion energy.
It is important to stress that the actual magnitude of a considered NS kick is somewhere between zero and the maximum value derived from the formulae of \citet{wjm13},
depending on the explosion asymmetry which develops stochastically. The statistical probability distribution of this stochastic asymmetry
must be determined using numerous explosion simulations.
Finally, it has been suggested that multi-dimensional effects in the convective burning shells at the onset of core collapse
can have an effect on the NS kick amplitude \citep[e.g.][]{bh96,am11,mvhj16}, i.e. pre-collapse asymmetries in the progenitor star could influence 
the asymmetry parameter of Eq.~(\ref{eq:vkick}) (see below) within the framework of the hydrodynamic kick mechanism. 

\begin{figure}[t]
 \centering
\includegraphics[width=0.70\columnwidth,angle=-90]{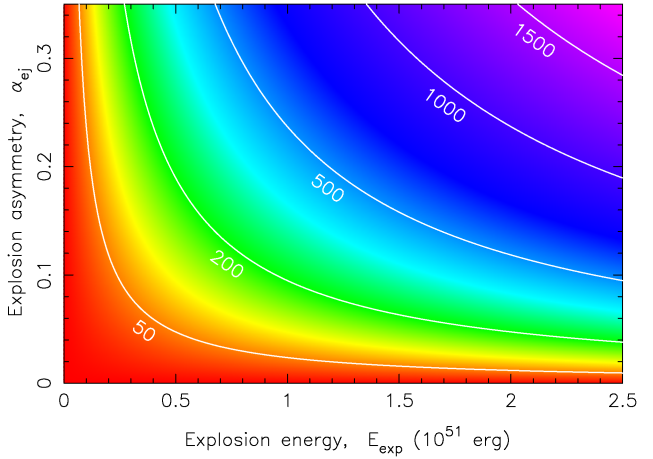}
 \caption{Dependence of the estimated NS kick magnitude, $w$ on the SN explosion energy, $E_{\rm exp}$ and the momentum-asymmetry parameter, $\alpha _{\rm ej}$, see Eq.~(\ref{eq:vkick}).
          Hyperbolas for constant values of $w=50,\,200,\,500,\,1000$ and $1500\;{\rm km\,s}^{-1}$ are shown.}
 \label{fig:janka}
\end{figure}

Recently, \citet{jan17} summarized the current theoretical understanding of how NS kicks come about using the {\it gravitational tug-boat} mechanism 
in asymmetric neutrino-driven core-collapse SN explosions. 
He derived a simple proportionality between the kick velocity, $w$, and the energy of the explosion, $E_{\rm exp}$:
\begin{equation}
    w = 211\;{\rm km\,s}^{-1}\;C\,
    \left(\frac{\alpha_{\rm ej}}{0.1}\right)
    \left(\frac{E_{\rm exp}}{10^{51}\;{\rm erg}}\right)
    \left(\frac{M_{\rm NS}}{1.5\;M_\odot}\right)^{-1} ,
\label{eq:vkick}
\end{equation}
where $C$ is a constant typically of order unity, $M_{\rm NS}$ is the (baryonic) NS mass, and $\alpha_{\rm ej}$ is the momentum-asymmetry parameter of the explosion ejecta, whose statistics needs to
be determined by hydrodynamic explosion modeling. The more asymmetric and more powerful the SN explosion is, the larger the NS kick can be (Fig.~\ref{fig:janka}). 
For example, according to \citet{jan17}, and references therein, core-collapse SNe of relatively massive iron cores have $E_{\rm exp}\simeq 3\times 10^{50}-2.5\times 10^{51}\;{\rm erg}$ and $\alpha_{\rm ej}=0-0.33$,
yielding kick values in a broad range between $w=0-1000\;{\rm km\,s}^{-1}$, whereas progenitors of electron capture SNe (EC~SNe) 
and low-mass iron core-collapse SNe (Fe~CCSNe) --- often relevant for ultra-stripped SNe, see below --- typically 
have $E_{\rm exp}\simeq 10^{50}\;{\rm erg}$ and $\alpha_{\rm ej}\la0.03$, resulting in very small kicks of $w\le 10\;{\rm km\,s}^{-1}$.

We now continue discussing the kicks expected for ultra-stripped SNe.
As we shall see in Section~\ref{sec:sim}, the possibility of such small kicks is indeed in fine agreement with constraints from many observed DNS systems.

\subsubsection{Ultra-stripped SN kicks}\label{subsubsec:ultra}
The concept of ultra-stripped SNe was recently addressed in \citet{tlm+13,tlp15}. These are SNe where a naked helium star experiences
extreme stripping by a close-orbit compact object prior to its explosion. Ultra-stripped SNe are therefore
particularly relevant for the second SN in DNS systems (Fig.~\ref{fig:vdHcartoon}).
Their expected optical signatures are compatible with observations of rapidly decaying SN light curves, such as SN~2005ek \citep{dsm+13,tlm+13,mmt+17}. 

The flavour of ultra-stripped SNe can be either an EC~SN, from a collapsing ONeMg core \citep{nom87}, or an Fe~CCSN.
The momentum kick imparted on a newborn NS via an EC~SN is always expected to be small.
This follows from detailed simulations which imply: i) explosion energies significantly smaller than those inferred for
standard Fe~CCSNe \citep{kjh06,dbo+06}; and ii) short timescales to revive the stalled SN shock,
compared to the timescales of the non-radial hydrodynamic instabilities which are required to produce strong
anisotropies, e.g. \citet{plp+04,jan12}.
Simulations of EC~SNe by \citet{kjh06,dbo+06} predict explosion energies of about (or even less than) $10^{50}\;{\rm erg}$, and
thus most likely kick velocities below $50\;{\rm km\,s}^{-1}$ (Fig.~\ref{fig:janka}).

Whereas Fe~CCSNe are certainly able to produce large kicks \citep[e.g.][]{jan17}, small NS kicks have been suggested to originate from 
CCSNe with small iron cores \citep[][and references therein]{plp+04}. For such small Fe~CCSNe the situation is similar to that of EC~SNe. 
In both cases, SN simulations \citep{gj17} suggest fast explosions where 
non-radial hydrodynamical instabilities (convectively driven or from the standing accretion shock) are unable to grow, leading to somewhat small kick velocities. 

Using binary stellar evolution arguments, \citet{tlp15} identified two factors which also imply that in ultra-stripped SNe the NS kicks may by small.
Firstly, from their modelling of the progenitor stars of ultra-stripped SNe, they have demonstrated that the expected amount of ejecta is extremely small ($\sim\!0.1\;M_{\odot}$)
compared to standard SN explosions in which several $M_{\odot}$ of material is ejected. This may likely lead to a weaker gravitational tug on the proto-NS \citep[caused by asphericity of the
ejecta, e.g.][]{jan12,jan17} and thus a small kick. 
Secondly, the binding energies of the envelopes of their final progenitor star models are often only a few $10^{49}\;{\rm erg}$, such that even a weak outgoing shock can quickly
lead to their ejection, potentially before large anisotropies can build up.

To mimic the outcome, but avoiding detailed binary stellar evolution calculations of Case~BB RLO, \citet{sys+15} performed axisymmetric hydrodynamical simulations
of neutrino-driven explosions of ultra-stripped Fe~CCSNe using the stellar evolution outcomes of single, evolved CO-stars.
All their models exhibited successful explosions driven by neutrino heating. Their diagnostic explosion energy, ejecta mass and nickel mass, were typically 
$10^{50}\;{\rm erg}$, $0.1\;M_{\odot}$ and $0.01\;M_{\odot}$, respectively, i.e. compatible with observations of rapidly decaying 
light curves such as SN~2005ek \citep{dsm+13,tlm+13}. Moreover, their calculated kick velocities were typically less than $50\;{\rm km\,s}^{-1}$,
and sometimes even below $\sim\!10\;{\rm km\,s}^{-1}$, in agreement with the simulations discussed in \citet{jan17} and references therein. 

\subsection{Relation between NS mass and SN kick?}\label{subsec:kick_mass}
While the binary stellar evolution arguments of \citet{tlp15} for small kicks hold best for the lowest mass NSs formed in their models, 
it is possible, in principle, that even rather massive NSs are produced from ultra-stripped progenitors, provided that the whole (most of the) metal core collapses.
Indeed, \citet{tlp15} predict that ultra-stripped SNe may potentially produce young NSs in DNS systems with a mass within the entire range $1.10-1.80\;M_{\odot}$.
On the other hand, it is also expected that more massive pre-SN metal cores produce larger iron cores and thus more ``normal" Fe~CCSNe with larger explosion
energies and therefore larger kicks. 

\subsubsection{Theoretical arguments for a $M_{\rm NS}-w$ relation}
While the relation in Eq.~(\ref{eq:vkick}) is supported by numerical SN simulations \citep[see references discussed by][]{jan17}, a dependence of the NS kick on the total SN ejecta mass or NS mass 
is more indirect and enters through possible correlations of these quantities with $E_{\rm exp}$. \citet{jan17} provides arguments why collapse events of stellar cores with low compactness
(which are associated with EC~SNe, low-mass Fe~CCSNe, and ultra-stripped SNe with small metal cores) can be expected to produce NSs with considerably smaller kicks than SNe from more ordinary iron cores. 
Since stellar cores with low compactness also typically produce lower-mass NSs \citep[see][]{ujma12,ejw+16,sew+16}, a relation between SN kick and NS mass seems plausible. 
We note that although in Eq.~(\ref{eq:vkick}) $w$ appears to scale inversely proportionally to $M_{\rm NS}$, the values of
$E_{\rm exp}$ and $\alpha _{\rm ej}$ are systematically smaller for explosions leading to small values of $M_{\rm NS}$, such that
a correlation between $w$ and $M_{\rm NS}$ is indeed expected in general.  

\begin{figure}[t]
  \begin{center}
     \includegraphics[width=0.72\columnwidth, angle=-90]{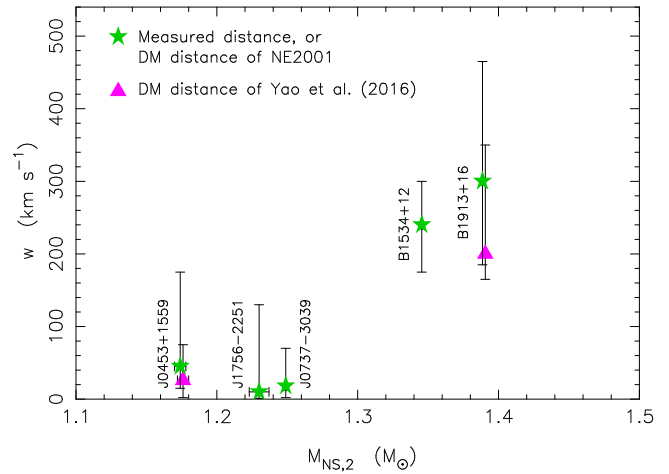}
    \vspace{0.2cm}
%   \caption{Kick velocity as a function of NS mass for the second SN explosion; see Table~\ref{table:kick_threshold} for further information.
    \caption{Kick velocity as a function of NS mass for the second SN explosion in the observed DNS systems (Table~\ref{table:kick_threshold}). 
       For each DNS system, we applied two distance models, cf. Section~\ref{subsec:vsys}. For clarity, the pink triangles are plotted systematically with an artificial 
       mass offset of $+0.002\;M_{\odot}$.
       The more massive NSs (PSR~B1534+12 and B1913+16) seem to have received significantly larger kicks compared to the low-mass NS systems 
       (see Sections~\ref{subsec:kick_mass}, \ref{subsec:1534} and \ref{subsec:1913+16}).}
  \label{fig:kick_mass}
  \end{center}
\end{figure}

\begin{table}[b]
  \caption[]{Young NS masses and resulting NS kicks in DNS systems}
  \begin{center}
  \begin{tabular}{lcccc}
        \hline
        \hline
        \noalign{\smallskip}
        \noalign{\smallskip}
                                  & $M_{\rm NS,2}$    &  $w_{\rm peak}$       &  $w_{\rm interval}$     & $e$\\
           Pulsar                 & ($M_{\odot}$)     &  $({\rm km\,s}^{-1})$ &  $({\rm km\,s}^{-1})$   & \\
           \noalign{\smallskip}
           \hline
           \noalign{\smallskip}
           B1913+16       & 1.389            & 300       & 185$-$465  & 0.617\\
           B1534+12       & 1.346            & 240       & 175$-$300  & 0.274\\
           J1906+0746     & 1.291            &  70       & 0$-$1500   & 0.085\\
           J0737$-$3039   & 1.249            &  18       & 0$-$70     & 0.088\\
           J1756$-$2251   & 1.230            &  10       & 0$-$130    & 0.181\\
           J0453+1559     & 1.174            &  45       & 15$-$175   & 0.113\\
        \noalign{\smallskip}
        \hline
        \hline
  \end{tabular}
\end{center}
%\begin{flushleft}
       {\bf Notes} --- DNS systems listed in order of measured masses of the second-born (young) NSs. 
       The values of $w_{\rm peak}$ (third column) denote the peak values of the solutions to their inferred kick velocity distribution with an interval of $w_{\rm interval}$ (fourth column),
       obtained from our analysis presented in Section~\ref{sec:sim}. Note that the peak values were estimated using flat input distributions
       of pre-SN parameters and do not reflect weighted distributions based on binary stellar evolution. The orbital eccentricities are listed in the fifth column.
       See Table~\ref{table:DNS} for further characteristics.
%\end{flushleft}
\label{table:kick_threshold}
\end{table}

\subsubsection{Observational arguments for a $M_{\rm NS}-w$ relation}
In this regard, it is interesting to note that the DNS systems where the second-born NS has a relatively large mass $\ga 1.33\;M_{\odot}$ are also those systems where 
it is evident that a large kick must have been at work, cf. Fig.~\ref{fig:kick_mass} and Table~\ref{table:kick_threshold}. 
The evidence for these large kick values is deduced from present post-SN orbital parameters of DNS systems, cf. Section~\ref{sec:sim}.
In Fig.~\ref{fig:kick_mass}, we have excluded PSR~J1518+4904 due to the relatively large error bar on its NS masses, and PSR~J1906+0746 because of the lack of a 
constraint on the kick velocity (we find solutions for $w=0-1400\;{\rm km\,s}^{-1}$, although with a peak at a low velocity of $w\simeq 70\;{\rm km\,s}^{-1}$). 
For the second-born NSs with smaller masses, we note that the inferred kicks are quite small (or could potentially have been small).

Our hypothesis of a possible correlation between $M_{\rm NS}$ and $w$ is still based on small number statistics. 
The number of observed DNS systems with precise NS mass measurements is limited. 
There could also be a transition region between small kicks and large kicks, possibly related to some element of 
stochasticity involved in the development of the explosion asymmetry \citep{wjm13} which may mask any correlation. 

To summarize our findings for NS kick magnitudes in the second SN in forming DNS systems, we conclude that while
it is difficult to estimate kick magnitudes, all of the above arguments taken together allow us to speculate that most ultra-stripped SNe (EC~SNe and, at least, 
Fe~CCSNe with relatively small metal cores) generally lead to the formation of NSs with small kicks.
Occasionally, however, large NS kicks are also at work in DNS systems. 

To explain this, we advance the hypothesis that 
for Fe~CCSNe the kick magnitude may increase with the mass of the iron core of the exploding star, and thus the mass of the resulting NS. 
Since NS masses are affected by binary interactions of the progenitor system, kicks imparted on NSs in binaries are indeed expected to be different from 
those imparted on NSs born in isolation.
Hence, if this hypothesis is correct we can also use the inferred kick magnitudes from observations as a diagnostic tool for probing the evolutionary state of the stellar
progenitor immediately prior to the second SN event. 

\subsection{Relation between NS mass and eccentricity?}\label{subsec:eccentricity_mass}
\begin{figure}[t]
  \begin{center}
     \includegraphics[width=0.72\columnwidth, angle=-90]{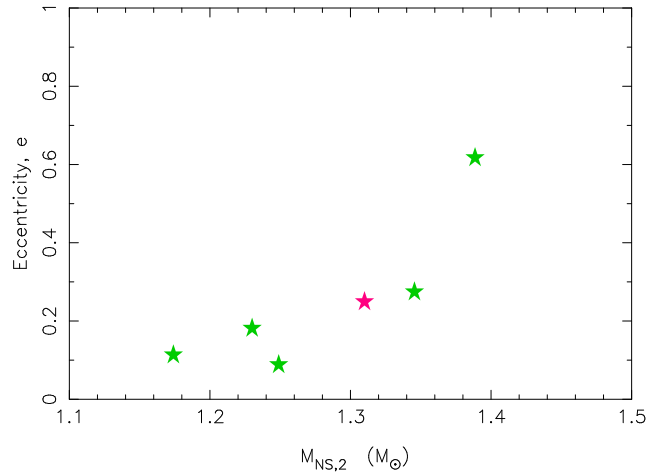}
    \vspace{0.2cm}
    \caption{Eccentricity as a function of second-born NS mass; see Table~\ref{table:kick_threshold}.
       PSR~J1518+4904 is marked in magenta color, due to an uncertainty in its mass (G.~Janssen, priv.~comm.).
       Although a correlation may seem to be present from the present data, we expect a large scattering in this
       diagram as more DNS sources are discovered.}
  \label{fig:mass-ecc}
  \end{center}
\end{figure}

Given our hypothesis of a ($M_{\rm NS},\,w$)--correlation for the second SN in DNS systems, one might also expect a ($M_{\rm NS},\,e$)--correlation, because 
systems which experience larger kicks are more likely to obtain larger eccentricities. However, as we have also demonstrated in this paper,
different kick directions result in a large spread in post-SN eccentricities, even for relatively small kicks. 
Therefore, although the six DNS systems with precisely measured masses (Table~\ref{table:kick_threshold}) do show some hint of such a correlation (cf. plot in Fig.~\ref{fig:mass-ecc}),
we would expect a rather large spread in the ($M_{\rm NS,2},\,e$)--plane, as more DNS sources are discovered, albeit some average trend may persist. 
Observational selection effects related to eccentricities of DNS systems are discussed further in Section~\ref{subsec:kick_direction}.
 
\subsection{Nature of ultra-stripped SNe in DNS progenitors}\label{subsec:ec_vs_fe}
There are two main reasons to believe that the majority of ultra-stripped SNe in DNS progenitor systems are Fe~CCSNe rather than EC~SNe.
As discussed in detail in \citet{plp+04,phlh08,tyu13,jhn+13,jhn14,dgs+15,wh15,jrp+16,rbv+17}, the ZAMS mass range of isolated stars producing EC~SNe is restricted to
a narrow interval (possibly with a width of less than $1\;M_{\odot}$) located somewhere within the range $8-11\;M_{\odot}$.
The corresponding mass window for producing EC~SNe from helium stars in close binaries is restricted to a width of $\sim\!0.2\;M_{\odot}$ \citep{tlp15}
and only for helium stars with a mass of $2.6\le M_{\rm He,i}/M_{\odot}\le2.9$ (depending on the orbital period).
Therefore, from a statistical point of view, since helium stars are formed in binaries with a broad range of masses, it would be somewhat unlikely that 
more than a few of the observed DNS systems experienced an EC~SN 
(keeping in mind that both low-mass Fe~CCSNe and EC~SNe are expected to produce small kicks, cf. Section~\ref{subsubsec:ultra}). 
Furthermore, it has been demonstrated \citep{tlm+13,sys+15,mmt+17} that ultra-stripped helium donor stars with a final metal core mass just above the interval 
producing EC~SNe ($\simeq 1.37-1.43\;M_{\odot}$) do actually produce Fe~CCSNe. The initial helium star mass range for producing Fe~CCSNe possibly extends up to $\sim\!8-10\;M_{\odot}$ 
before BHs form, depending on e.g. their wind mass-loss rates. 

\citet{nsy16} recently suggested that a relationship between the moment of inertia and the binding energy of NSs
can be used in DNS systems formed via ultra-stripped SNe to distinguish between EC~SNe and Fe~CCSNe.

Finally, we notice that the observational classification of ultra-stripped SNe can be of both Type~Ib and Type~Ic \citep{tlp15}, 
depending on the amount of helium remaining in the envelope as well as the production and mixing of nickel during the explosion. 

\subsection{Selection effects from kicks in the DNS population}\label{subsec:selection_kick}
Binary systems which experience larger SN kicks may have a smaller probability of surviving the second SN. This provides a selection effect, favouring systems descending from small NS kicks 
to be overrepresented among the observed DNS systems. 
It is therefore natural to ask whether we can use the observed sample of DNS systems to probe the SN explosions.

\begin{figure}[t]
  \begin{center}
     \includegraphics[width=0.72\columnwidth, angle=-90]{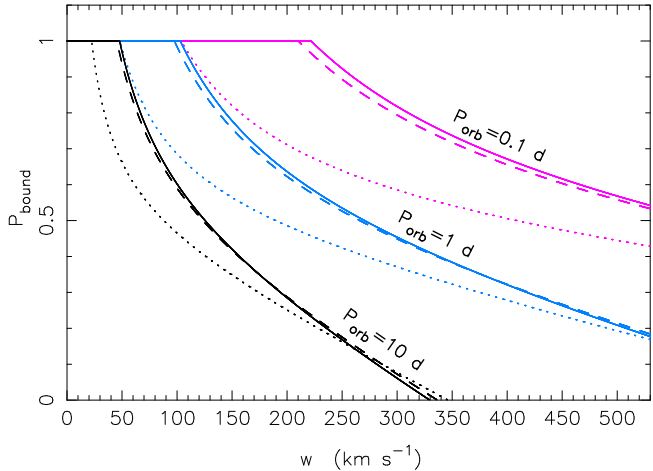}
    \vspace{0.2cm}
    \caption{Probability for surviving the second SN in a binary as a function of the imparted NS kick velocity, $w$, and for
       three different pre-SN orbital periods of 0.1, 1 and 10~days --- see Eq.~(\ref{eq:bound}).
       The solid and the dashed lines are for ultra-stripped SNe with $(M_{\rm He,f}/M_{\odot},\Delta M/M_{\odot})=(1.50,0.30)$ and $(1.80,0.45)$, respectively, 
       and the dotted line is for a less stripped SN $(3.00,1.45)$. In all cases we assumed a first-born NS mass of $M_{\rm NS,1}=1.40\;M_{\odot}$.}
  \label{fig:SN_probability2}
  \end{center}
\end{figure}

In Fig.~\ref{fig:SN_probability2} we have quantified this selection effect. 
It is seen that for tight pre-SN systems with $P_{\rm orb}=0.1\;{\rm days}$, the probability for surviving an ultra-stripped SN
remains at 100\% all the way up to $w\sim 220\;{\rm km\,s}^{-1}$, whereas for wide pre-SN system with $P_{\rm orb}=10\;{\rm days}$
the survival probability already declines from 100\% at $w\sim 50\;{\rm km\,s}^{-1}$, and reaches zero, with no systems surviving, for $w> 330\;{\rm km\,s}^{-1}$. 
The survival probability for a more massive exploding star ($M_{\rm He,f}=3.0\;M_{\odot}$, dotted line) is generally smaller because more mass is lost in the SN 
--- this amount ($\Delta M$) is the sum of baryonic material ejected in the SN shell and the mass loss carried away by neutrinos due to the
release of gravitational binding energy when the NS forms. 

Since the majority of the observed DNS systems seem to have had a pre-SN orbital period of $0.1-1\;{\rm days}$ (based on the observed distribution of
$P_{\rm orb}$ for DNS systems and our simulations in Section~\ref{sec:sim}), the survival probability remains within roughly 50\% even for large kicks up to $400\;{\rm km\,s}^{-1}$.
Therefore, we conclude that although there is indeed some selection at work for large kicks, the effect is rather limited.
Moreover, we would expect a much broader distribution of eccentricities among the DNS population in case a large fraction of DNS
systems almost become disrupted --- in particular, for wide-orbit systems where gravitational damping has little effect.

The possibility that a fair fraction of bound DNS systems leave our Galaxy as a result of very large kicks also seems unlikely 
given that the local escape velocity in the Galactic rest frame is quite large, about $550\;{\rm km\,s}^{-1}$ \citep{srh+07}.
Nevertheless, further population synthesis studies, e.g. of disrupted binaries \citep{lma+04}, might help shed more light on the kick magnitudes.  

\begin{figure}[t]
  \begin{center}
     \includegraphics[width=0.72\columnwidth, angle=-90]{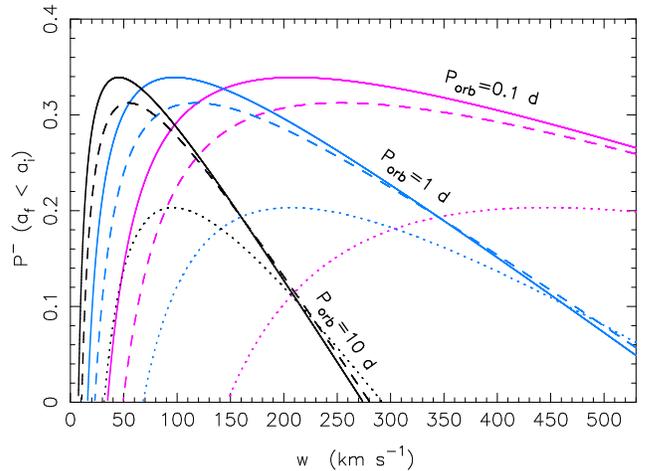}
    \vspace{0.2cm}
    \caption{Probability of the post-SN orbit shrinking as a consequence of the SN explosion with a kick value, $w$ (i.e. $(a_{\rm f}/a_{\rm i}) < 1$, 
             cf. Eqs.~\ref{eq:a_ratio_SN} and \ref{eq:prob_crit_a}) for the same systems plotted in Fig.~\ref{fig:SN_probability2}.}
  \label{fig:SN_probability_shrink}
  \end{center}
\end{figure}

For a measure of the post-SN merger timescale of a DNS system, $\tau_{\rm gwr}$ (Section~\ref{sec:mapping}), it is often useful to know the probability of the orbit shrinking as a consequence of the SN, 
i.e.  $(a_{\rm f}/a_{\rm i}) < 1$, leading to small values of $\tau_{\rm gwr}$. In Fig.~\ref{fig:SN_probability_shrink}, we have plotted these probabilities using Eq.~(\ref{eq:prob_crit_a}) 
for the same systems that exploded as in Fig~\ref{fig:SN_probability2}. It is seen that the probability of obtaining a post-SN system with a reduced
semi-major axis peaks at $\sim\!34$\% (see below). For tight pre-SN systems, e.g. $P_{\rm orb}=0.1\;{\rm days}$, it is obvious that for small kick values ($w\la50\;{\rm km\,s}^{-1}$)
the post-SN systems will always become wider ($P^{-}_{\rm a}=0$). For very wide-orbit pre-SN systems, e.g. $P_{\rm orb}=10\;{\rm days}$, it is the large
kicks (here $w\simeq 280-330\;{\rm km\,s}^{-1}$) which always cause the post-SN systems to widen. In these cases, even if the kick is directed backwards toward the companion star, 
the kick ``overshoots'' the newborn NS into a wide orbit. For kicks above $330\;{\rm km\,s}^{-1}$, the pre-SN systems with $P_{\rm orb}=10\;{\rm days}$ are all disrupted 
as shown in Fig.~\ref{fig:SN_probability2}. 

Comparing the solid, dashed and dotted curves in Fig.~\ref{fig:SN_probability_shrink}, we see that the more ultra-stripped the exploding star is,
the more likely it is to reduce its orbit as a consequence of the SN explosion.
The location and the values of the peak probabilities of $P^{-}_{\rm a}$ can easily be derived from Eq.~(\ref{eq:prob_crit_a}).
We find that the location of the peak is at $(w/v_{\rm rel})=\sqrt{\Delta M/M}$, and its value is given by:
\begin{equation}\label{eq:prob_crit_a_max}
  P^{-}_{\rm a}({\rm max}) = \frac{1}{2}\,\left[1-\sqrt{\frac{\Delta M}{M}}\right] \; .
\end{equation}
As an example, for the solid lines in Fig.~\ref{fig:SN_probability_shrink} we have $(M_{\rm He,f}/M_{\odot},\Delta M/M_{\odot})=(1.50,0.30)$ and $M_{\rm NS,1}=1.40\;M_{\odot}$, which yields
$P^{-}_{\rm a}({\rm max}) = 0.34$. 

As we argued in Section~\ref{subsec:PorbEcc}, the wider the orbit of the exploding star the less stripped it will be prior to its collapse and thus the larger is $\Delta M$.
This leads to smaller probabilities for decreasing the orbit and therefore production of a DNS system which merges within a Hubble time. However, the situation is more
complicated since the kick magnitude is likely to be correlated with $\Delta M$, as argued in the previous subsections, which affects the probability
of the post-SN orbit decreasing (Fig.~\ref{fig:SN_probability_shrink}).
Furthermore, the post-SN eccentricity is also important for $\tau _{\rm gwr}$ (see Section~\ref{sec:mapping}). 

\subsubsection{Selection effects from the first SN explosion}
\begin{figure}[t]
  \begin{center}
     \includegraphics[width=0.72\columnwidth, angle=-90]{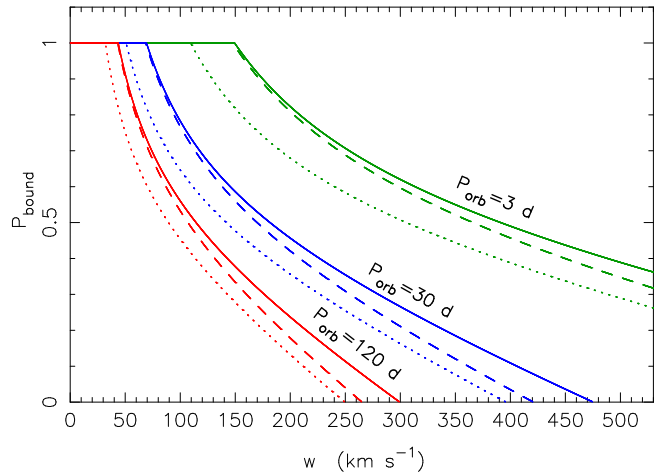}
    \vspace{0.2cm}
    \caption{Probability for surviving the first SN in a binary as a function of the imparted NS kick velocity, $w$, and for
       three different pre-SN orbital periods of 3, 30 and 120~days --- see Eq.~(\ref{eq:bound}).
       The solid, dashed and dotted lines are for Type~Ib SNe with $(M_{\rm He,f}/M_{\odot},M_2/M_{\odot})=(4.0,22.0)$, $(2.0,15.0)$ and $(4.0,12.0)$, respectively, 
       We assumed a newborn NS mass of $M_{\rm NS,1}=1.40\;M_{\odot}$.}
  \label{fig:SN_probability1}
  \end{center}
\end{figure}

Finally, to complete the discussion of NS kicks and selection effects, and their implications for DNS production, we show in Fig.~\ref{fig:SN_probability1} examples of the survival probabilities
from the {\it first} SN explosion, i.e. the probability that the post-SN systems will remain bound (and later evolve to HMXB sources).
The solid, dashed and dotted lines are for different Type~Ib SNe with typical values of $(M_{\rm He,f}/M_{\odot},M_2/M_{\odot})$ expected for the first SN. 
It is seen that the probability of surviving the first SN is strongly dependent on the pre-SN orbital period (as is also the case for the second SN).
This means that, from a pure kinematics point of view (disregarding the previous binary evolution), close-orbit HMXBs, like the RLO-HMXBs, are much more likely 
to form compared to wide-orbit HMXBs. In particular, we notice that there is a significant bias against the production of those HMXBs in 
orbits wide enough to survive the subsequent CE phase, which is needed to produce a DNS system (Section~\ref{subsec:HMXBevol}).
For example, no binaries with pre-SN $P_{\rm orb}=120\;{\rm days}$ will survive a kick of $w>300\;{\rm km\,s}^{-1}$. 
This selection effect must be kept in mind when trying to reconstruct the kick magnitudes based on observations of HMXB systems. 

For HMXBs to avoid merging in the CE phase, the systems must be in a fairly wide orbit prior to the HMXB phase. Based on a population synthesis study, \citet{vt03} 
concluded that this argument means that DNS systems can indeed only survive small kicks in the first SN in order to 
avoid disruption \citep[indeed,][discovered that there is a population of wide-orbit BeHMXBs with nearly circular orbits, indicating that in these systems the NS received small kicks]{prps02}. 
On the other hand, after the CE and spiral-in phases the orbits of the surviving systems will always be tight and therefore these systems will often be able to 
withstand even very large kicks in the second SN explosion.

A detailed population synthesis study to quantify the above-mentioned effects in more detail, and using weighted input distributions from binary stellar evolution, is
currently ongoing (Kruckow~et~al., in prep.).

%%%%%%%%%%%%%%%%%%%%%%%%%%%%%%%%%%%%%%%%%%%%%%%%%%%%%%%%%%%%%%%%%%%%%%%%%%%
\section{Mapping observed DNS systems to simulated post-SN DNS systems}\label{sec:mapping}
When comparing observational data of DNS systems with theoretically derived properties, 
it is important to take into account selection effects. As discussed already in Section~\ref{subsec:selection},
radio pulsars in highly relativistic (tight) orbit DNS systems with $P_{\rm orb}$ less than a few hours are more difficult to detect
than radio pulsars in wider orbits. Furthermore, as pointed out above, the DNS systems in close orbits evolve with time. This means that
the post-SN parameters applicable to DNS systems at birth ($P$, $P_{\rm orb}$ and eccentricity), are often somewhat 
different from their current observed values. 

\begin{figure}[t]
 \centering
 \includegraphics[width=0.95\columnwidth,angle=0]{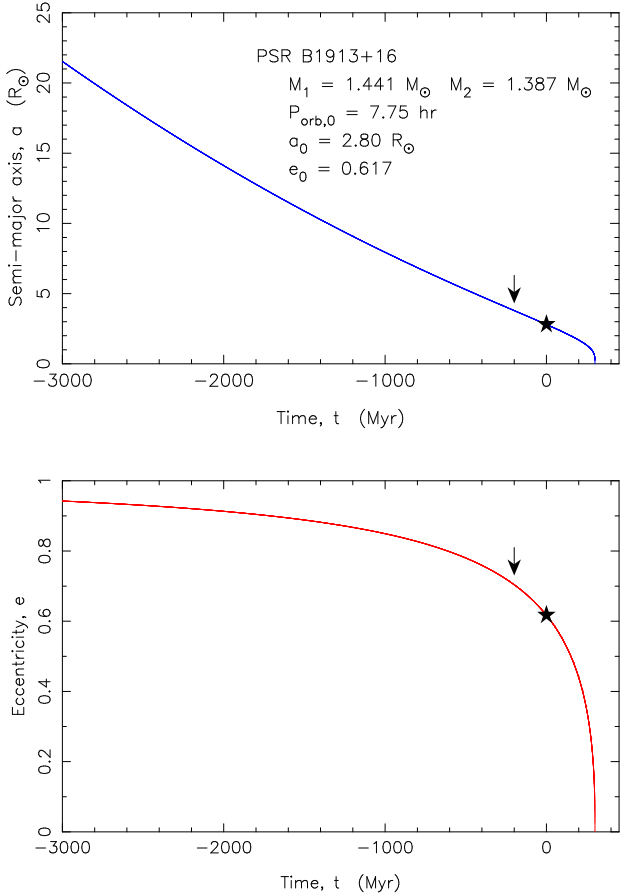}
 \caption{Past ($t<0$) and future ($t>0$) evolution of the semi-major axis (top panel) and eccentricity (bottom panel)
          of PSR~B1913+16. The current system is shown with a solid star, located at $t=0$.
          Constraints on the spin evolution (Fig.~\ref{fig:1913_spin}), however,
          yields a maximum age of this system of about 217~Myr (marked with an arrow).
          The system will merge in 301~Myr.} 
 \label{fig:1913_past_future}
\end{figure}

As an example, consider PSR~B1913+16 (the Hulse-Taylor pulsar). In Fig.~\ref{fig:1913_past_future}, we have plotted the
past and the future evolution of this system with respect to its semi-major axis and orbital eccentricity,
using the well-known quadrupole formalism of GR \citep{ein18,pet64,ll71,bla14}. 
This lowest-order secular orbital evolution is easily calculated from changes in the elements of the relative
orbit of two point masses resulting from GW damping and the rates of change are \citep{pet64}:
\begin{equation}
  \left<\frac{da}{dt}\right> = -\frac{64}{5}\frac{G^3 M^2 \mu}{c^5 a^3 (1-e^2)^{7/2}}\,\left(1+\frac{73}{24}e^2+\frac{37}{96}e^4\right)
\end{equation}
\begin{equation}
  \left<\frac{de}{dt}\right> = -\frac{304}{15}\frac{G^3 M^2 \mu\;e}{c^5 a^4 (1-e^2)^{5/2}}\,\left(1+\frac{121}{304}e^2\right)
\end{equation}
which can be transformed into an expression for the merger time, $\tau_{\rm gwr}$ as a function of the initial values:
\begin{eqnarray}\label{eq:mergertime}
  \tau_{\rm gwr} (a_0,e_0) & = & \frac{12}{19}\frac{C_0^{4}}{\beta}\\ \nonumber
                           & \times & \int_{0}^{e_0}\frac{e^{29/19}[1+(121/304)e^{2}]^{1181/2299}}{(1-e^{2})^{3/2}}\,de \;,
\end{eqnarray}
where the constants are given by: 
\begin{equation}
C_0=\frac{a_0(1-e_0^{2})}{e_0^{12/19}}\,[1+(121/304)e_0^{2}]^{-870/2299}\; , 
\end{equation}
and
\begin{equation}
\beta =\frac{64G^3}{5c^5}M^2\mu \; .
\end{equation}
Eq.~(\ref{eq:mergertime}) cannot be solved analytically and must be evaluated numerically, unless we consider  
circular orbits which can easily be solved analytically: $\tau _{\rm gwr}^{\rm circ} = a_0^4/4\beta$.
The merger timescale is very dependent on both $a_0$ and $e_0$. Tight and/or eccentric orbits
spiral-in much faster than wider and more circular orbits 

\begin{figure}[h]
 \centering
 \includegraphics[width=0.72\columnwidth,angle=-90]{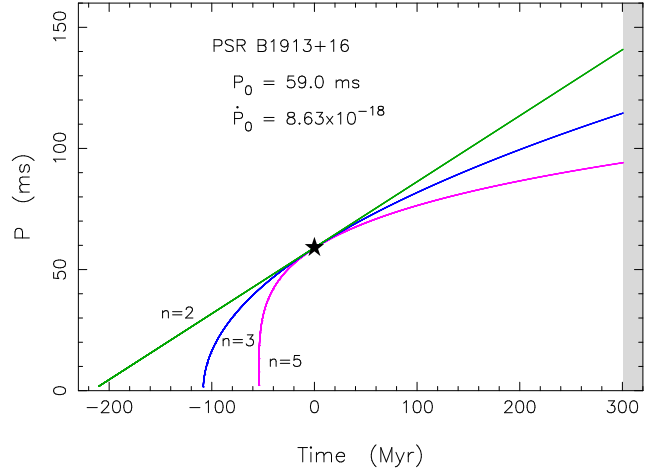}
 \caption{Past and future spin evolution of the recycled pulsar in PSR~B1913+16,
          calculated for three different values of the braking index, $n=\{2,3,5\}$.
          The current system is shown with a solid star, located at $t=0$.
          The system will merge in 301~Myr.} 
 \label{fig:1913_spin}
\end{figure}

The current location of PSR~B1913+16 is plotted in Fig.~\ref{fig:1913_past_future} by a solid star at $t=0$ ($a_0=2.80\;R_{\odot}$, $e_0=0.617$).
The system is doomed to merge in 301~Myr and will become a strong Galactic GW source at that time.
For comparison, the estimated Galactic merger rate of DNS systems is of the order $1-100\;{\rm Myr}^{-1}$ \citep{vt03,aaa+10}.
If this system had formed 3~Gyr ago, the semi-major axis and eccentricity at birth would have been about $21\;R_{\odot}$
and 0.95, respectively --- i.e. considerably larger than observed today.
The question is, how can we place limits on the age of this system to learn about its post-SN birth properties and constrain the SN explosion physics. 
To answer this question, we need to look at the spin evolution of the recycled NS.

For magnetodipole spin-down of a pulsar (or spin-down due to plasma currents in the magnetosphere and a pulsar wind), 
the braking torque, $N\propto B^2$ \citep{mt77,lk04}. In more general terms, one can express the braking torque as 
$N=I\,\dot{\Omega}\propto \Omega^{n}$, where $n$ is the braking index and $\Omega =2\pi/P$.
By integrating the deceleration law: $\dot{\Omega}=-K\Omega ^n$, one obtains the following solution
at time $t$ \citep[positive in the future; negative in the past, e.g.][]{ltk+14}: 
\begin{equation}
  P = P_0\left[ 1+(n-1) \frac{\dot{P}_0}{P_0}\,t\right] ^{1/(n-1)} 
\label{eq:spinevol}
\end{equation}
\begin{equation}
  \dot{P} = \dot{P}_0 \left( \frac{P}{P_0}\right) ^{2-n} \; .
\end{equation}

In Fig.~\ref{fig:1913_spin}, we have plotted the past and the future spin evolution of PSR~B1913+16, using three
different values of the braking index, $n=\{2,3,5\}$, see e.g. \citet{tlk12} and references therein for a discussion.
Given the relatively large B-field of this pulsar ($B_0\simeq 7\times 10^{9}\;{\rm G}$), the braking torque acting on it is substantial.
Hence, even in the most conservative case, assuming evolution with a constant $\dot{P}$ ($n=2$),
the maximum age of PSR~B1913+16 is less than about 217~Myr (twice the characteristic age, cf. Eq.~\ref{eq:spinevol}), given that the spin period after recycling 
must have been larger than zero. Hence, returning to Fig.~\ref{fig:1913_past_future} (see the arrows) we find that the maximum
values of the semi-major axis and the eccentricity after the SN in this case would be $a=3.88\;R_{\odot}$ (corresponding to $P_{\rm orb}=12.6\;{\rm hr}$) and $e=0.710$, respectively.
However, there is little evidence for pulsars generally evolving with a such a small {\it long-term} braking index of $n=2$. \citet{tk01} argued in favour
of long-term evolution with $n>3$ \citep[see further support in][]{jk17} and showed as an example that a pulsar born with $P_0=10\;{\rm ms}$ and $\dot{P}_0=10^{-12.5}\;{\rm s\,s}^{-1}$ 
would spin down to $P=10\;{\rm s}$ in only 1~Myr, in case it evolved with a constant $n=2$ (see also Eq.~\ref{eq:spinevol}). This is clearly not the case for radio pulsars in general, and  
therefore we shall assume a standard evolution with $n=3$. 
In this case, the maximum age of the system since recycling is given exactly by the characteristic age of the recycled pulsar (see Eq.~\ref{eq:spinevol} for $n=3$ and solve for $P=0$). 
For PSR~B1913+16, this yields a maximum age of $t=\tau _0 = 108.7\;{\rm Myr}$ and thus upper limits on the post-SN parameters of $a=3.34\;R_{\odot}$ 
(corresponding to $P_{\rm orb}=10.1\;{\rm hr}$) and $e=0.670$.
In the following section, we take into account this uncertainty in the age of a given DNS system when analysing its SN explosion properties.

Finally, for each DNS system, we need to relate the observed projected 2D velocity ($v_{\perp}=v_{\rm T}^{\rm SBB}$) in the plane of the sky to its 
underlying 3D systemic velocity ($v_{\rm sys}=v^{\rm LSR}$).
In general, assuming an isotropic orientation of $\vec{v}_{\rm sys}$ and given $v_{\perp}$, we find from a simple integration that the probability that $v_{\rm sys}$ is less than 
a certain value $v_{\rm sys}^{\prime}$ is given by:
\begin{equation}
  P(v_{\rm sys}<v_{\rm sys}^{\prime})=1-\sqrt{1-\psi ^2}
  \label{eq:v_proj}
\end{equation}
where $\psi$ is the ratio: $v_{\perp}/v_{\rm sys}^{\prime}$.
However, rather than the 3D systemic velocity of the DNS system relative to the Solar system barycenter, we are interested in the 3D systemic velocity in its local standard of rest, $v^{\rm LSR}$. 
These velocities were calculated via the method explained in Section~\ref{subsec:vsys}.
In the simulations below, we use the derived velocities ($v^{\rm LSR}$) from Table~\ref{table:DNS_v_LSR} to help constrain the SN explosion and the pre-SN binary parameters.

%%%%%%%%%%%%%%%%%%%%%%%%%%%%%%%%%%%%%%%%%%%%%%%%%%%%%%%%%%%%%%%%%%%%%%%%%%%
\section{Simulations of the second SN in DNS progenitor systems}\label{sec:sim}
We apply MC techniques to simulate SN explosions (between 10 and 500~million trials for each system) to calculate the
kinematic properties of DNS systems surviving the second SN explosion.
From the outcome of these simulations, we can compare with any given observed DNS system and iterate on adjusting the pre-SN parameter space until the
outcome matches with the observed post-SN values within a certain error margin (including the evolution of $P_{\rm orb}$ and eccentricity since the
formation of the DNS system, if necessary, cf. Section~\ref{sec:mapping}). 

Our simulations take their basis in a five-dimensional phase~space.
The input parameters are:
 the pre-SN orbital period, $P_{\rm orb,i}$;
 the final mass of the (stripped) exploding star, $M_{\rm He,f}$;
 the magnitude of the kick velocity imparted on the newborn NS, $w$;
 and the two angles defining the direction of the kick velocity, $\theta$ and $\phi$.
A sixth input parameter is the mass of the first-born NS, $M_{\rm NS,1}$. However, the SN simulation results are not very dependent on the small variations of this parameter.

For almost all DNS systems, the total mass, $M$ is known from measurements of the periastron advance of the radio pulsar:
\begin{equation}
  \dot{\omega} = 3\,T_{\odot}^{2/3}\;\left(\frac{P_{\rm orb}}{2\,\pi}\right) ^{-5/3}\;\frac{M^{2/3}}{1-e^2}\;\,
\end{equation}
assuming that this effect is caused purely by the effects of GR to first post-Newtonian order \citep{rob38,tw82}, and where
$T_{\odot}\equiv GM_{\odot}c^{-3}=4.925490947\;\mu {\rm s}$. 
Hence, we can use this post-SN total mass to constrain the combination of $M_{\rm NS,1}$ and $M_{\rm He,f}$ prior to the SN explosion,
by requiring that $M_{\rm He,f}$ cannot be less than $M_{\rm NS,2}$ (after correction for the released gravitational binding energy, cf. Section~\ref{subsec:theoPorbPspin}). 

To compute the outcome of an asymmetric SN in a binary system, analytic 
expressions for the general solutions have been found both for systems which remain bound \citep{hil83} 
and for systems which disrupt \citep{tt98}. 
We follow the notation and the methodology outlined in these papers as summarized in  Section~\ref{sec:kicks}. 
The formulae of \citet{tt98} are solutions for the general case and reproduce the formulae of \citet{hil83} for bound systems.

In Figs.~\ref{fig:app0453} to \ref{fig:app1930} in the Appendix, we show the results of our simulations of each observed DNS system in six panels.
The upper-left panel shows the post-SN values of $P_{\rm orb}$ and eccentricity. The chosen solutions are plotted as
small black dots (making up a square) centered on the observed values of $(P_{\rm orb},e)$ with an accepted error margin of $\pm$3\% in both
$P_{\rm orb}$ and $e$. All constraints on the observed parameters of the DNS system are also written in this panel, as well as
the number of simulated SN explosions, $N$.

The upper-right panel shows the distribution of the resulting 3D systemic velocities. These velocities can be combined with the merger time
for the DNS system to estimate the offset distance of the future DNS merger (short~GRB) from its birth place in the Galactic disk \citep[e.g.][]{bsp99,pb02,vt03,fbf10,krz+10,bhp16b}.
Assuming, as a first-order approximation, constant motion throughout a galactic potential, an upper limit for this distance can be simply estimated as: $d_{\rm pc} \simeq v_{\rm km/s} \cdot \tau _{\rm gwr,Myr}$. 
The recent discovery of significant abundances of $r$-process elements (e.g. Eu) in an ultra-faint dwarf galaxy \citep[Reticulum~II,][]{rmb+16} 
requires very small kick velocities of $w<50\;{\rm km\,s}^{-1}$ for the second SN explosion \citep{bhp16}, if these elements are produced by a DNS merger retained in such a small galaxy. 
For ultra-faint dwarf-like galaxies, the escape velocity can be quite small (down to $\sim\!15\;{\rm km\,s}^{-1}$). However, a number of the known Galactic DNS systems 
could, in principle, have formed with such small 3D systemic velocities, see our simulations discussed below and the plots in Figs.~\ref{fig:app0453}--\ref{fig:app1930}. 

The central panels show the pre-SN orbital period, $P_{\rm orb,i}$ (left) and the mass of the exploding star, $M_{\rm He,f}$ (right).
The intervals of the trial values are also stated in the legend, and these intervals must be broader than (and cover the entire range of) the solution values.
The mass of the exploding star, $M_{\rm He,f}$ is typically restricted between $1.50-7.0\;M_{\odot}$. The lower limit is determined by observational
constraints on $M_{\rm NS,2}$, taking the released gravitational binding energy from the core collapse into account. We assumed a conservative upper limit of $7.0\;M_{\odot}$ 
for $M_{\rm He,f}$, since naked metal cores which were stripped by winds and RLO and still have a final mass above this limit will form a BH rather than a NS.
It is, however, quite possible that BHs are also formed from less massive exploding stars. 

The lower-left panel shows the applied kick velocities for all successful solutions; and the lower-right panel shows the two associated kick angles.
The red circles indicate kick directions parallel ($\phi=90^{\circ},\,\theta=90^{\circ}$) and anti-parallel ($\phi=-90^{\circ},\,\theta=90^{\circ}$) to the pre-SN orbital angular momentum vector, 
see Section~\ref{subsec:kick_direction}.

We now proceed by discussing the solutions for all the relevant SN parameters for each of the known DNS systems listed in Table~\ref{table:DNS}. The corresponding plots are shown in the Appendix. 

\subsection{PSR J0453+1559}\label{subsec:0453}
PSR~J0453+1559 was recently discovered by \citet{msf+15}. It is outstanding compared to the rest of the DNS population in that it has 
a large mass ratio between the two NS component masses of $q=1.33$.
The proper motion has been measured for this system, yielding a transverse velocity and thus a handle on its 3D systemic LSR velocity (Table~\ref{table:DNS_v_LSR}).
From our SN simulations we are then able to place relatively tight constraints on e.g. the kick magnitude of the second SN.  

The simulated sample of solutions for this system is plotted in Fig.~\ref{fig:app0453}, based on the NE2001 DM distance of 1.07~kpc. It is seen that there is a large probability for a relatively small imparted kick
velocity of $w\la100\;{\rm km\,s}^{-1}$ during the second SN (for an assumed upper limit on the resulting systemic velocity, $v_{\rm sys}<85\;{\rm km\,s}^{-1}$, Table~\ref{table:DNS_v_LSR}),
although the tail of solutions extends to relatively large kick values up to $w\simeq 175\;{\rm km\,s}^{-1}$.
For the mass of the exploding star (producing the second-born NS) we find solutions for $M_{\rm He,f}<3.2\;M_{\odot}$.
The pre-SN orbital period is found to be $2.5\la P_{\rm orb,i}/{\rm days}\la 4.5$.

Unfortunately, the distance to this pulsar remains rather uncertain which makes it difficult to place tighter constraints 
on the kick magnitude. The kick is most likely to be $w< 100\;{\rm km\,s}^{-1}$ (although it could be somewhat larger).
Using the Galactic electron distribution model of \citet{ymw16} yields a DM distance of only 0.52~kpc. As a result, the systemic velocity of the system becomes smaller
and the imparted kick is found to be $w< 75\;{\rm km\,s}^{-1}$. An upper limit for the mass of the exploding star is in this case $M_{\rm He,f}<2.1\;M_{\odot}$.
A parallax measurement would be able to resolve the distance discrepancy and thus allow for a tighter constraint on both $w$ and $M_{\rm He,f}$. 
Any corrections for gravitational damping in this system are negligible due to its wide orbit ($P_{\rm orb}=4.07\;{\rm days}$).
 
\subsection{PSR J0737$-$3039}\label{subsec:0737}
This DNS system is the famous double pulsar \citep{bdp+03,lbk+04,ksm+06}.
It has been intensively timed since its discovery in 2003 and, due to its fortunate geometry (orbital inclination angle of $89^{\circ}$),
its orbital and kinematic parameters are very well constrained. In particular, the misalignment angle of the first-born, recycled pulsar has been measured to be $\delta <3.2^{\circ}$ \citep{fsk+13}, and 
from VLBI observations \citep{dbt09} its distance ($\sim\!1150\pm 200\;{\rm pc}$) and proper motion ($4.37\pm 0.65\;{\rm mas\,yr}^{-1}$) have been revealed. 
These measurements result in a small 2D transverse velocity of $\sim\!25\;{\rm km\,s}^{-1}$ and correcting for the unknown radial component and applying its location
in the Galactic potential we find an estimated 3D systemic LSR velocity of $\sim\!33^{+24}_{-15}\;{\rm km\,s}^{-1}$ (Table~\ref{table:DNS_v_LSR}). 

The resulting kick velocity found from our simulated solutions, where we restrict ourselves to only select systems with $v_{\rm sys}=18-57\;{\rm km\,s}^{-1}$, 
is therefore $w\la70\;{\rm km\,s}^{-1}$, as shown in Fig.~\ref{fig:app0737}. As also seen in this figure, the kick magnitude distribution peaks at a value of only $\sim\!18\;{\rm km\,s}^{-1}$.
A caveat to this conclusion is that the DNS system also received a (small) systemic velocity from the first SN, which was 
probably of the order of $10-20\;{\rm km\,s}^{-1}$, cf. Section~\ref{subsec:velocities}.
A close inspection of Fig.~\ref{fig:app0737} shows that there are no solutions for $v_{\rm sys}\le 32\;{\rm km\,s}^{-1}$.
The reason for this is the finite amount of mass loss, $\Delta M$.

For a comparison, we plot in Fig.~\ref{fig:app0737_2} our simulated solutions without applying any constraints on $v_{\rm sys}$.
The difference in the widths of the pre-SN solution parameter space is striking, especially for the value of the final mass of the exploding star, $M_{\rm He,f}$.
  
Further important conclusions on the formation of PSR~J0737$-$3039 can immediately be drawn from our simulations. 
Applying the constraint $v_{\rm sys}<57\;{\rm km\,s}^{-1}$ provides an almost {\it unique} solution to the pre-SN progenitor binary of PSR~J0737$-$3039 (Fig.~\ref{fig:app0737}); 
an argument that was first made by \citet{ps05}. 
This can be understood from the observations of the small proper motion of this system which results in $(w/v_{\rm rel})\ll1$, and thus
a firm solution is obtained from the observed values of $P_{\rm orb}$ and $e$, cf. Eqs.~(\ref{eq:aratio_symm}) and (\ref{eq:ecc_symm}).
Similarly, it automatically follows that the misalignment angle in this case must be small (Eq.~\ref{eq:misalign}), in agreement with observations.
For 94\% of our solutions, the pre-SN binary had an orbital period of $P_{\rm orb,i}=0.085\pm 0.005\;{\rm days}$ and the mass of the (ultra-stripped) exploding star must have been $M_{\rm He,f}=1.56\pm 0.06\;M_{\odot}$. 
Interestingly enough, the very first calculation of a helium~star--NS binary system leading to an ultra-stripped SN \citep{tlm+13} had pre-SN values of
$P_{\rm orb,i}=0.070\;{\rm days}$ and $M_{\rm He,f}=1.50\;M_{\odot}$, and is thus basically a {\it solution} to the immediate progenitor of PSR~J0737$-$3039. 
For this particular system, \citet{tlm+13} found that the nature of the SN would be an Fe~CCSN of Type~Ic. From their binary stellar evolution modelling, 
we can therefore also deduce that the pre-SN system is the descendant of a helium~star--NS binary with an initial (post-CE) orbital period of $\sim\!0.1\;{\rm days}$ and a helium star ZAMS mass of $\sim\!2.9\;M_{\odot}$.

In the discussion above, we have neglected any influence from gravitational damping on this system. The fact that we also observe the young
pulsar in this system, means that we can set an upper limit to the age of this binary. The characteristic spin-down age of pulsar~B (the young NS)
is about 50~Myr, which is also a typical radio pulsar lifetime for a non-recycled pulsar \citep{lk04}. Unless the braking index of this pulsar is
significantly less than 3, this will serve as a proxy for the upper limit of its true age. 
However, in this particular binary, the magnetosphere of pulsar~B, and thus its spin-down activity, is likely to be affected by the
relativistic wind from pulsar~A (the recycled pulsar, releasing spin-down energy at a rate of 3600 times more than pulsar~B), as also
indicated by observations \citep{mll+04,lyu04,bkm+12}. Hence, the true age of pulsar~B could possibly be more than 100~Myr \citep{lfs+07}.

For a DNS age of 100~Myr, we calculate corresponding post-SN parameters of $P_{\rm orb}=0.138\;{\rm days}$ and $e=0.119$ (see location of red star),  
and the solution would be that the pre-SN binary had an orbital period of $P_{\rm orb,i}=0.11\pm 0.01\;{\rm days}$ and that the mass of the exploding star must 
have been $M_{\rm He,f}=1.59\pm 0.09\;M_{\odot}$. The distribution of valid kick velocities remains roughly similar to the one shown in Fig.~\ref{fig:app0737}.

Since the exact age of PSR~J0737$-$3039 remains unknown, and could in principle take any value between 0 and 100~Myr (or even more), we conclude that this DNS system 
formed from a pre-SN binary with $P_{\rm orb,i}=0.10\pm0.02\;{\rm days}$ and $M_{\rm He,f}=1.59\pm 0.09\;M_{\odot}$, as well as a small imparted kick, $w\la 70\;{\rm km\,s}^{-1}$.

A final note on this double pulsar system is that it is often being used to constrain the empirical Galactic DNS merger rate. The outcome is dependent
on the beaming geometry of both pulsars~A and B. As an example, \citet{kpm15} derived a value of $21^{+28}_{-14}\;{\rm Myr}^{-1}$.  

\subsection{PSR J1518+4904}\label{subsec:1518}
Since the discovery of this DNS system \citep{nst96}, its component masses have remained difficult to constrain \citep{jsk+08}.
The total mass of the system ($M=2.72\;M_{\odot}$) remains a firm result
and we use that for our SN simulations of this system. Assuming that the mass of the second-born NS, $M_{\rm NS,2}$ falls within the range of such measured 
masses in other DNS systems ($1.17-1.39\;M_{\odot}$, Table~\ref{table:obs_char}), and using MC simulations with a flat probability distribution for $M_{\rm NS,2}$
in this range, the mass of the first-formed NS is then simply evaluated from $M_{\rm NS,1}=M-M_{\rm NS,2}$. The results of our SN simulations are shown in Fig.~\ref{fig:app1518}. 
The solutions yield values of the pre-SN orbital period between 4 and 12~days. The solutions for the mass of the exploding star in the second SN is  
restricted to $<2.8\;M_{\odot}$, and with a preference toward smaller values. Similarly, we only find solutions for a relatively
small kick velocity ($w< 100\;{\rm km\,s}^{-1}$) with a peak\footnote{We repeat that this peak is based on flat input distributions of various pre-SN parameters, 
neglecting previous binary stellar interactions. A full study requires population synthesis.} 
near $w\simeq 40\;{\rm km\,s}^{-1}$. 

Ongoing intensive data analysis, including data obtained in recent observation campaigns, yields preliminary
values of $M_{\rm NS,1}=1.41\;M_{\odot}$ and $M_{\rm NS,2}=1.31\;M_{\odot}$ for the (mildly) recycled radio pulsar and its young NS companion, respectively (G.~Janssen, priv.~comm.).
The reported error bars are still rather large, roughly $\pm0.08\;M_{\odot}$.
Redoing our MC simulations with these mass constraints does not introduce any significant changes to our results presented in Fig.~\ref{fig:app1518}. 
Any corrections for gravitational damping in this system, during the time interval between its formation and now, are negligible due to its wide orbit ($P_{\rm orb}=8.63\;{\rm days}$).

\subsection{PSR B1534+12}\label{subsec:1534}
PSR~B1534+12 was discovered by \citet{wol91} and studied further in detail by e.g. \citet{sttw02,fst14}.
The system parameters are well-constrained (Table~\ref{table:DNS}) --- not only because of its precisely measured masses, but also via its determined misalignment angle, $\delta=27\pm3\;^{\circ}$
and its proper motion. The results from our SN simulations are shown in Fig.~\ref{fig:app1534}.   
The mass of the exploding star is constrained to be $M_{\rm He,f}< 3.7\;M_{\odot}$. More noteworthy, the kick in the second SN is shown to have been large: $w\simeq 175-300\;{\rm km\,s}^{-1}$. 
This makes this binary special in the sense that it represents one out of only two DNS systems (besides PSR~B1913+16) for which there is evidence for a large kick in an ultra-stripped SN.

The reason that a large kick is required in this system, is the relatively large measured value of $\delta$. Without any constraints on this angle, solutions are found for a wide 
range of kick values: $50\la w \la 425\;{\rm km\,s}^{-1}$, illustrating again the importance of measuring $\delta$. 
The small error bar on the measurement of $\delta$ explains why
the direction of this kick is seen to be confined to a relatively small area on the sphere surrounding the exploding star (lower-right panel).
Corrections for gravitational damping in this system (since its formation) is rather limited due to the relatively small characteristic
age of the recycled pulsar ($\tau = 248\;{\rm Myr}$) and somewhat small eccentricity ($e=0.274$), see the red solid star in the upper-left panel.

\subsection{PSR J1753$-$2240}\label{subsec:1753}
There are no mass constraints from observations of this DNS system \citep{kkl+09}.
Assuming that the total mass of the system is within the interval of the total masses of all other known DNS systems ($M=2.57-2.88\;M_{\odot}$),
and similarly that the mass of the second-born NS, $M_{\rm NS,2}$ is also within the range ($1.17-1.39\;M_{\odot}$) measured in other DNS systems,
we simulated the possible solutions for the pre-SN parameters and the NS kick. The results are shown in Fig.~\ref{fig:app1753}.
As can be seen, we cannot derive interesting constraints for the origin of this DNS system.
The kick is likely to have been small ($<100\;{\rm km\,s}^{-1}$), but we also find solutions for kicks up to $w=400\;{\rm km\,s}^{-1}$.
The wide orbit of this system excludes any gravitational damping of significance.

\subsection{PSR J1755$-$2550}\label{subsec:1755}
This interesting DNS candidate was discovered by \citet{ncb+15}. The nature of the system is still unclear.
The spin period of the radio pulsar is 315~ms which makes it quite likely to be a young pulsar. 
The young nature of this pulsar was recently confirmed by a measurement of $\dot{P}=2.43\times10^{-15}\;{\rm s\,s}^{-1}$ \citep{nkt+17}. 
Indeed, if the NS had been even mildly recycled, we would have expected a smaller spin period, given its orbital period of 9.7~days (cf. Section~\ref{subsec:PorbPspin}).
The compact object composition of the companion star remains unknown. It could be either a DNS system or a WD+NS system where the NS formed after the WD \citep{ts00}.
From the mass function it is known that $M_2 > 0.40\;M_{\odot}$ and given the unknown orbital inclination angle, the probability of a WD+NS binary is somewhat 
larger than that of a DNS system, see discussions in \citet{nkt+17}. On the other hand, the eccentricity of this system ($e=0.089$) is similar to many DNS systems.
The wide orbit of the system excludes any gravitational damping of significance, so this cannot be the explanation for its small eccentricity.
Assuming that the system is indeed a DNS system, we show in Fig.~\ref{fig:app1755} our results from the simulations of the second SN.
As in the case of PSR~J1753$-$2240 discussed above, we have no constraints on the total mass nor on the value of $M_{\rm NS,2}$, 
and we follow the same procedure as for PSR~J1753$-$2240.
The kick is likely to have been small ($<50\;{\rm km\,s}^{-1}$), but we also find solutions for kicks up to almost $w=400\;{\rm km\,s}^{-1}$.

\subsection{PSR J1756$-$2251}\label{subsec:1756}
PSR~J1756$-$2251 \citep{fkl+05} is very well constrained from precise measurements of its NS masses ($M_{\rm NS,1}=1.341\;M_{\odot}$, $M_{\rm NS,2}=1.230\;M_{\odot}$)
and the misalignment angle ($\delta < 34^{\circ}$) of the recycled pulsar \citep{fsk+14}. The proper motion measurement in declination, however, has been difficult
to measure for this pulsar, since it is located near the ecliptic plane. \citet{fsk+14} derived a best estimate of $\mu _\delta = 5.5\pm 7.0\;{\rm mas\,yr}^{-1}$
which we combine with its proper motion in right ascension, $\mu _\alpha =-2.42\pm 0.08\;{\rm mas\,yr}^{-1}$ to estimate its transverse velocity     
and a determination of its 3D systemic velocity of $v^{\rm LSR}=41\;{\rm km\,s}^{-1}$ ($v^{\rm LSR}_{1\,\sigma}=21-71\;{\rm km\,s}^{-1}$, cf Table~\ref{table:DNS_v_LSR}). 

This DNS system shares many properties with the double pulsar (PSR~J0737$-$3039) and can be seen as a wider-orbit version of it.
The SN simulations reveal a similar solution for the origin of the second-born NS as an ultra-stripped exploding star with a mass of $1.72\pm0.22\;M_{\odot}$.
The kick velocity is found to be $w< 130\;{\rm km\,s}^{-1}$, with a main peak in solutions toward $w\rightarrow 0$. 
It is interesting to note that this DNS system is the only one for which half of the  
solutions require a kick in a forward direction, i.e. $\theta < 90^{\circ}$. This is partly due to the relatively small 3D systemic velocity inferred for of this system. 

Using the NE2001 electron distribution model yields a DM distance of $d=0.73\;{\rm kpc}$, whereas
applying the model of \citet{ymw16} results in a distance four times larger ($d=2.81\;{\rm kpc}$). Such a large distance would significantly increase our derived
interval of possible kick velocities for this system. 
However, a firm upper value for the distance of 1.2~kpc can be derived using the measured orbital period decay \citep{fsk+14}. 
This example illustrates again the need for precise distance determinations via parallax measurements.

The orbital period of this system of $P_{\rm orb}=0.320\;{\rm days}$, and its eccentricity of $e=0.181$, means that some
gravitational damping may have been at work since its formation. Assuming an age of $t=\tau=443\;{\rm Myr}$, yields initial
values of $P_{\rm orb}=0.353\;{\rm days}$ and $e=0.198$ right after the second SN (see the red solid star in the upper-left panel in Fig.~31).
However, this potential slight change in initial values does not impose any effects on the constraints of the pre-SN parameters and the kick velocity 
derived here.

\subsection{PSR J1811$-$1736}\label{subsec:1811}
PSR~J1811$-$1736 \citep{cks+07} is a wide-orbit, highly eccentric  DNS system with $P_{\rm orb}=18.8\;{\rm days}$ and $e=0.828$. Its total mass is $2.57\;M_{\odot}$,
whereas its individual NS mass components remain unknown. Assuming that the mass of the second-born NS, $M_{\rm NS,2}$ is within the range of observed masses
in other DNS systems ($1.17-1.39\;M_{\odot}$) we simulated the possible solutions for the pre-SN parameters and the NS kick. The results are shown in Fig.~\ref{fig:app1811}.
As can be seen, there is a large range of possible values for all pre-SN parameters and the kick, and we conclude that we cannot derive interesting
constraints for the origin of this DNS system. The wide orbit of PSR~J1811$-$1736 excludes any gravitational damping of significance.

\subsection{PSR J1829+2456}\label{subsec:1829}
This DNS system was discovered by \citet{clm+04}. The total mass of the system is $2.59\;M_{\odot}$ and the second-born NS has a mass $M_{\rm NS,2}>1.22\;M_{\odot}$.
Assuming as usual $M_{\rm NS,2}\le 1.39\;M_{\odot}$, our SN simulations yield the results shown in Fig.~\ref{fig:app1829}.
Also in this case, there is a large range of possible values for all pre-SN parameters and the kick, and thus we cannot derive interesting
constraints for the origin of this DNS system. The relatively wide orbit ($P_{\rm orb}=1.18\;{\rm days}$), and the small eccentricity ($e=0.139$), of the system
means that its orbital parameters have hardly changed its orbital parameters since its formation.

\subsection{PSR J1906+0746}\label{subsec:1906}
This is a remarkable binary since the observed radio pulsar is the young, second-born, NS \citep{lsf+06}. The evidence is a combination of a large B-field
(i.e. $\dot{P}=2.03\times 10^{-14}\;{\rm s\,s}^{-1}$) and a relatively slow spin period ($P=144\;{\rm ms}$), cf. Table~\ref{table:DNS} and Fig.~\ref{fig:PPdot}.
These values are typical for a young, non-recycled NS. The NS component masses are measured to be $M_{\rm NS,1}=1.322\;M_{\odot}$ and $M_{\rm NS,2}=1.291\;M_{\odot}$, respectively \citep{vks+15}.
Since the companion star remains unseen, this system could, in principle, also be a WD+NS system. ONeMg~WD masses of $1.322\;M_{\odot}$ are certainly possible.
We show our simulations of the (second) SN in Fig.~\ref{fig:app1906}.
However, we cannot make any interesting constraints on the pre-SN parameters of the system. A remarkable result though is that this binary could, in principle, 
have survived a NS kick velocity of about $w=1500\;{\rm km\,s}^{-1}$. A proper motion measurement would be able to constrain the kick velocity. 
Since we treat the first-formed compact object as a point mass in these simulations,
the results are equally valid for a WD+NS system where the NS forms after the WD \citep{ts00}.

The close orbit of this system ($P_{\rm orb}=0.166\;{\rm days}$) implies that gravitational damping since birth could be of significance.
However, this NS must be quite young given its high value of $\dot{P}=2.03\times 10^{-14}\;{\rm s\,s}^{-1}$ ($\tau = 114\;{\rm kyr}$), 
and therefore its present orbital parameters resembles very well its birth parameters.
\citet{ok10} argued that this DNS system makes quite a significant contribution to the empirical estimate of the Galactic DNS merger rate.

\subsection{PSR J1913+1102}\label{subsec:1913+11}
\citet{lfa+16} recently announced the discovery of this DNS system with $P_{\rm orb}=0.206\;{\rm days}$ and $e=0.090$.
The total mass of the system is reported to be $M=2.88\;M_{\odot}$. The NS masses are expected to be measured soon,
but until then we assume that the mass of the second-born NS, $M_{\rm NS,2}$ is within the usual range of $1.17-1.39\;M_{\odot}$, as measured in other DNS systems.
In Fig.~\ref{fig:app1913+1102}, we show our simulated possible solutions for the pre-SN parameters and the NS kick. 
As can be seen, we cannot derive interesting constraints for the origin of this DNS system.
The orbit of the binary could have been affected significantly by gravitational damping since its formation.
The main reason for this is the large characteristic age of the system ($\tau = 2690\;{\rm Myr}$), which potentially allows for
a large orbital decay on a Gyr timescale. In the upper-left panel, we have plotted its birth location with a red solid star ($P_{\rm orb}=0.438\;{\rm days}$
and $e=0.189$) for an assumed age $t=\tau$. These initial values are almost independent of the exact NS masses 
(here we simply assumed $M_{\rm NS,1}=1.58\;M_{\odot}$ and $M_{\rm NS,2}=1.30\;M_{\odot}$, motivated by typical values of $M_{\rm NS,2}$, cf. Table~\ref{table:obs_char}).

\subsection{PSR B1913+16}\label{subsec:1913+16}
The Hulse-Taylor pulsar \citep{ht75,kra98,wnt10} is an unusual case since this DNS system must have been accompanied by a rather
large kick during the second SN explosion \citep[e.g.][]{cw84,bw86,wkk00,wkh04}.
We confirm this result from our analysis of its 3D systemic velocity (Section~\ref{subsec:vsys}) and the SN simulations presented in Fig.~\ref{fig:app1913}.
We find kick solutions for the entire interval of $w=185-465\;{\rm km\,s}^{-1}$. This result (as well as that of the other DNS system with a large kick, PSR~B1534+12) is  
in excellent agreement with the findings of \citet{wwk10}. For the mass of the exploding star, however, we cannot constrain its mass  
to better than $M_{\rm He,f}<4.4\;M_{\odot}$. Indeed, the possibility of a somewhat more massive exploding star, compared to some
of the DNS systems which experienced small kicks (e.g. PSR~J0737$-$3039 and PSR~J1756$-$2251), may be a clue to understanding
the large kick in this system (cf. discussions in Section~\ref{subsec:kick_mass}).

A caveat to this conclusion is that the distance to PSR~B1913+16 is uncertain. The \citet{ymw16} DM distance of 5.25~kpc is only half the
distance of 9.8~kpc provided by \citet{wsx+08} and \citet{wh16}, based on the Galactic model of \citet{rmb+14}. Adapting this smaller distance results
in typical kick velocities that are about $50-100\;{\rm km\,s}^{-1}$ smaller than those quoted above, although with a minimum kick velocity remaining 
at a relatively large value of $w=165\;{\rm km\,s}^{-1}$. Perhaps more interesting and important, the mass of the exploding star can in this case be constrained to be $M_{\rm He,f}<2.6\;M_{\odot}$. 

The torus of possible kick angles in the ($\phi,\theta$)--plane is a result of the constraint from the measured misalignment angle, $\delta = 18\pm6^{\circ}$ \citep{fsk+13}.
The combination of a small orbital period of $P_{\rm orb}=0.323\;{\rm days}$ and a large eccentricity of $e=0.617$, means that 
gravitational damping is significant in this system, as discussed in Section~\ref{sec:mapping}. We illustrate this by marking the birth location of PSR~B1913+16 in the
($P_{\rm orb},e$)--plane (upper-left panel in Fig.~\ref{fig:app1913}) with a red solid star for a DNS age of $t=108.7\;{\rm Myr}$.
The SN simulations leading to these birth properties are shown in Fig.~\ref{fig:app1913_birth}. The differences in the solutions for the pre-SN
parameters and the second NS kick are very limited compared to the solutions shown in Fig.~\ref{fig:app1913}.
For an age of $t=108.7\;{\rm Myr}$, we do find solutions for a slightly more massive exploding star (up to $4.8\;M_{\odot}$).
However, the most likely kick remains at roughly $200-450\;{\rm km\,s}^{-1}$.

\subsection{PSR J1930$-$1852}\label{subsec:1930}
This important very wide-orbit DNS system was discovered by \citet{srm+15}. The binary has $P_{\rm orb}=45\;{\rm days}$ and
a marginally recycled pulsar spinning at $P=185\;{\rm ms}$.
The total mass of this system is $2.59\;M_{\odot}$ and the second-born NS has a mass $M_{\rm NS,2}>1.30\;M_{\odot}$.
Assuming as usual $M_{\rm NS,2}\le 1.39\;M_{\odot}$, our SN simulations yield the results shown in Fig.~\ref{fig:app1930}.
Also in this case, there is a large range of possible values for all pre-SN parameters and the kick, and thus we cannot derive interesting
constraints for the origin of the DNS system. The kick was most likely rather small ($w<100\;{\rm km\,s}^{-1}$), although we find a small tail
of possible solutions up to $290\;{\rm km\,s}^{-1}$. Gravitational damping is not relevant for this system. 
Assuming an age of 50~Myr or 100~Myr reveals an initial spin period of the marginally recycled pulsar of 154~ms or 115~ms, respectively,
assuming a spin evolution with a braking index of $n=3$ (cf. Eq.~\ref{eq:spinevol}). The orbital parameters and the spin of this marginally recycled pulsar
are discussed further in \citet{tlp15}.

%%%%%%%%%%%%%%%%%%%%%%%%%%%%%%%%%%%%%%%%%%%%%%%%%%%%%%%%%%%%%%%%%%%%%%%%%%%%%%%%%%%%%%%%%%%
\section{Ramifications from SN simulations}\label{sec:ramifications}
\subsection{Resulting NS and pre-SN Properties}\label{subsec:SNsim_pre}
The extreme stripping of helium stars during Case~BB RLO, following the CE phase, leads to relatively low masses of the second exploding star in DNS systems 
--- typically of values $M_{\rm He,f}\le 2.5\;M_{\odot}$ for a broad range of initial values of orbital periods and helium star masses \citep{tlp15}.
The solutions for the second SNe presented in Figs.~\ref{fig:app0453}--\ref{fig:app1930} are based purely on kinematics. 
Many of the solutions with massive exploding stars (up to $\sim\!7\;M_{\odot}$) are therefore not physical, since binary stellar evolution would,
in most cases, prevent such massive exploding stars. Only in a few cases is it possible that very massive helium stars (WR-stars) may produce exploding stars
with such large masses if their stellar wind mass-loss rate is small \citep[e.g.][and references therein]{ywl10}. 
To illustrate the strong dependence on stellar wind mass-loss rate, \citet{tlp15} calculated that the outcome of a binary system with a NS and a $10\;M_{\odot}$ helium star 
could be an exploding star with any mass between $3-7\;M_{\odot}$. However, if such a binary system is in a close orbit, GW radiation will force the system
into RLO while the helium star is still massive, causing the system to merge upon RLO. Alternatively, for a wide-orbit system, the binary may expand significantly as 
a result of the strong wind such that the NS escapes being recycled via Case~BB RLO. 

Since some of the quantities plotted in Figs.~\ref{fig:app0453}--\ref{fig:app1930} are correlated with the mass of the exploding star 
(e.g. the magnitude and required direction of the kick velocities leading to a given DNS solution), care must be taken when drawing conclusions
from these SN simulations. Full population synthesis is needed to disentangle cross-dependent quantities and enable more trustworthy probability distributions
of various parameters at play. 

\subsection{Direction of the NS kick}\label{subsec:kick_direction}
\begin{figure}[t]
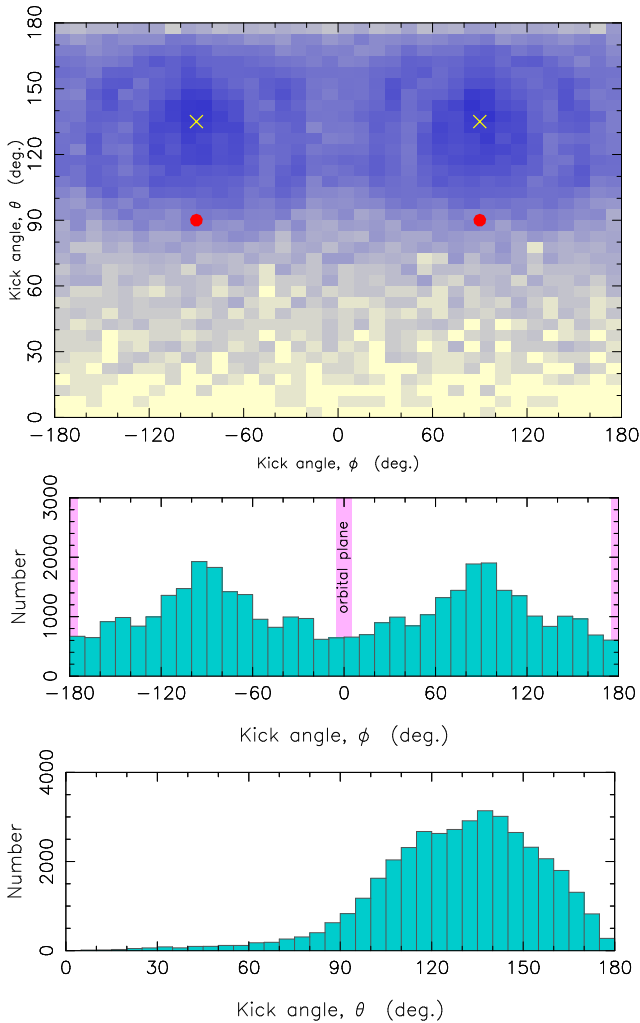

  \begin{center}
     \includegraphics[width=0.72\columnwidth, angle=-90]{Figures/angle_dist.ps}
     \includegraphics[width=0.42\columnwidth, angle=-90]{Figures/angle_dist_histo.ps}
     \includegraphics[width=0.42\columnwidth, angle=-90]{Figures/angle_dist_histo_theta.ps}
    \vspace{0.2cm}
    \caption{Distribution of kick directions for all simulated DNS systems shown in Figs.~\ref{fig:app0453}--\ref{fig:app1930}.
       In this plot, 3000 kick directions ($\theta,\phi$) were simulated for the second SN of each system and stacked together.  
       The darker the blue-scale colour (log-10 scale, spanning 3~dex), the more solutions in a given direction.
       The red circles indicate kick directions parallel ($\phi=90^{\circ},\,\theta=90^{\circ}$) and anti-parallel ($\phi=-90^{\circ},\,\theta=90^{\circ}$)
       to the pre-SN orbital angular momentum vector. The yellow crosses mark $\theta=135^{\circ}$, and $\phi=90^{\circ}$ or $\phi=-90^{\circ}$.
       The central and bottom panels show the distribution of $\phi$ and $\theta$-angles, projected from the upper panel.
       An apparent preference is seen for a kick direction out of the orbital plane of the pre-SN binary, as well as in a backward direction (yellow crosses).
       However, this anisotropy may well be a combined artifact of input priors and a selection effect --- see text. 
       Fig.~\ref{fig:angle_dist_iso} gives a verification of the isotropy for the input (trial) kicks used in our SN simulations.}
  \label{fig:angle_dist}
  \end{center}
\end{figure}
In Fig.~\ref{fig:angle_dist}, we have plotted the directions of all kick velocities leading to a solution for the simulations of the second SN
of the DNS systems shown in Figs.~\ref{fig:app0453}--\ref{fig:app1930}. We restricted the number of solutions to 3000 for each system and 
stacked together the ($\phi,\theta$)--solution planes on top of each other. The darker the blue-scale colour (log scale), the more solutions in a given direction.
A remarkable apparent preference is seen for a kick direction out of the plane of the pre-SN binary. The pre-SN orbital plane is along $\phi=0^{\circ}$ and $\phi=\pm 180^{\circ}$
and the kick solutions clearly cluster in directions out of the plane ($\phi\simeq 90^{\circ}$ and $\phi\simeq -90^{\circ}$).
Furthermore, the kicks leading to the solutions show a significant preference for a direction backwards ($\theta > 90^{\circ}$).
This latter result is perhaps not surprising since the probability of the system remaining bound is larger for a backward kick, cf. Section~\ref{sec:kicks}.
For a verification of an isotropic input (trial) distribution of kick velocity directions, see Fig.~\ref{fig:angle_dist_iso} in the Appendix. 

Given the above discussion on kinematic solutions versus realistic (from a binary stellar evolution point of view, Section~\ref{subsec:SNsim_pre}) masses of the 
exploding stars ($M_{\rm He,f}$), we performed an additional simulation of kick distributions in the observed DNS systems, in which we only kept solutions with $M_{\rm He,f}<2.5\;M_{\odot}$.
The outcome of this experiment remains fairly identical to the result shown in Fig.~\ref{fig:angle_dist}. The only minor difference is that the peak of the kick directions
is located at a slightly smaller value of $\theta \simeq 125^{\circ}$, and that more solutions allow for kicks in a forward direction ($\theta < 90^{\circ}$).
The latter result is expected for smaller masses of the exploding stars.

Further motivated by the puzzle of an apparent NS kick anisotropy, we performed additional investigations and found that this anisotropy can be understood from
a combination of an input prior and a selection effect. 

\subsubsection{Kick anisotropies and input priors}
Firstly, we discuss how our application of unconstrained input distributions of kick magnitudes (i.e. without upper limits) can affect the solutions of kick angles for those DNS systems 
in which there are no 
observational constraints on the proper motion. These DNS systems include PSRs J1753$-$2240, J1755$-$2550, J1829+2456, J1906+0746, J1913+1102 and J1930$-$1852,
which have small ($e \simeq 0.1$) or intermediate ($e\simeq 0.3-0.4$) eccentricities, and for which our SN simulations revealed kick solutions up to several $100\;{\rm km\,s}^{-1}$
(in a few cases, even above $1000\;{\rm km\,s}^{-1}$). In order for the simulated progenitor systems to remain bound and reproduce the
observed DNS systems, the kick directions must preferentially be in the regions near the yellow crosses in Fig.~\ref{fig:angle_dist}. 
The reason for this is that to compensate for the large applied kick magnitudes, 
the kick directions are limited in order to reproduce the small post-SN eccentricities (cf. Figs.~\ref{fig:ecc_angle} and \ref{fig:ecc_angle2}),
which are needed to match most of the observed DNS systems. 
If instead we impose limitations on the maximum kick magnitude in the second SN in these systems, i.e. treating the kick magnitudes as elicited priors, 
and only allow kicks up to some limited maximum, say $w=50-60\;{\rm km\,s}^{-1}$, then the apparent kick anisotropy along $\phi$ disappears. 

If assuming that nature produces an isotropic distribution of kick directions, we can therefore reverse the above argument and claim further evidence for
small kicks in ultra-stripped SNe in general (but not always, as demonstrated by PSRs B1913+16 and B1534+12).

\subsubsection{Kick anisotropies and selection effects}
Secondly, we must beware of the observational selection effect \citep{cb05} wherein high-eccentricity and tight-orbit DNS systems will be removed from 
the observable sample shortly after their birth due to their small merger timescales. From Eq.~(\ref{eq:mergertime}), we find that
any DNS system with $e>0.694$ will merge at least 10 times faster than a circular DNS systems with the same post-SN $P_{\rm orb}$ and NS masses.
This effect also gives an explanation for the apparent kick anisotropy since DNS systems with small eccentricities remain observable on
a longer timescale and, importantly, since they have small eccentricities their kick angles are restricted to a certain area in the 
($\phi,\,\theta$)--plane near the yellow crosses in Fig.~\ref{fig:angle_dist} (see Fig.~\ref{fig:ecc_angle}, and Fig.~\ref{fig:ecc_angle2} in the Appendix for the case of
an ultra-stripped SN in a tight binary).

Any finding of a preference for a certain kick direction would otherwise have been interesting and should be kept in mind when comparing to other studies indicating, for example,  
a preference for a kick along the spin axis of isolated radio pulsars \citep[][and references therein]{nsk+13}. 
We find that for DNS systems, the kick direction is not particularly favoured along the pre-SN orbital angular momentum vector 
(marked by red circles in Figs.~\ref{fig:angle_dist}, \ref{fig:app0453}--\ref{fig:app1930}), 
which is likely to be parallel to the spin-axis of the exploding star. 
When comparing to measurements of post-SN binaries, it is uncertain, however, whether or not the spin-axis of the exploding star is tossed in a new 
(possibly random) direction as a result of the SN \citep{sp98,fklk11}, in which case all past memory will be lost. This seems to be the case, for example, 
for both of the known young pulsars in DNS systems: PSR~J0737$-$3039 \citep{bkk+08} and J1906+0746 (Desvignes~et~al., in~prep.). 

Finally, it should be noted that if such tossing of the spin axis during core collapse also applies to the formation of BHs, then this might
explain the small effective inspiral spin parameters, $\chi _{\rm eff}$ inferred for the first four LIGO events \citep{aaa+17} without the need for
a dynamical formation channel to explain their origin. 

To summarize, we find that the apparent kick anisotropy (in a diagonal backward direction, cf. Fig.~\ref{fig:angle_dist}), obtained from all of the kinematic solutions to the 
known DNS systems, is most likely caused by allowing for (unrealistic?) large kicks in the second SN. 
Furthermore, an observational selection effect favoring small eccentricities seems to be at work, which also affects the apparent kick angles.
Additional investigations and statistical analysis is needed to confirm our preliminary findings. We caution that we are still
dealing with small number statistics in terms of known DNS systems and that it is unclear how exactly to disentangle the obtained results 
from the probability bias for survival and observability.

%%%%%%%%%%%%%%%%%%%%%%%%%%%%%%%%%%%%%%%%%%%%%%%%%%%%%%%%%%%%%%%%%%%%%%%%%%%%%%%%%%%%%%%%%%%
\section{DNS mergers}\label{sec:merger}
In the previous sections, we have investigated in detail the formation of DNS systems and also discussed their further evolution.
We end this paper with a short discussion of the final fate of close DNS binaries.

\subsection{Short GRBs}
Merging DNS and NS+BH binaries are considered the most promising sources to explain the phenomenology of short~GRBs \citep{pac86,elps89,npp92,ffp+05,rgb+11,ber14}. 
These short GRBs (sGRBs) are possibly followed by a so-called ``kilonova'' or ``macronova'' \citep{lp98b,kul05,mmd+10,bk13,th13,tlf+13,bfc13,fm15,met16,gnp17}. 
According to the standard scenario for the generation of sGRBs, an accretion-powered jet from a remnant BH torus system is formed within $10-100\;{\rm  ms}$ after the 
merger event \citep[e.g.][]{npp92,jerf99,rgb+11,joj+16}. 
Moreover, the energy release should cease once the torus has been accreted on a timescale of $\le 1\;{\rm s}$, which is consistent with the duration of the sGRB prompt emission of less than $\sim\!2\;{\rm s}$. 

Observations by the Swift satellite, however, have recently revealed long-lasting ($10^2-10^5\;{\rm s}$) X-ray activity in a large fraction of sGRB events, which is indicative
of a continuous energy release from a central engine on timescales orders of magnitude larger than the typical torus accretion timescale \citep{rot+10,rom+13,gowr13,gow14,lzl+15}.
This late X-ray activity is difficult to explain with the accretion of the torus or by the prolonged afterglow radiation produced by the interaction of the jet with the ambient medium \citep{kz15}. 
\citet{rk15} and \citet{sc16} recently suggested a solution to this issue. If a large fraction of DNS mergers result in the formation of a meta-stable or long-lived NS rather than a BH, extraction of
rotational energy via magnetic spin-down from such an object \citep[typically a millisecond magnetar,][]{uso92} could power the observed long-lasting X-ray activity.
Such a model will not work for the NS+BH scenario for sGRBs and would therefore exclude a large fraction of sGRBs from originating from NS+BH mergers. 

\subsection{Heavy $r$-process elements}
The merger process of DNS systems is a vital source of chemical enrichment of the interstellar medium with heavy $r$-process elements such as, for example, 
gold \citep[e.g.][]{elps89,frt99,wsn+14,ros15,jbp+15,his+15,wfmm16,bhp16b}. 
These elements were previously thought to be synthesized primarily in SNe.
The total mass of the ejected baryonic matter from DNS mergers, including the yield of heavy $r$-process elements, is typically of the order $0.01\;M_{\odot}$,
depending on the mass ratio between the two merging NSs. Until recently, all observed DNS systems had a mass ratio, $q\simeq 1$. The discovery of PSR~J0453+1559 \citep{msf+15} revealed the first known
DNS system with a more asymmetric pair of NS masses (1.56 and $1.17\;M_{\odot}$) and $q>1.3$.
This mass asymmetry is very important as it may possibly increase (see below) the yield of expelled heavy $r$-process elements, when the lighter NS fills its Roche~lobe and becomes tidally disrupted. 

In \citet{tlp15}, the modelling of DNS progenitor systems resulted in second-born NS masses in the interval
$1.1-1.8\;M_{\odot}\;(\pm0.1\;M_{\odot})$. The fact that the lightest observed first-born NSs in DNS systems have masses $<1.2\;M_{\odot}$ 
means that, in principle, we might even expect a future detection of a DNS merger with a mass ratio $q\ge 1.5$.
Such a DNS merger would produce $r$-process rich ejecta with larger asymmetry and a different direction-dependent composition than those expected for a DNS merger with $q\simeq 1$ \citep{jbp+15};
and possibly, but not necessarily, a larger yield \citep[depending on the NS masses and the NS EoS,][]{jbp+15,llp+16}.  

\subsection{Merger rates}
The Galactic merger rate of DNS systems has been estimated for almost four decades \citep[e.g.][]{cvs79}. 
As we have discussed in the previous sections, the standard formation scenario of DNS binaries involves a number of highly uncertain aspects of binary 
interactions \citep[e.g.][]{vt03,tv06,bkr+08,dbf+12}.
The main uncertainties include, in particular, the treatment of CE evolution \citep{ijc+13,ktl+16}, SN kicks \citep{jan12,jan17} and the efficiency of accretion
and spin-up from mass transfer \citep[e.g.][and this paper]{tlp15}. Together, these processes lead to large uncertainties in the expected merger rates of several orders of magnitude. 
Currently, the DNS merger rate is predicted to be about $1-100\;{\rm Myr}^{-1}$ per Milky~Way Equivalent Galaxy \citep{aaa+10}.

The vast majority\footnote{It is possible that a few merging DNS systems originate from wide binaries in which the NS is shot directly into a tight/eccentric orbit,
i.e. similarly to the direct-SN mechanism discussed by \citet{kal98}.
\citet{vt03} find that this route may account for up to 5\% of all merging DNS systems.}, almost all, of DNS systems which merge within a Hubble time due to GW radiation are produced by ultra-stripped SNe \citep{tlp15}.
And because some relatively wide-orbit DNS systems are also produced via ultra-stripped SNe, the observed rate of ultra-stripped SNe is expected to be larger than the rate of DNS mergers. 
Furthermore, ultra-stripped SNe may also originate in binary
systems with a BH or a WD accretor, and thus the total rate of ultra-stripped SNe is estimated to be one in every 100--1000 SNe \citep{tlm+13}. 

After the recent success in detecting GW signals from merging BHs by
advanced LIGO \citep{aaa+16,aaa+16b,aaa+17}, it is also expected that GWs from merging DNS systems will be detected in the near future with LIGO and other
detectors like VIRGO and KAGRA. Thus, it will soon be possible to obtain DNS merger rate constraints from GW observations. This should place a lower
limit to the rate of ultra-stripped SNe. Comparing the rates of ultra-stripped SNe inferred from high-cadence optical surveys and DNS mergers from GW detectors, we will be able to test
whether the ultra-stripped SN channel is actually the major path for forming DNS systems, and thus constrain binary stellar evolution using GWs.

%%%%%%%%%%%%%%%%%%%%%%%%%%%%%%%%%%%%%%%%%%%%%%%%%%%%%%%%%%%%%%%%%%%%%%%%%%%%%%%%%%%%%%%%%%%
\section{Summary of DNS formation and conclusions}\label{sec:summary}
DNS systems are produced in the Galactic disk from the interaction of two OB-stars in a close binary system.
To produce a DNS system, the minimum ZAMS mass of the primary star must be about $10\pm2\;M_{\odot}$ (depending on metallicity and orbital period), whereas the secondary star in some cases initially can 
have a ZAMS mass as low as $5\pm1\;M_{\odot}$ \citep[e.g.][and K.~Belczynski, priv.~comm.]{ts00,vt03}, depending on its efficiency in accreting 
material from the primary star (which will bring it above the SN threshold mass). 

To form a DNS system, the binary system must survive several phases of mass transfer without coalescing and remain bound after both of the two SN explosions (Fig.~\ref{fig:vdHcartoon}).
Furthermore, to produce a DNS merger (sGRB) the evolution has to lead to a final DNS system with an orbital period short enough (and/or eccentricity large enough) to merge within a Hubble time
as a result of orbital damping from GW radiation.

One obstacle that hinders the formation of DNS systems is that many of the initially close-orbit systems are expected to coalesce early during their evolution \citep[e.g.][]{sdd+12}, or slightly later when the more massive star evolves
off the main sequence, expands, and starts to transfer mass to its companion star \citep{wlb01}. 
To avoid this problem, it has been common practice to model the formation of DNS systems starting from relatively wide-orbit binaries and first let the systems evolve through a stable RLO 
and then, after the formation of the first NS, continue through a CE, following the HMXB phase (see Fig.~\ref{fig:vdHcartoon}). 
This practice has become the standard scenario for modelling the production of DNS systems \citep[][and references therein]{tv06}. However, there are currently
no self-consistent hydrodynamical simulations for successfully modelling the spiral-in of NSs inside a massive envelope, leading to envelope ejection. As a result,  
this leads to large uncertainties in both the estimated number of post-CE systems and their orbital separations \citep{ijc+13,ktl+16}. 
Another issue that is often ignored is that the fate of a massive star in a binary depends on when it loses its hydrogen rich envelope \citep{blb99,bhl+01,wl99}.

As an alternative formation channel, the so-called double-core scenario has been suggested \citep{bro95,dps06}. In this scenario, both of the NS progenitors evolve beyond helium core burning 
before experiencing a CE~phase in which the cores of both stars spiral in, thereby producing a close binary of two evolved helium/carbon-oxygen stars that subsequently collapse to form NSs. 
However, this scenario requires significant fine-tuning since the initial masses of the two stars have to be very close to each other (with an initial ZAMS mass ratio of $0.96 < q < 1$). 

Finally, there is a third formation channel to DNS systems which involves dynamical interactions in dense cluster environments \citep{ps91,gpm06}.
This dynamical channel, however, is expected to be significantly less frequent compared to the standard formation channel of DNS systems. 
Some simulations predict that this channel will account for less than 1\% of all detected DNS mergers \citep{bkl14}. One reason is that dense cluster simulations 
show that the segregation of NSs (and hence the formation of DNS systems via encounters) is not efficient on a Hubble timescale 
--- especially at low metallicities \citep[][see figs.~6 and 7]{ban17}.  
In the following, we continue summarizing our results for the standard formation channel.

The extreme stripping of helium donor stars, via Case~BB RLO in post-HMXB/post-CE systems, prior to the second SN, leads to almost naked metal cores
which will produce EC or Fe~CCSNe with little ejecta mass. \citet{tlm+13} demonstrated that in extreme cases, post-CE mass stripping in a
helium star--NS binary can produce an Fe~CCSN progenitor star with a total mass of only $\sim\! 1.50\;M_{\odot}$ prior to collapse and an envelope mass of barely $0.05\;M_{\odot}$.
The resulting explosion leads to a Type~Ic SN with an ejecta mass in the range $0.05-0.20\;M_{\odot}$.  
Through synthetic light curve calculations, they found that SN~2005ek \citep{dsm+13} is a viable candidate for such an event. 
\citet{tlp15} followed up with systematic calculations of helium star--NS binaries leading to ultra-stripped pre-SN stars with masses of $1.45-3.5\;M_{\odot}$.
They argued that it is expected that the second-born NSs left behind will have gravitational masses in the range $1.1-1.8\;M_{\odot}\;(\pm 0.1\;M_{\odot})$, 
depending on the mass cut during the explosion and the yet unknown EoS of NS matter and its associated release of gravitational binding energy \citep{ly89}. 
\citet{tlp15} also concluded that the observed spectral classes of ultra-stripped SNe can be either Type~Ib or Type~Ic, depending on the amount of helium left 
after the stripping of the envelope, as well as the amount and mixing of synthesized nickel in the ejecta.

From light-curve and spectral modelling of ultra-stripped SNe, \citet{mmt+17} recently argued that, in addition to SN~2005ek, SN~2010X and PTF10iuv are
also good candidates for ultra-stripped SNe. Therefore, each of these SNe could potentially be an event which gave birth to a DNS system. 
Given the current ambitious observational efforts to search for peculiar and weak optical transients,
it seems probable that many more ultra-stripped SNe with diminutive ejecta will be detected within the coming years. 

Our main conclusions from the present investigation of DNS formation can be summarized as follows:
\begin{enumerate}[leftmargin=*]
\item  With a focus on the combination of the latest observational data and recent progress in the physics and qualitative understanding of processes related to the production of DNS systems, 
       we have investigated various aspects of massive binary evolution from the pre-HMXB stages to accretion, CE evolution and SN explosions. A main uncertainty in our formation picture is related to
       the transition from the different subpopulations of HMXBs to DNS systems (Fig.~\ref{fig:corbet}) --- for example, related to their in-spiral in a CE and the spin-up
       of the first-born NS before the second SN explosion.
       We argued that few, if any, of the presently known Galactic HMXBs will survive their subsequent CE evolution and eventually produce a DNS system.
       This result is not in conflict with population statistics given that the active radio lifetime of mildly recycled pulsars in DNS systems is often a factor of $>100$ larger
       than the X-ray lifetime of their HMXB progenitors.

\item From our analysis of five potential phases of accretion onto the NS in HMXB systems (the first-born NS in DNS systems), we conclude that by the time these NSs are 
      observable as mildly recycled radio pulsars they have accreted at most $\sim\!0.02\;M_{\odot}$. This means that their present masses closely resemble their birth masses. 
      This is important for extrapolating the pre-SN stellar properties and probing the explosions leading to the first-born NSs, in comparison to the second-born (young) NSs in DNS systems. 
      The first-born NSs are noted to have larger (birth) masses by an amount of the order $\sim\!0.1\;M_{\odot}$ (Fig.~\ref{fig:NSmass}).
      An explanation for this difference in masses is that in the second SN, the progenitor of the exploding star is being stripped significantly deeper by its NS companion. 

\item The span of precisely measured NS masses in DNS systems is at present $1.17-1.56\;M_{\odot}$ ($\pm 0.01\;M_{\odot}$). Compared to masses of MSPs
      with WD companions \citep{ato+17}, the observed NS mass distribution in DNS systems is lacking the tail of massive NSs ($\ge1.7\;M_{\odot}$). Such masses are, however, 
      under certain circumstances, even possible for the second-born NS (the one often formed in an ultra-stripped SN), depending on the mass of its helium star progenitor \citep{tlp15}. 
      Furthermore, some HMXBs (the potential progenitors of DNS systems) like Vela~X-1 have NS masses of about $1.9-2.1\;M_{\odot}$ \citep{bkv+01,fbl+15}. Therefore, the lack of massive
      NSs in presently known DNS systems might simply be a matter of small number statistics or it could reflect the possibility that sufficiently massive helium stars
      rarely form via this formation channel. A full population synthesis is needed to investigate this further, or we simply await the recorded mass spectrum
      from near future LIGO/VIRGO detections of DNS mergers. 

\item The observed DNS systems show a clear trend between $P_{\rm orb}$ and $P$ of the first-born (recycled) pulsars (Figs.~\ref{fig:corbet} and \ref{fig:PorbP}). 
      We find that the empirical data can approximately be fitted to the following birth relation: $P\approx 36\pm14\;{\rm ms}\;(P_{\rm orb}{\rm /days})^{0.40\pm0.10}$ (Eq.~\ref{eq:PspinPorb}), 
      although the spread in data points is large, especially for small $P_{\rm orb}$. 
      The fact that a correlation is present at all is another indication that kicks imparted on the second-born NSs cannot be too large in general.
      To further test the validity of such a correlation it is important to find more wide-orbit DNS systems like 
      PSR~J1930$-$1852 \citep{srm+15} which has $P_{\rm orb}=45\;{\rm days}$, by far the widest DNS system known.
      As expected from binary stellar evolution, this radio pulsar has the slowest spin period (185~ms) which proves its past history with only marginal recycling \citep{tlp15}. 

\item Using binary stellar evolution arguments, we find that general correlations between $P_{\rm orb}$, $P$ and eccentricity should, in principle, be present in DNS systems 
      in the case of {\it symmetric} SNe and for systems evolving from similar helium star donors.
      We calculated such correlations on the basis of previous theoretical modelling \citep{tlp15} of progenitors of ultra-stripped SNe.
      However, the combination of different masses of the exploding stars and, in particular, the addition of (even small) kicks in random directions will easily tend to
      camouflage any such correlations (cf. Figs.~\ref{fig:PorbP-theory2}, \ref{fig:ecc} and \ref{fig:ecc_spin}), although some trends are still present
      in the observational data.
      
\item We discussed kicks added to newborn NSs and revisited the arguments to distinguish between the following three cases: i) isolated NSs (or very-wide orbit NS binaries),
      ii) the first SN explosion in a close binary from a stripped star, and finally, iii) the second SN explosion in a close binary from an ultra-stripped star.
      We find that the following picture is emerging where  
      the first case often seems to produce large kicks (on average, $\left<w\right>\simeq 400-500\;{\rm km\,s}^{-1}$), whereas the first SN in a close binary (leading to HMXBs) results in significantly
      smaller average kicks, and the second SN (ultra-stripped SNe) is often, but not always, accompanied by kicks of even less than $50\;{\rm km\,s}^{-1}$.
      Indeed, the majority of the observed DNS systems can best be explained with small kicks in the second SN \citep[e.g.][]{bp16}, although in a couple of cases a larger kick ($200-400\;{\rm km\,s}^{-1}$)
      must have been at work \citep[see also][]{fk97,wkk00}. The explanation for this diversity in kick magnitudes among DNS systems could either reflect some stochasticity
      involved in the development of the explosion asymmetries \citep{wjm13}, or be related to the mass of the metal core of the exploding star (see below).

\item From current DNS data, we find that all second-born NSs with a mass $\ga 1.33\;M_{\odot}$ seem to be associated with rather large kicks, whereas 
      below this mass limit they can be related to small kick events ($w\la 50\;{\rm km\,s}^{-1}$; see Table~\ref{table:kick_threshold} and 
      Fig.~\ref{fig:kick_mass}). We have presented theoretical arguments in favour of such a possible correlation between NS mass and 
      kick magnitude (Section~\ref{subsec:kick_mass}). 
      The nature of the SNe leading to the latter (low-mass, small kick) NSs is either EC~SNe or Fe~CCSNe from stars with small iron cores. Given the somewhat narrow mass window for 
      producing EC~SNe \citep{plp+04}, it is likely that Fe~CCSNe dominate the formation rate of low-mass NSs.
      Detailed population synthesis studies are needed to answer this (Kruckow~et~al.~2017, in prep.). 

\item Based on observational constraints of a few key parameters (orbital periods, eccentricities, NS masses, systemic velocities and misalignment angles) 
      we have simulated the kinematic effects of the second SN in all observed DNS systems. As a result, we have been able to probe the solutions to place limits
      on e.g. the masses of the exploding stars and the kick velocity distribution. For DNS systems which are well constrained from observations,
      e.g. PSR~J0737$-$3039, we are able to find an almost {\it unique} solution for the pre-SN progenitor system.
      We find that for this particular system, the second-born NS (pulsar~B) originates from an exploding star with a mass of $1.59\pm 0.09\;M_{\odot}$ 
      and with an orbital period of $0.10\pm0.02\;{\rm days}$. A similar small progenitor mass was derived by \citet{ps05} based on kinematic arguments.
      Binary stellar evolution models by \citet{tlm+13} have confirmed that indeed such small progenitor star masses (down to $\sim\!1.50\;M_{\odot}$) are possible from ultra stripping via Case~BB RLO.
      For PSR~J1756$-$2251 we can constrain the mass of the second exploding star to be $1.72\pm 0.22\;M_{\odot}$. 
      However, for most of the other DNS systems we can only produce a possible range of the solution parameters. The reason for this is large differences in the
      outcome of SNe due to kicks with different directions. 
      In particular, it would be desirable to measure more transverse velocities (i.e. proper motions and distances) and misalignment angles to help constrain these systems better.
      The inferred transverse velocities still suffer from (in some cases relatively large) uncertainties in the DM distance estimates. 
      Improved distances of nearby DNS systems can also be derived from future determinations of $\dot{P}_{\rm orb}$. 

\item One outcome of our SN simulations reproducing the observed DNS systems is an apparent preference for a NS kick direction 
      out of the orbital plane of the pre-SN binary (Fig.~\ref{fig:angle_dist}). However, we argue that this anisotropy is caused by a combination of considering
      all valid kinematic solutions, irrespective of any physical limits on the applied kick magnitudes of ultra-stripped SNe in general, and an observational selection bias in favour of
      low-eccentricity DNS systems.
      Further investigations and statistical analysis on this topic are needed. 

\item In the following decade, the completion of the SKA and the newly constructed FAST radio telescopes are anticipated to boost the number of known DNS systems 
      --- for the SKA possibly by a factor of 5 to 10 \citep{kbk+15}. Such an increase in sources will greatly improve the population statistics and thus our knowledge of DNS formation.
      SKA will also improve the pulsar timing significantly, which is important for deriving NS masses, distances and proper motions. 
      In addition, and also from near-future observations, the rate of ultra-stripped SNe can be compared to the rate at which GW detectors (LIGO, VIRGO, KAGRA, IndIGO) will detect  
      DNS mergers. This will help us to verify, or falsify, our hypothesis that ultra-stripped SNe occur in (almost) all DNS mergers originally produced in the Galactic disk.
      Finally, it will be interesting to compare the mass distribution of DNS mergers from future LIGO observations with our theoretical predictions. 
\end{enumerate}

\hspace{0.3cm}
\acknowledgements
We thank the organizers of and all participants in the MIAPP 2015 workshop on {\it The Many Faces of Neutron Stars}
for many stimulating discussions.
TMT and MUK acknowledge financial support by the Deutsche Forschungsgemeinschaft (DFG) Grant No.\ TA 964/1-1.
At Garching, the work was supported by the DFG through the Excellence Cluster ``Universe'' EXC 153 and by the
European Research Council through grant ERC-AdG No.\ 341157-COCO2CASA.
 
\bibliographystyle{apj}
\bibliography{tauris_refs}
%\end{document}
%%%%%%%%%%%%%%%%%%%%%%%%%%%%%%%%%%%%%%%%%%%%%%%%%%%%%%%%%%%%%%%%%%%%%%%%%%%%%%%%
\newpage
\appendix
\section{Simulated second SN of individual DNS systems} \label{sec:appendix}

\begin{figure}
 \centering
 \includegraphics[width=0.87\columnwidth,angle=0]{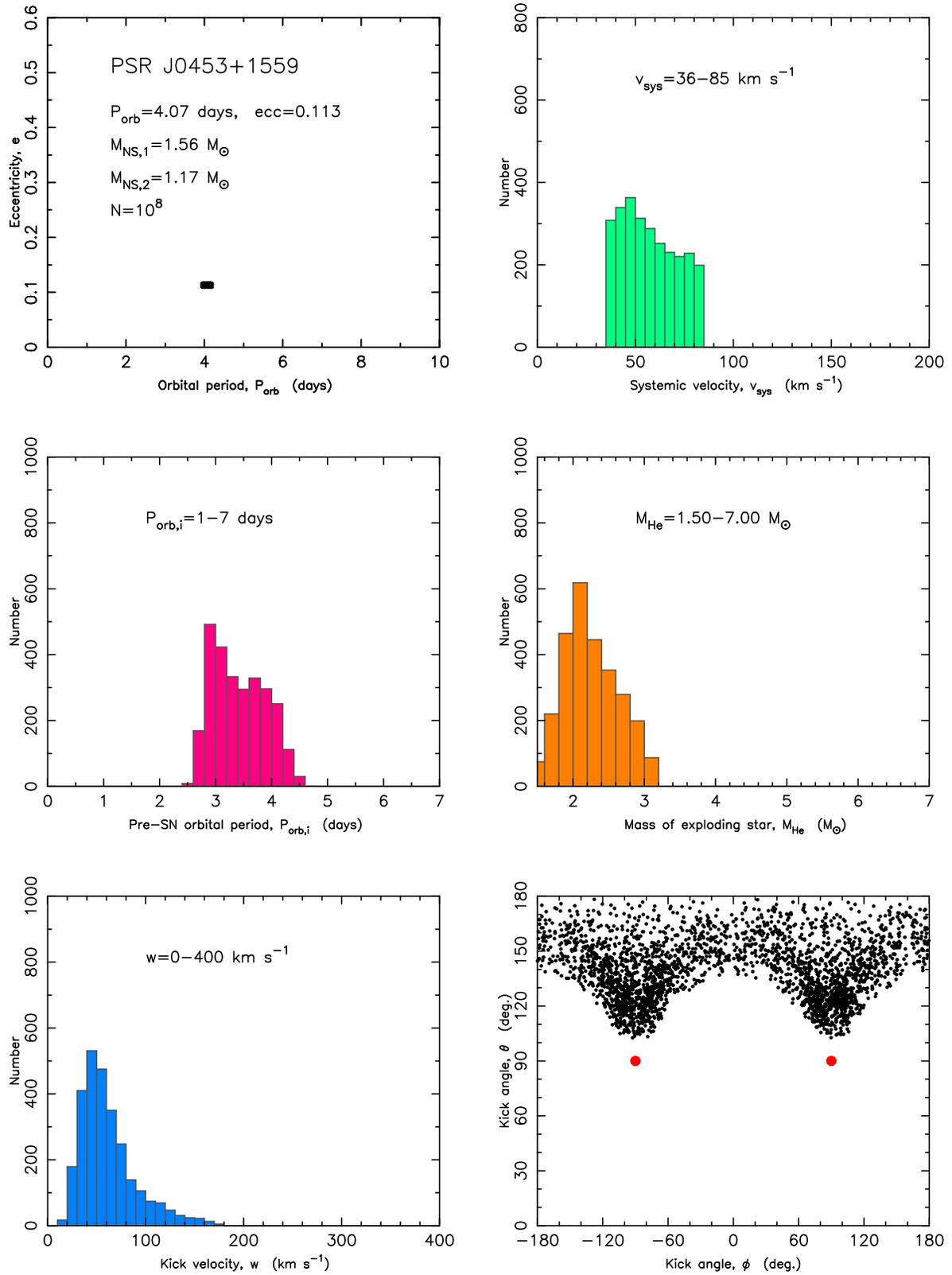}
 \caption{Properties and constraints on the formation of PSR~J0453+1559. See Table~\ref{table:DNS} and discussions in Section~\ref{subsec:0453}.}
 \label{fig:app0453}
\end{figure}

\begin{figure}
 \centering
 \includegraphics[width=0.87\columnwidth,angle=0]{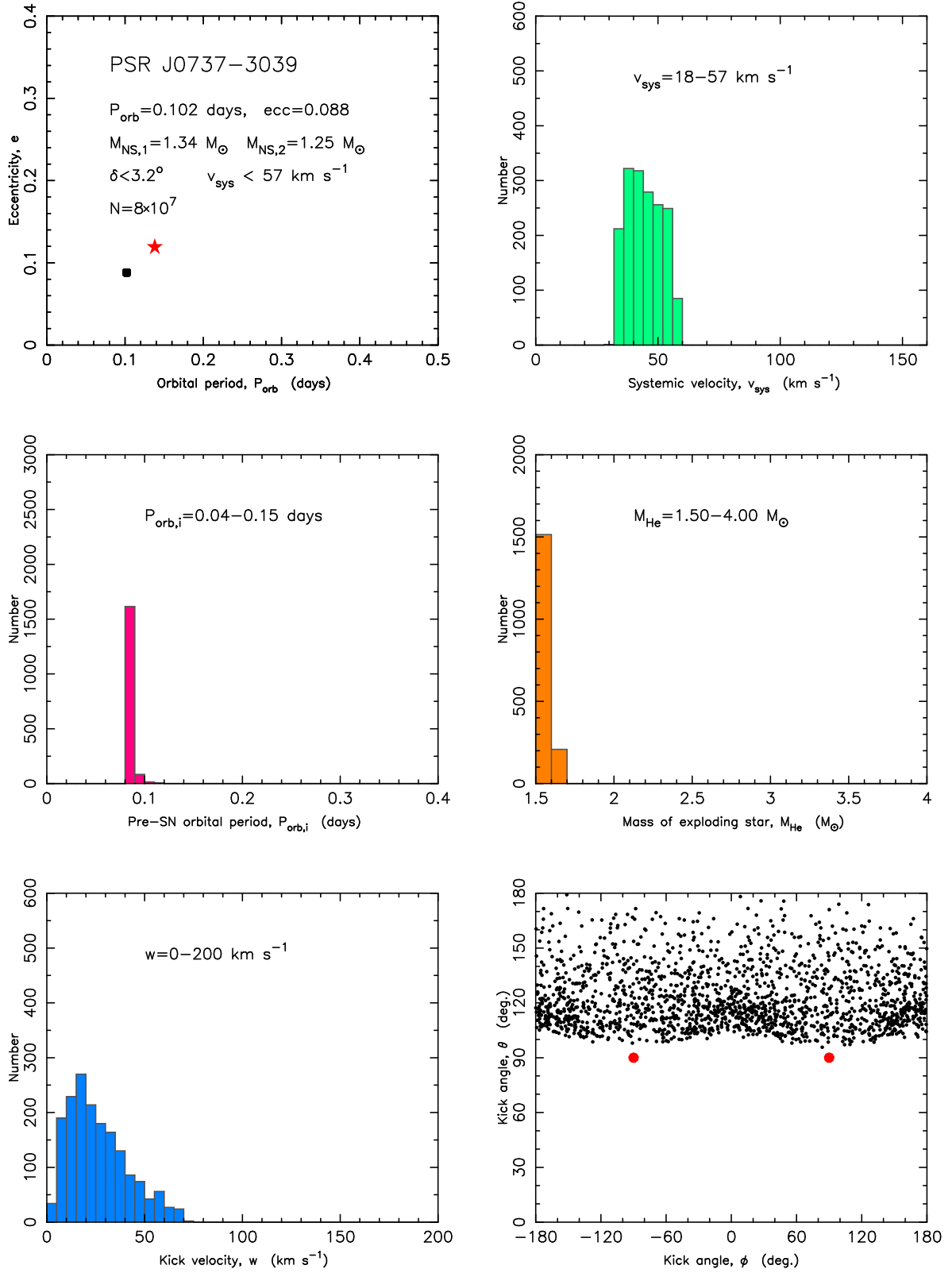}
 \caption{Properties and constraints on the formation of PSR~J0737$-$3039. See Table~\ref{table:DNS} and discussions in Section~\ref{subsec:0737}. 
          The red star shows the initial location of this DNS system if it formed 100~Myr ago.} 
 \label{fig:app0737}
\end{figure}

\begin{figure}
 \centering
 \includegraphics[width=0.87\columnwidth,angle=0]{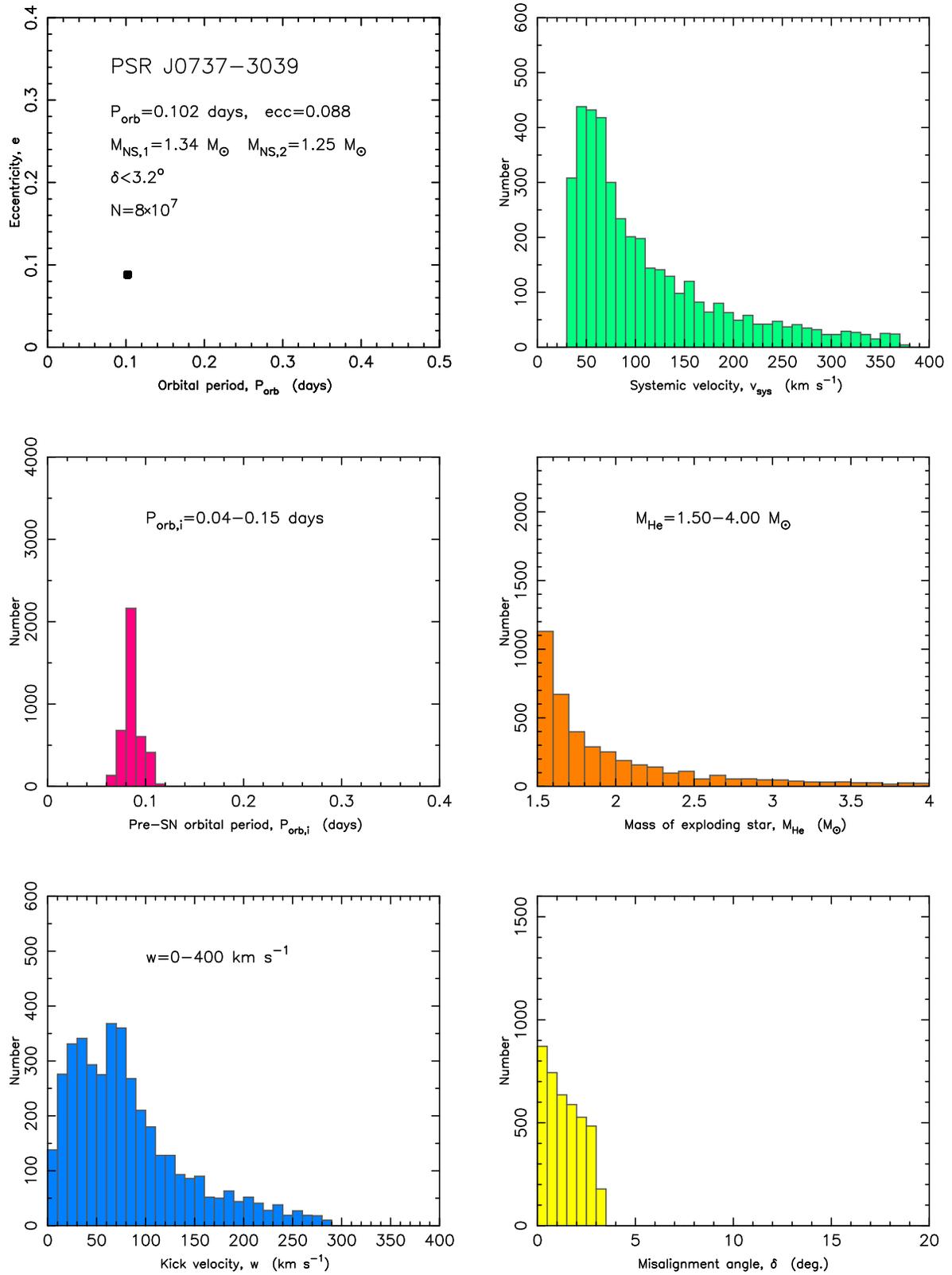}
 \caption{Properties and constraints on the formation of PSR~J0737$-$3039. See Table~\ref{table:DNS} and discussions in Section~\ref{subsec:0737}.
          In this simulation, the constraints on the systemic velocity have been ignored. The lower-right panel displays the distribution of misalignment angles.}
 \label{fig:app0737_2}
\end{figure}

\begin{figure}
 \centering
 \includegraphics[width=0.87\columnwidth,angle=0]{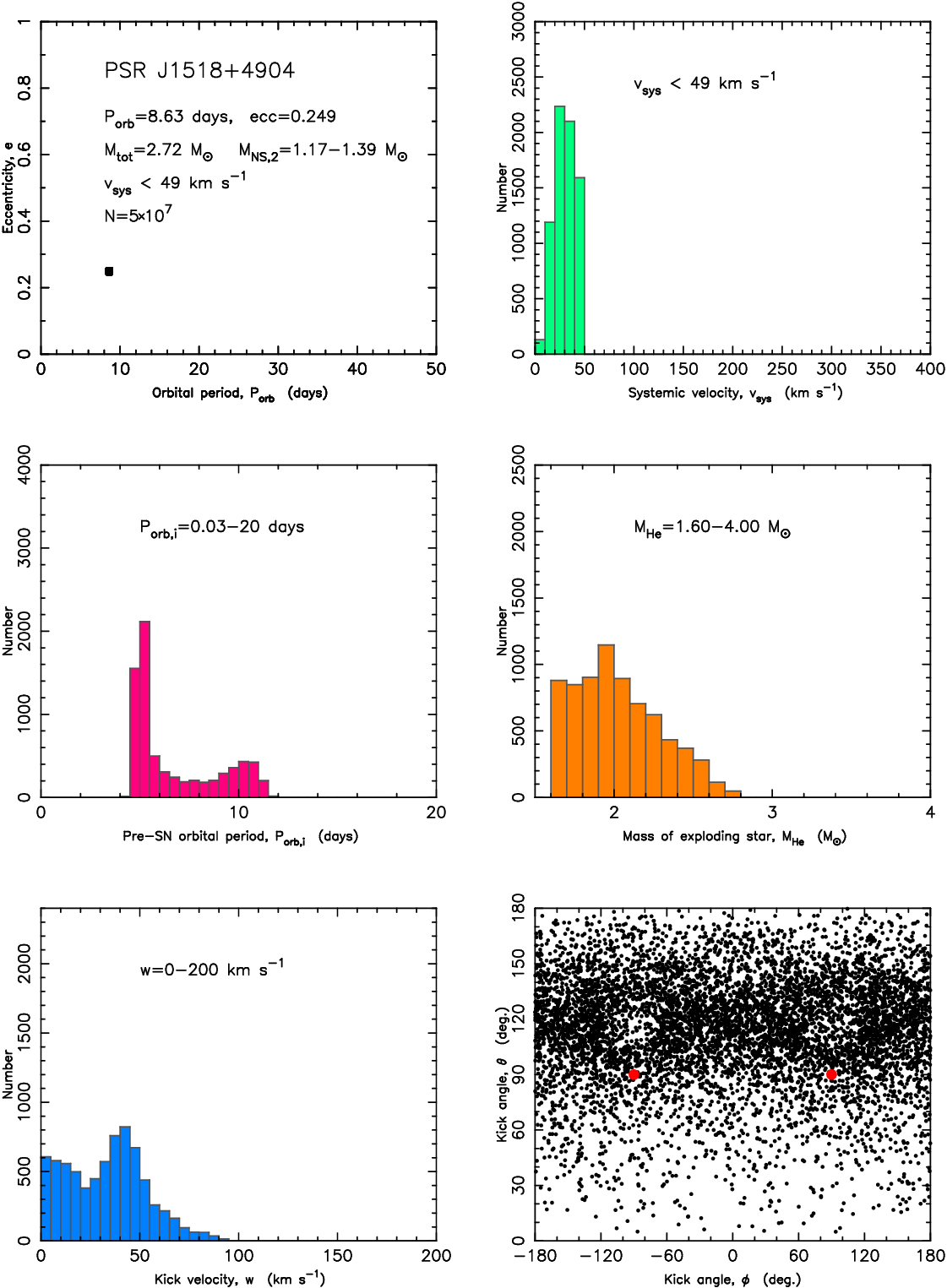}
 \caption{Properties and constraints on the formation of PSR~J1518+4904. See Table~\ref{table:DNS} and discussions in Section~\ref{subsec:1518}.}
 \label{fig:app1518}
\end{figure}

\begin{figure}
 \centering
 \includegraphics[width=0.87\columnwidth,angle=0]{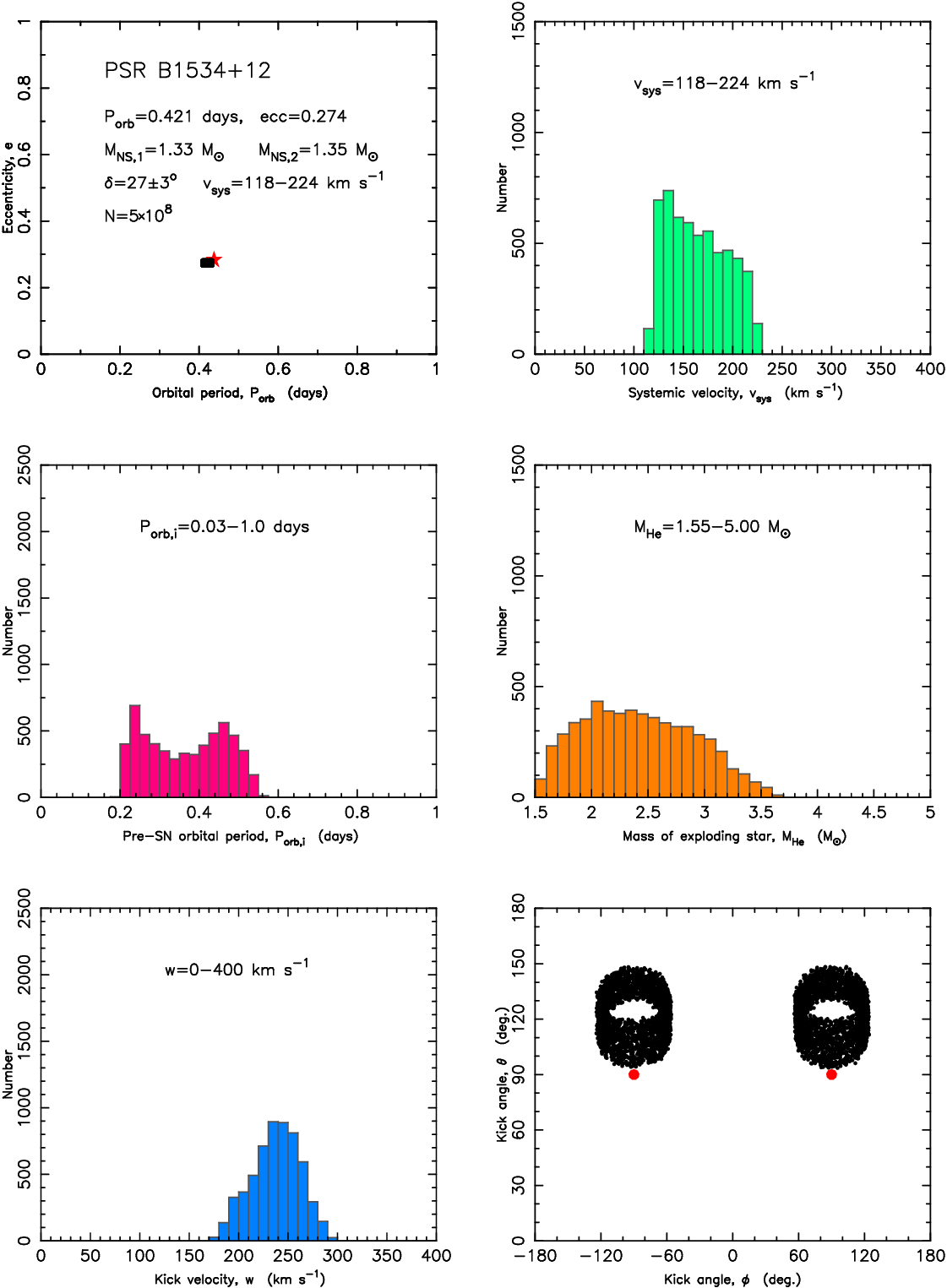}
 \caption{Properties and constraints on the formation of PSR~B1534+12. See Table~\ref{table:DNS} and discussions in Section~\ref{subsec:1534}.
          The red star shows the initial location of this DNS system for an age of $t=\tau=248\;{\rm Myr}$.}
 \label{fig:app1534}
\end{figure}

\begin{figure}
 \centering
 \includegraphics[width=0.87\columnwidth,angle=0]{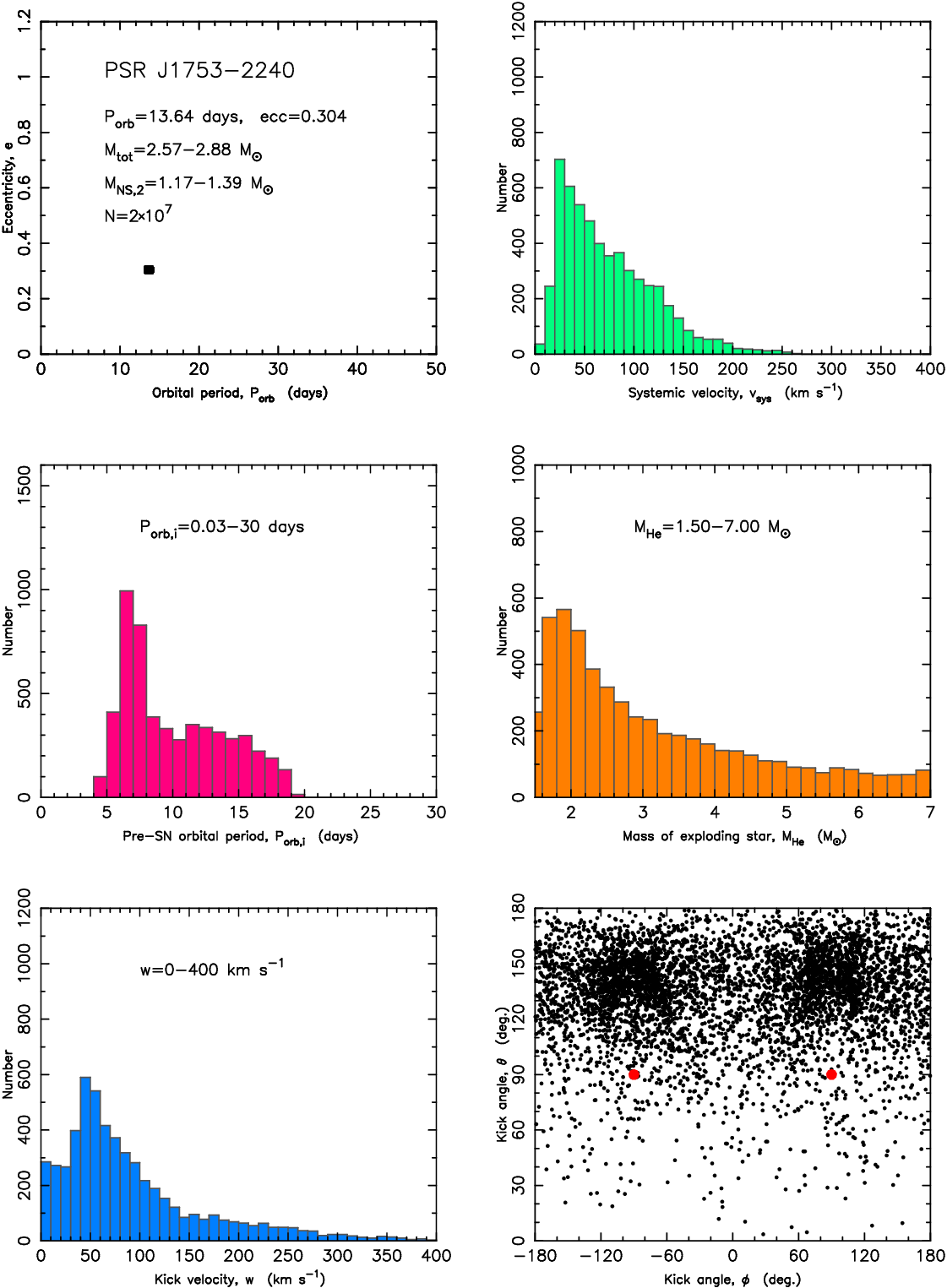}
 \caption{Properties and constraints on the formation of PSR~J1753$-$2240. See Table~\ref{table:DNS} and discussions in Section~\ref{subsec:1753}.}
 \label{fig:app1753}
\end{figure}

\begin{figure}
 \centering
 \includegraphics[width=0.87\columnwidth,angle=0]{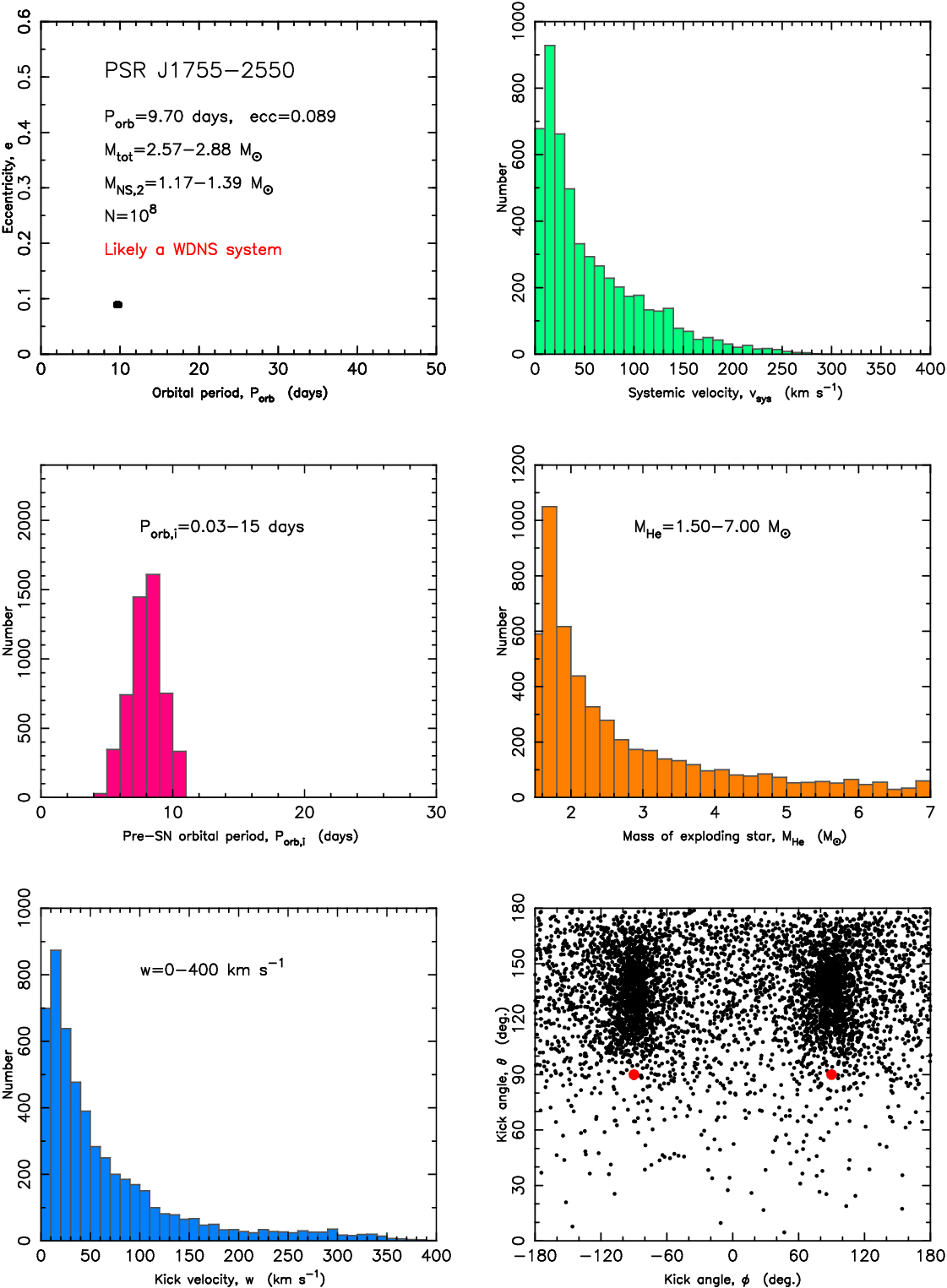}
 \caption{Properties and constraints on the formation of PSR~J1755$-$2550. See Table~\ref{table:DNS} and discussions in Section~\ref{subsec:1755}.}
 \label{fig:app1755}
\end{figure}

\begin{figure}
 \centering
 \includegraphics[width=0.87\columnwidth,angle=0]{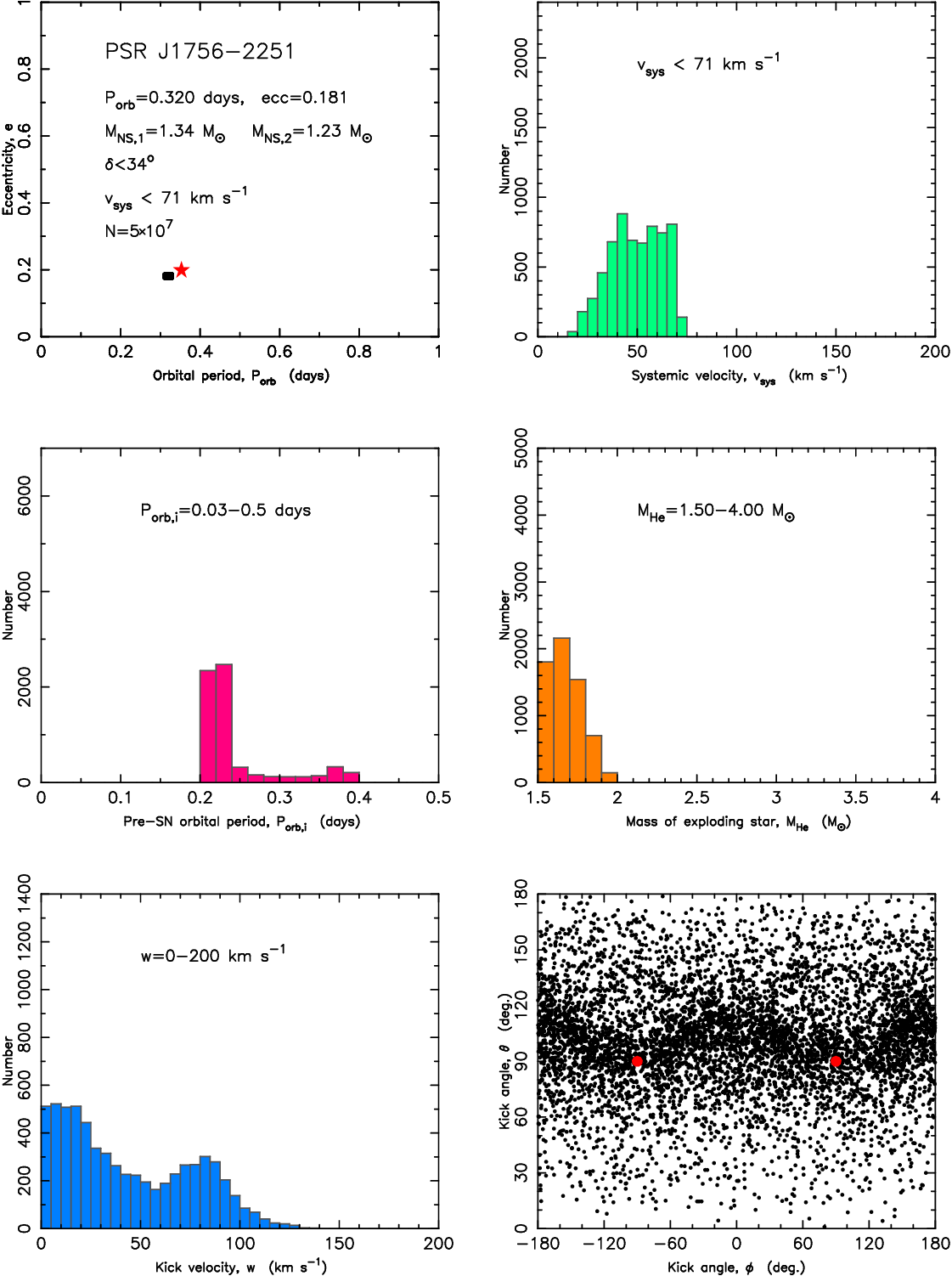}
 \caption{Properties and constraints on the formation of PSR~J1756$-$2251. See Table~\ref{table:DNS} and discussions in Section~\ref{subsec:1756}. 
          The red star shows the initial location of this DNS system for an age of $t=\tau=443\;{\rm Myr}$.}
 \label{fig:app1756}
\end{figure}

\begin{figure}
 \centering
 \includegraphics[width=0.87\columnwidth,angle=0]{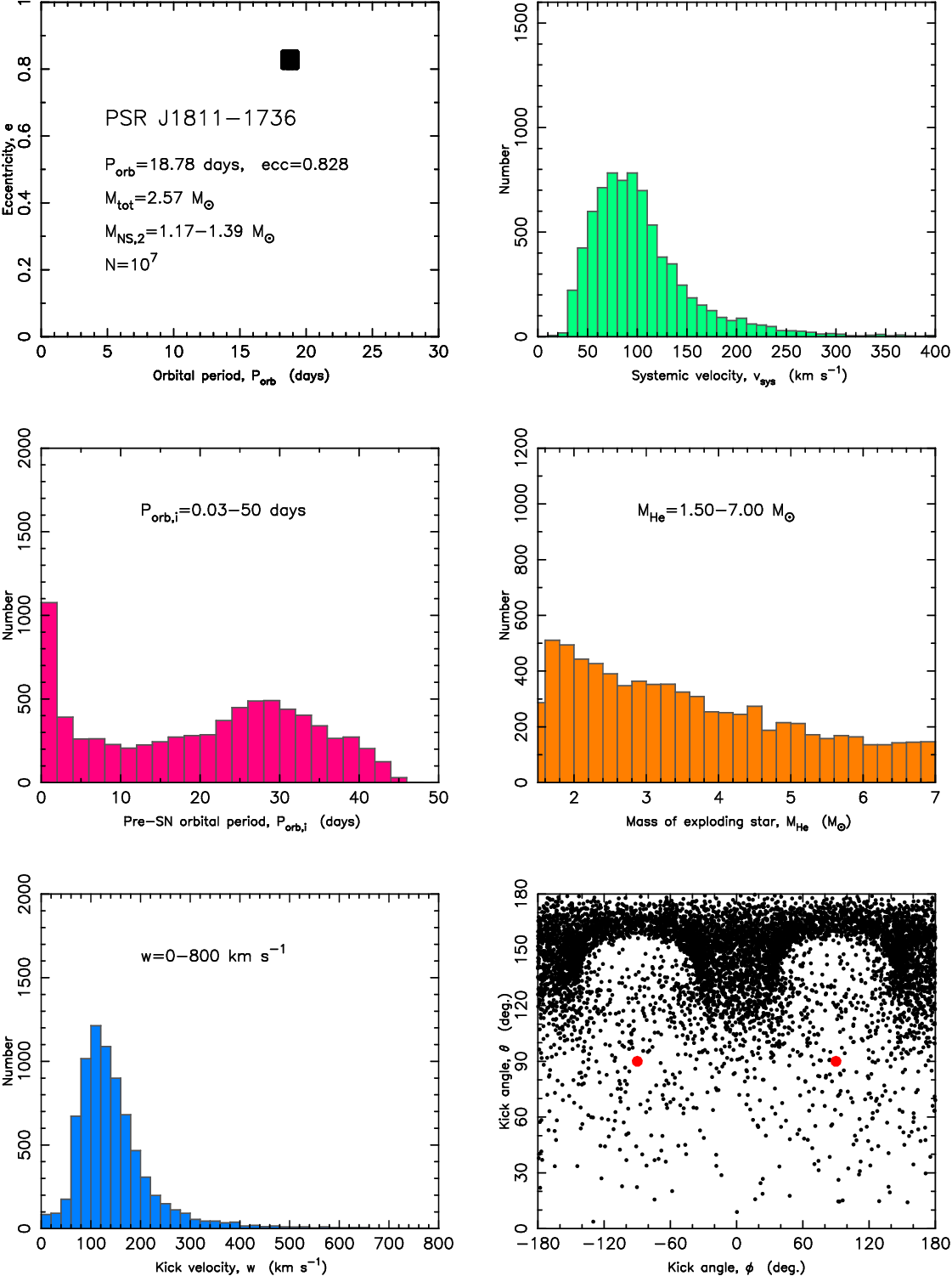}
 \caption{Properties and constraints on the formation of PSR~J1811$-$1736. See Table~\ref{table:DNS} and discussions in Section~\ref{subsec:1811}.}
 \label{fig:app1811}
\end{figure}

\begin{figure}
 \centering
 \includegraphics[width=0.87\columnwidth,angle=0]{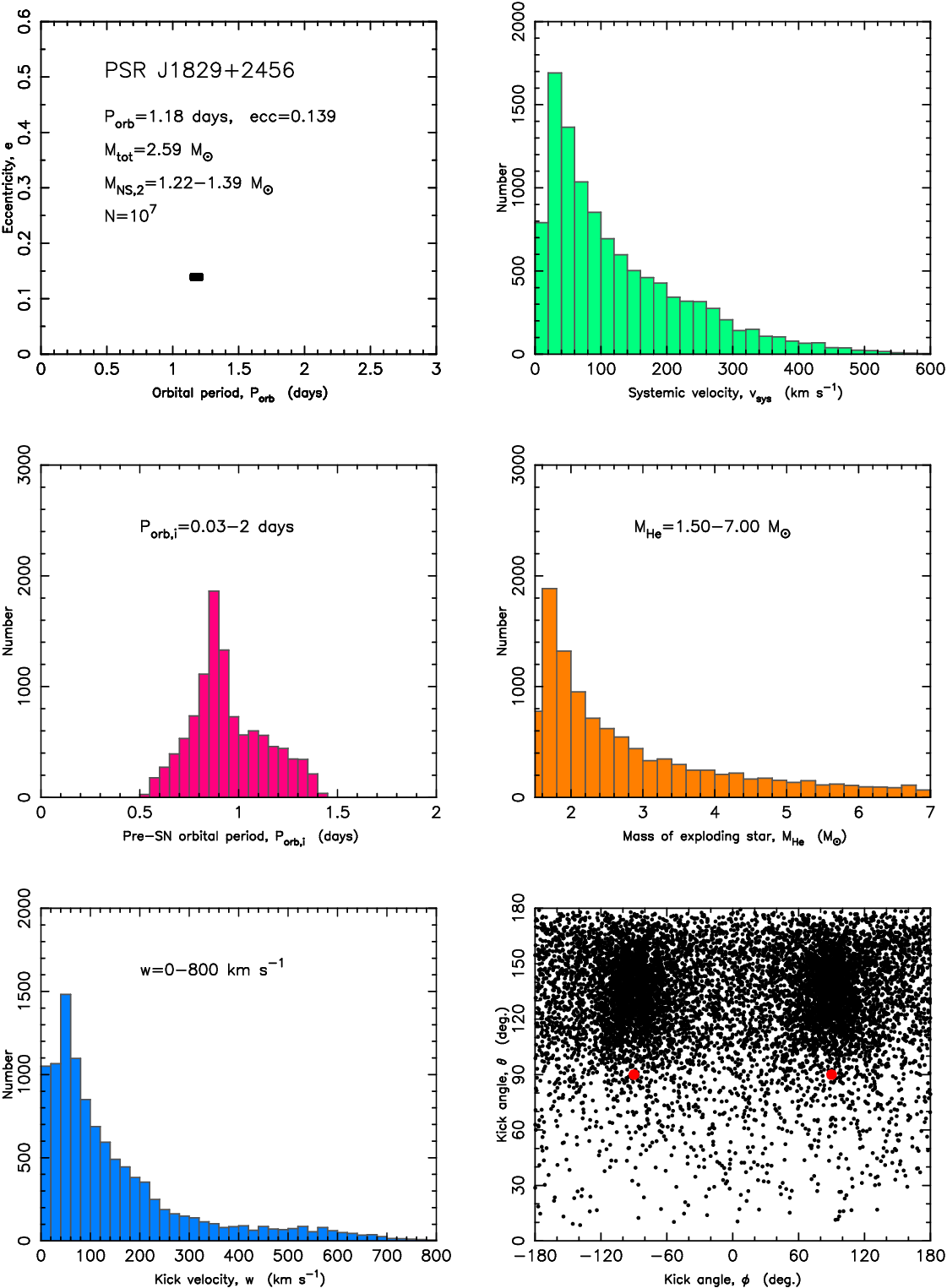}
 \caption{Properties and constraints on the formation of PSR~J1829+2456. See Table~\ref{table:DNS} and discussions in Section~\ref{subsec:1829}.}
 \label{fig:app1829}
\end{figure}

\begin{figure}
 \centering
 \includegraphics[width=0.87\columnwidth,angle=0]{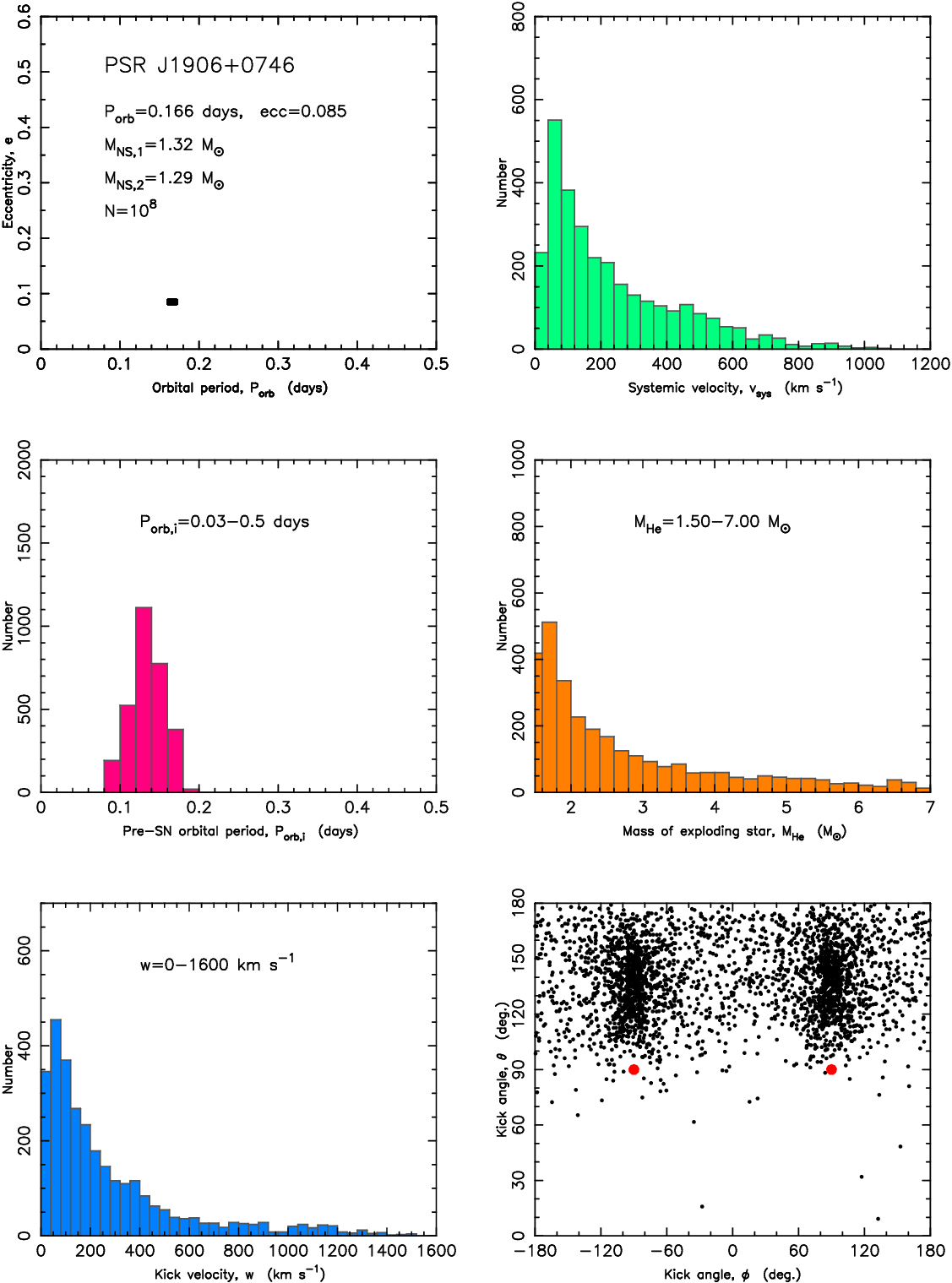}
 \caption{Properties and constraints on the formation of PSR~J1906+0746. See Table~\ref{table:DNS} and discussions in Section~\ref{subsec:1906}.}
 \label{fig:app1906}
\end{figure}

\begin{figure}
 \centering
 \includegraphics[width=0.87\columnwidth,angle=0]{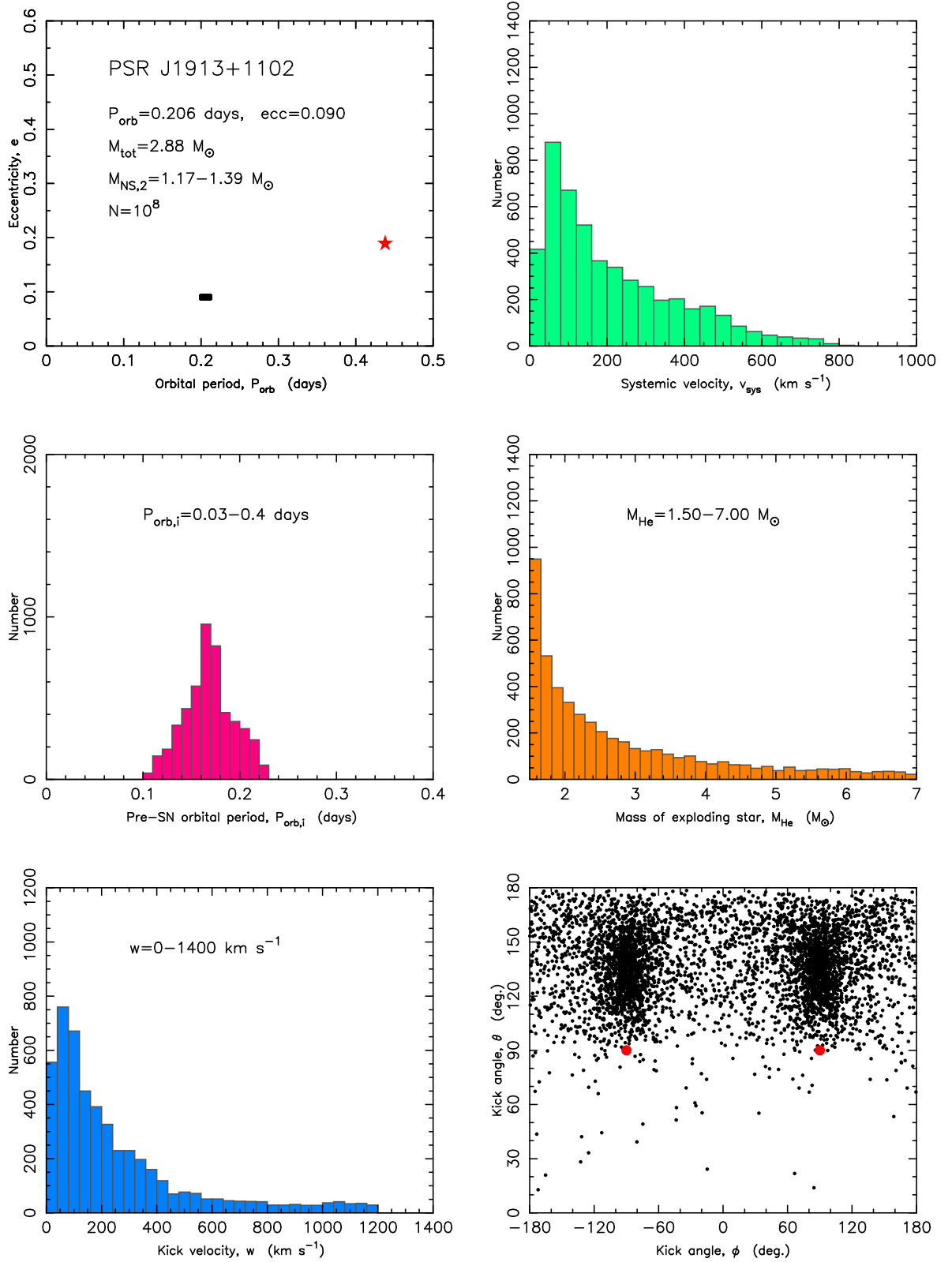}
 \caption{Properties and constraints on the formation of PSR~J1913+1102. See Table~\ref{table:DNS} and discussions in Section~\ref{subsec:1913+11}.
          The red star shows the initial location of this DNS system for an age of $t=\tau=2690\;{\rm Myr}$.}
 \label{fig:app1913+1102}
\end{figure}

\begin{figure}
 \centering
 \includegraphics[width=0.87\columnwidth,angle=0]{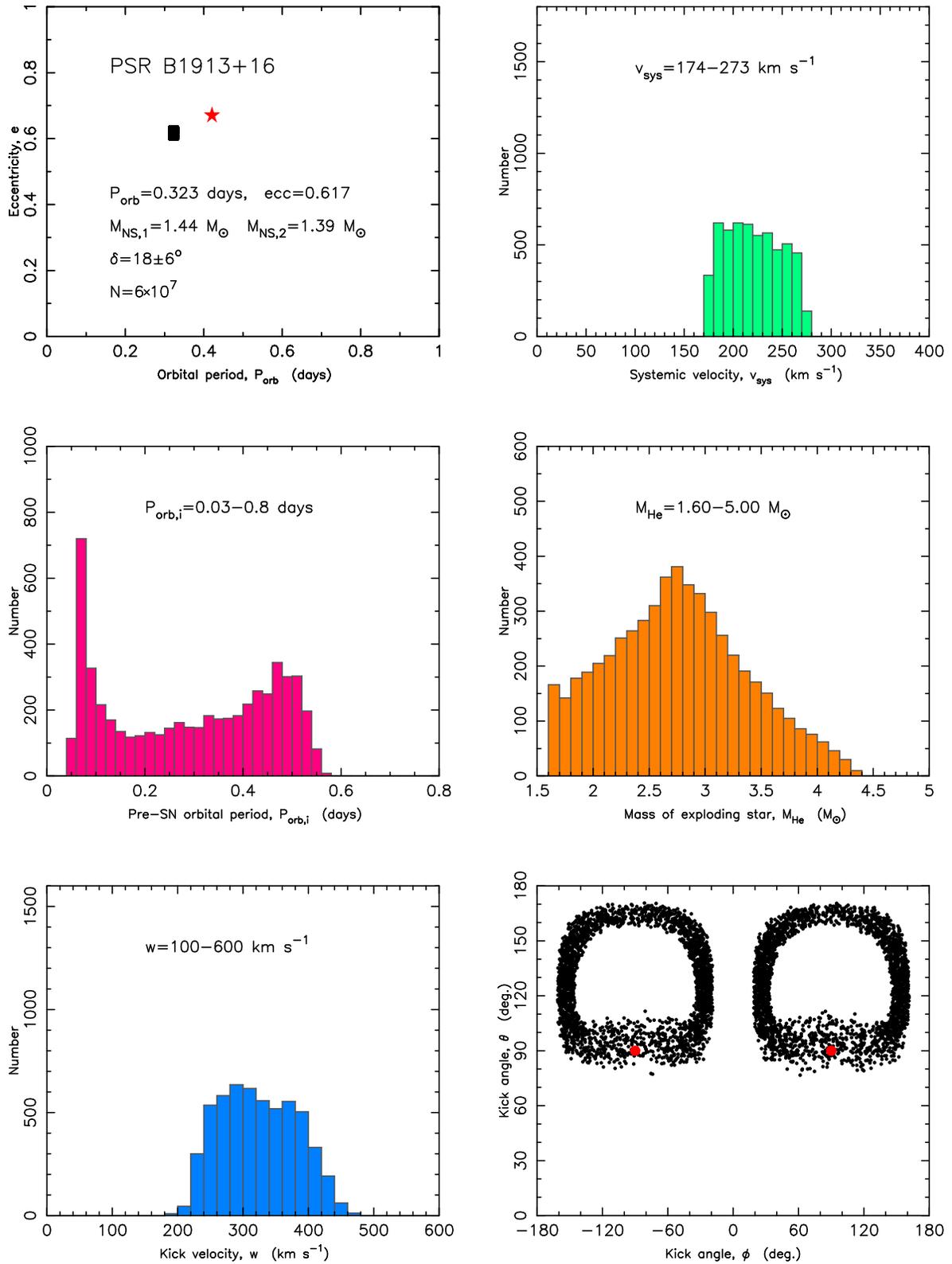}
 \caption{Properties and constraints on the formation of PSR~B1913+16. See Table~\ref{table:DNS} and discussions in Section~\ref{subsec:1913+16}.
          The red star shows the initial location of this DNS system after its formation, see Section~\ref{sec:mapping} for discussions and the 
          simulations in Fig.~\ref{fig:app1913_birth}.}
 \label{fig:app1913}
\end{figure}

\begin{figure}
 \centering
 \includegraphics[width=0.87\columnwidth,angle=0]{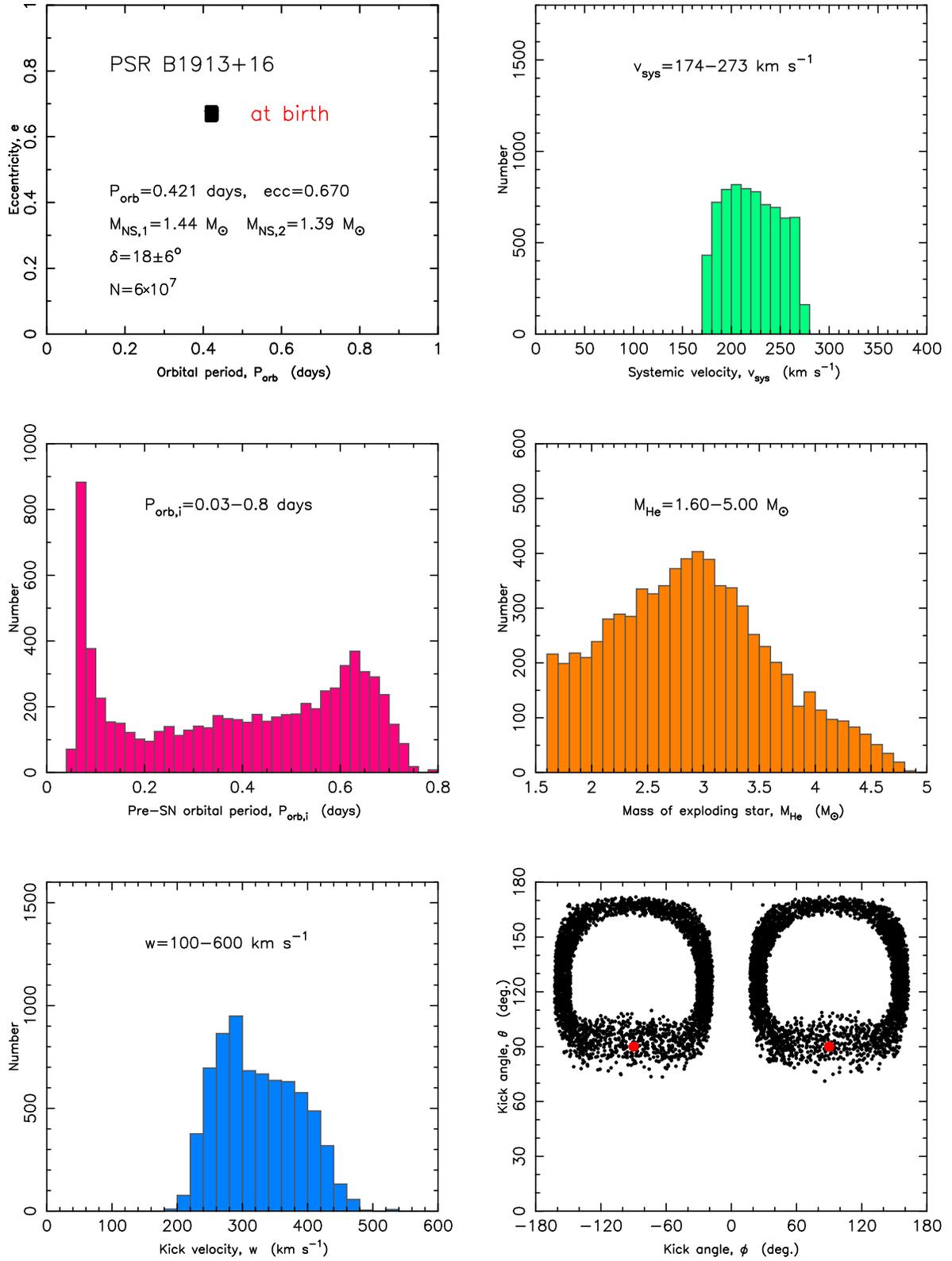}
 \caption{Properties and constraints on the formation of PSR~B1913+16, assuming the system formed 108.7~Myr ago. This is its maximum age if assuming evolution with a braking index of $n=3$.}
 \label{fig:app1913_birth}
\end{figure}

\begin{figure}
 \centering
 \includegraphics[width=0.87\columnwidth,angle=0]{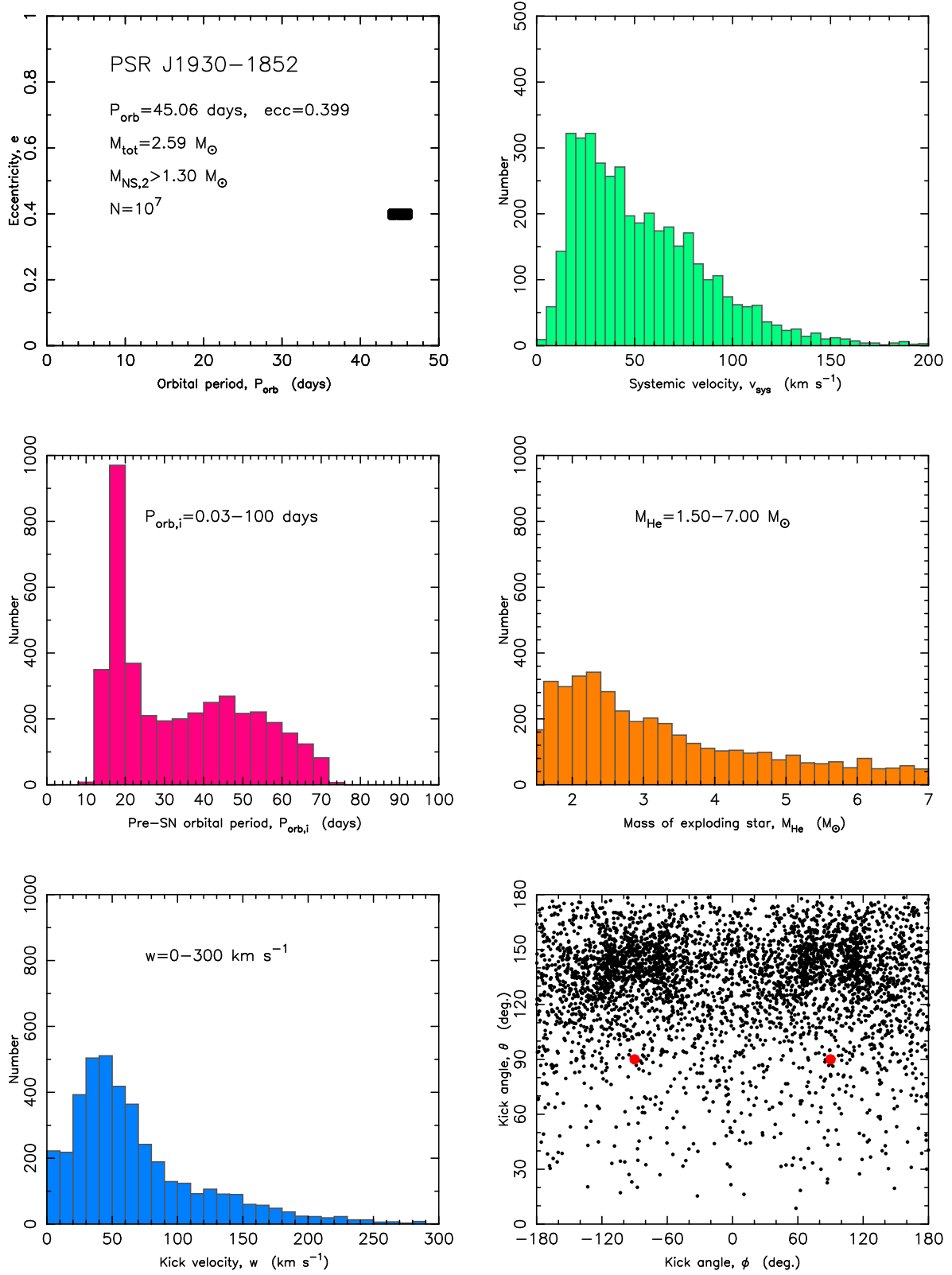}
 \caption{Properties and constraints on the formation of PSR~J1930$-$1852. See Table~\ref{table:DNS} and discussions in Section~\ref{subsec:1930}.}
 \label{fig:app1930}
\end{figure}

\begin{figure}
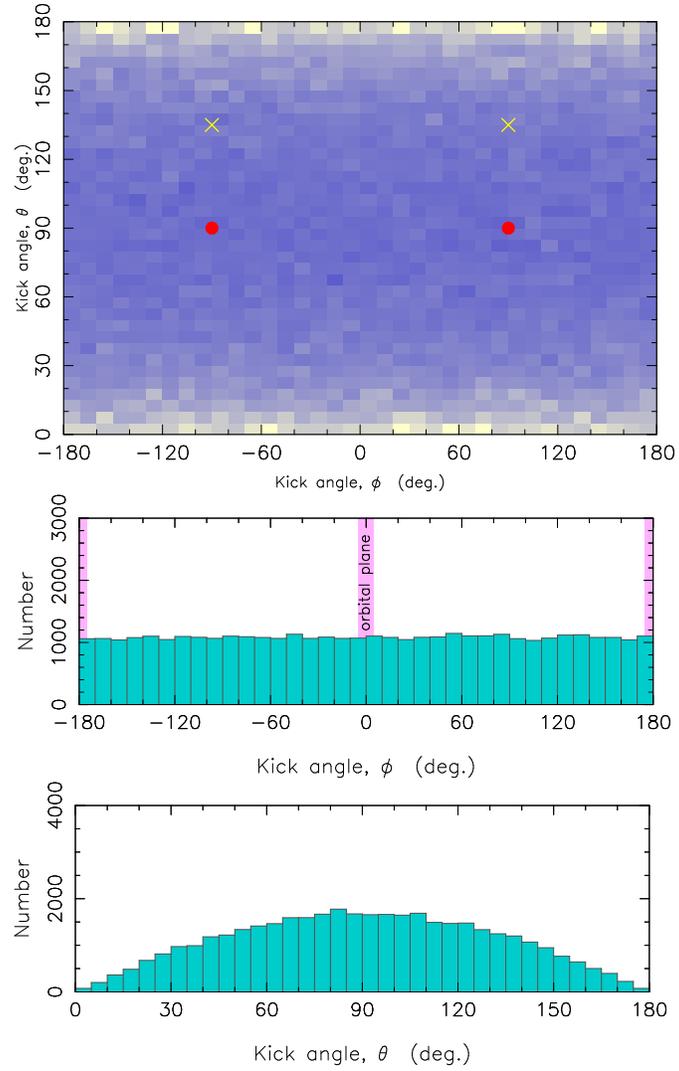

  \begin{center}
     \includegraphics[width=0.36\columnwidth, angle=-90]{Figures/angle_dist_iso.ps}\\
     \includegraphics[width=0.21\columnwidth, angle=-90]{Figures/angle_dist_histo_iso.ps}\\
     \includegraphics[width=0.21\columnwidth, angle=-90]{Figures/angle_dist_histo_theta_iso.ps}
    \vspace{0.2cm}
    \caption{Distribution of the trial kick velocity directions which are used for our SN simulations shown in Fig.~\ref{fig:angle_dist}.
             This plot verifies the random directions of input kick velocities (i.e. isotropy on the kick sphere).
             To obtain isotropy, the probability for the directions of $\theta$ and $\phi$ are weighted by $P(\theta)=\sin(\theta)$ and $P(\phi)={\rm constant}$, respectively. 
             The statistics of this plot is based on a total of 39\,000 SNe (similar number as in Fig.~\ref{fig:angle_dist} for simulations of 13 observed DNS systems). 
             The resulting statistical noise is very minor.}
  \label{fig:angle_dist_iso}
  \end{center}
\end{figure}

\begin{figure}[t]
 \centering
\includegraphics[width=0.76\columnwidth,angle=0]{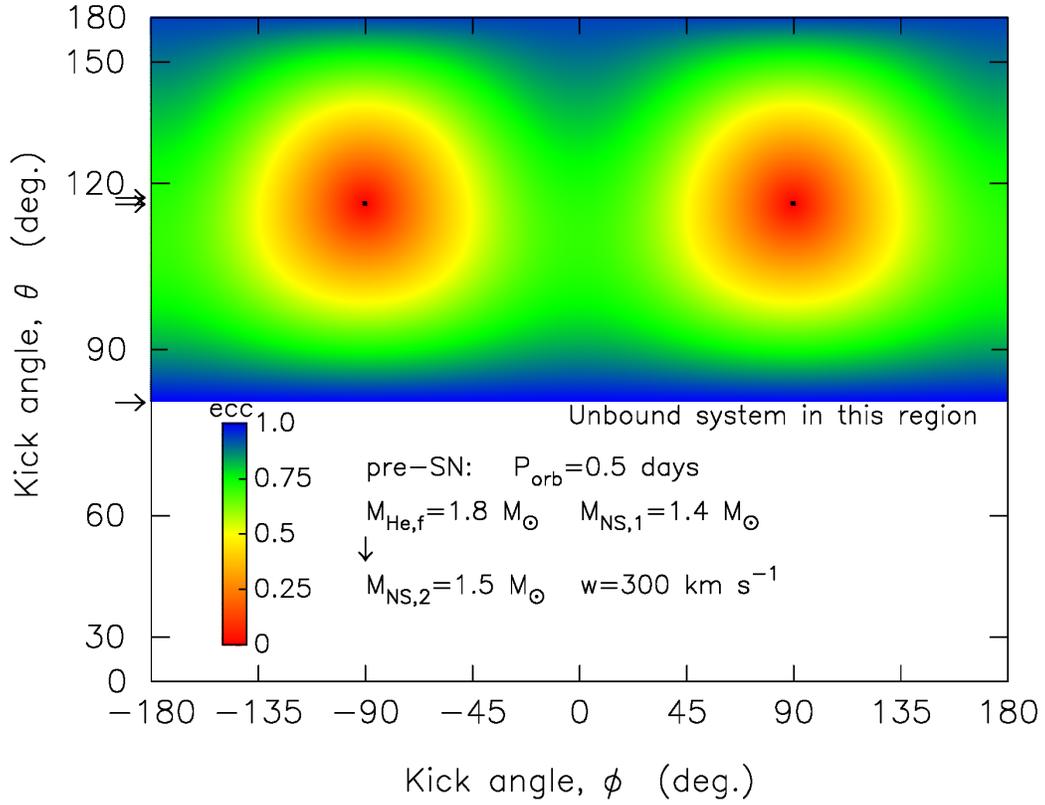}
 \caption{Dependence of the post-SN eccentricity on the direction of the kick for an ultra-stripped SN with a large kick of $w=300\;{\rm km\,s}^{-1}$ (See Fig.~\ref{fig:ecc_angle} for a comparison). 
          The assumed pre-SN parameters are: $P_{\rm orb,i}=0.5\;{\rm days}$, $M_{\rm He,f}=1.80\;M_{\odot}$ and the 
          companion star (first-born NS) is assumed to have a mass of $M_{\rm NS,1}=1.40\;M_{\odot}$.
          The resulting NS is produced with a mass of $M_{\rm NS,2}=1.50\;M_{\odot}$.
          Systems with $\theta \rightarrow 180^{\circ}$ will have very tight and eccentric post-SN orbits, and some of them
          will merge in less than 1~Myr as a result of GW radiation.} 
 \label{fig:ecc_angle2}
\end{figure}

\end{document}